%% file: IJMF.tex
\documentclass[3p,authoryear,preprint]{elsarticle}
\input{macros.tex}


\input{structure.tex}

\usepackage{import} %
\usepackage[]{siunitx} 
\usepackage{lipsum} 
\usepackage{fancyhdr}
\fancypagestyle{firststyle}
{
   \fancyhf{}
   \fancyfoot[L]{
    \footnotesize 
 This work is licensed under a Creative Commons Attribution-NonCommercial-NoDerivatives 4.0 International License.
      \includegraphics[width=0.2\textwidth,keepaspectratio]{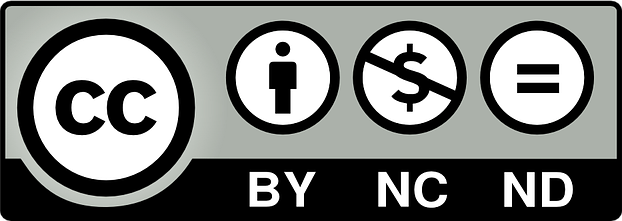}}
}

\newcommand\ddfrac[2]{\frac{\displaystyle #1}{\displaystyle #2}}
\newcommand\gommettes{0.30\textwidth}
\newcommand\timeseriewidth{0.08\textwidth}

\def\graphpath{./}
\graphicspath{{\graphpath}}

\usepackage{placeins}
\let\Oldsection\section
\renewcommand{\section}{\FloatBarrier\Oldsection}
\begin{document}
\begin{frontmatter}
\date{}

\title{Influence of the ambient pressure on the liquid accumulation and on the primary spray in prefilming airblast atomization}
\author[aff1]{G.~Chaussonnet}
\ead{geoffroy.chaussonnet@kit.edu}
\author[aff1]{S.~Gepperth}
\author[aff1]{S.~Holz}
\author[aff1]{R.~Koch}
\author[aff1]{H.-J.~Bauer}
\address[aff1]{Karlsruher Institut f{\"u}r Technolgie - Institut f{\"u}r Thermische Str{\"o}mungsmaschinen, Karlsruhe, Germany}

\begin{abstract}
The influence of the ambient pressure on the breakup process is investigated by means of PIV and shadowgraphy in the configuration of a planar prefilming airblast atomizer.
The ambient pressure is varied from 1 to 8~bar. Other investigated parameters are the gas velocity and the film loading.
From single-phase PIV measurements, it is found that the gas velocity in the vicinity of the prefilmer partly matches the analytical profile from the near-wake theory.
the characteristics of the liquid accumulation are extracted from the shadowgraphy images of the liquid phase directly downstream of the prefilmer.
Two different characteristic lengths, as well as the ligament velocity and a breakup frequency are determined. In addition, the droplets generated directly downstream of the liquid accumulation are captured. Hence, the spray Sauter Mean Diameter (SMD) and the mean droplet velocity are given for each operating point.
The novelty of this study is that a scaling law of these quantities with regard to ambient pressure is derived.
A correlation is observed between the characteristic length of the accumulation and the SMD, thus reinforcing the idea that the liquid accumulation determines the primary spray characteristics.
In this paper, a threshold to distinguish the zones between primary and secondary breakup is proposed based on an objective criterion.
It is also shown that taking non-spherical droplets into account significantly modifies the shape of the dropsize distribution, thus stressing the need to use shadowgraphy when investigating primary breakup.
Additionally, the ambient pressure and the velocity are varied accordingly to keep the aerodynamic stress $\rho_g U_g^2$ constant. This leads to almost identical liquid accumulation and spray characteristics. Hence, it is confirmed that the aerodynamic stress is a more appropriate parameter than the gas velocity or the ambient pressure to characterize prefilming airblast breakup.
Finally, SMD correlations from the literature are compared to the present experiment. Most of the correlations calibrated with LDA/LDT measurement underestimate the SMD. This highlights the need to use shadrowgraphy for calibrating primary breakup models.
\end{abstract}
\end{frontmatter}

\section{Introduction \label{sec_intro}}
\thispagestyle{firststyle}

\begin{figure}%
\centering
	\includegraphics[width=0.4\textwidth,keepaspectratio]{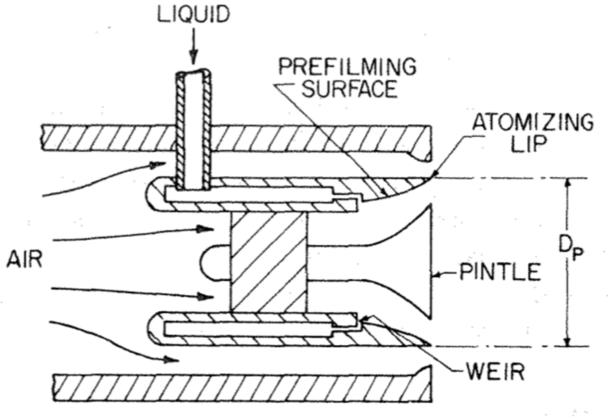}
	\caption{Sketch of a prefilming airblast atomizer, from \citet{Lefebvre:1989a}}
	\label{fig_intro_prefilming_lefebvre}
\end{figure}

In the perspective of reducing the CO\textsubscript{2} emissions, the use of fossil fuel must be optimized.
For combustors fed by a liquid fuel, the quality of the atomization plays an important role on the efficiency of the combustion \citep{bossard_droplet_1996}, on the pollutant emissions and on the soot production \citep{hayashi2011effects}. When the spray is composed of smaller droplets, the gas/liquid contact area is increased, thus promoting a quick vaporization and consequently a better mixing with the gas. This leads to a more homogeneous mixture that ensures a more stable flame \citep{meier_detailed_2007,franzelli_large_2012}.
The quality of the atomization is also important for transient conditions \citep{mishra_effect_2014} such as the variations of the engine load (take-off or approach phase for airplanes), ignition, or high altitude relight.
For instance, in the case of ignition, the presence of small droplets facilitates the propagation of the flame kernel.
Prefilming airblast atomizers (figure~\ref{fig_intro_prefilming_lefebvre}) are one of the mostly used devices in gas turbines. They provide constant spray characteristics over a wide range of air speed, fuel mass flow rate and ambient pressure \citep{Lefebvre:1989a}. 
In this type of device, the liquid fuel is fed at low velocity and is atomized by the high-speed air flow. In order to improve the momentum transfer, the liquid is disposed as a thin liquid film, which increases the contact interface between gas and liquid. The liquid film is driven by the air flow to the tip of the injector, referred to as the atomizing edge, where it accumulates. This liquid accumulation is then fragmented by the surrounding high-speed air flow \citep{gepperth2012ligament}.
Even though this type of injector is widely used, the fundamental mechanism of the liquid disintegration occurring at the atomizing edge is not fully understood. 
This lack of understanding prevents the development of reliable and universal models to be embedded in large scale simulations of complete combustion chambers.
The influence of global parameters such as the gas velocity, film loading, and surface tension was recently investigated by \citet{gepperth2012ligament,gepperth2010pre,gepperth2013analysis}.
In his PhD thesis, \citet{gepperth2019experimentelle} investigated the influence of ambient pressure on the disintegration process of the liquid accumulation in prefilming airblast atomization.
This paper uses the original results from \citet{gepperth2019experimentelle} and
compare them qualitatively to the breakup sequences found in similar configurations \mbox{\citep{zandian2017planar}}. Moreover, we propose here a mathematical model to predict the aerodynamic stress directly downstream the prefilmer. We also present a global scaling of the geometrical quantities of the liquid accumulation versus the ambient pressure. Finally we conduct a comparative study of standard correlations versus the present results.

\begin{figure}%
\centering
\linespread{1.0}
	\def \svgwidth {0.85\textwidth}
	{\scriptsize
	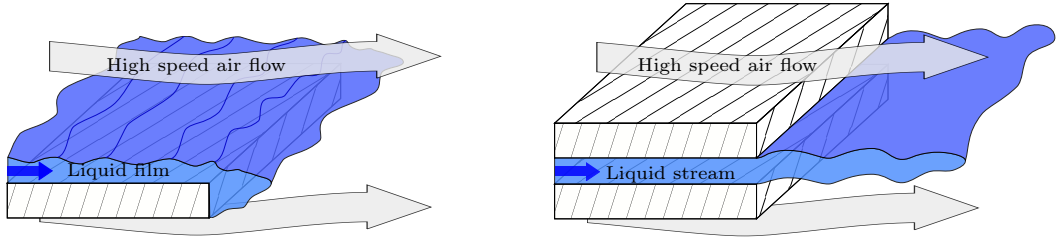
	}
	\caption{Left: accumulation breakup. Right: liquid sheet breakup}
	\label{fig_liquid_sheet_vs_accumulation}
\end{figure}

\begin{figure}%
\centering
	\includegraphics[width=\textwidth,keepaspectratio]{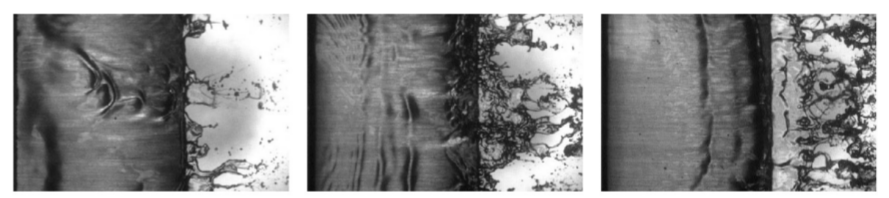}
	\caption{Transition from accumulation breakup to liquid sheet breakup by increasing the film loading\protect\footnotemark.}
	\label{fig_intro_regime_transition}
\end{figure}

Experimental studies of airblast atomizers were pioneered by \citet{rizkalla1975influence}, \citet{lefebvre1980airblast}, \citet{el1980airblast}, \citet{Rizk:1980, Rizk:1983}, \citet{sattelmayer1986internal} and \citet{aigner1988swirl}. In these studies, the main parameters influencing the mean drop size were identified as the gas velocity, the surface tension and the Air Liquid Ratio (ALR). It should be pointed out that \citet{sattelmayer1986internal}, for the first time, observed the liquid accumulation based on photographs.
Among the aforementioned works, two studies investigated the influence of ambient pressure on the mean drop size. \citet{Rizk:1983} investigated a plain airblast injector while \citet{rizkalla1975influence} studied the design of a prefilming airblast injector. In both studies a Light Diffraction Technique (LDT) was used to determine the SMD of the spray. The distances between the light beam and the nozzle exit were roughly 50 mm and 200 mm for \citet{rizkalla1975influence} and \citet{Rizk:1983}, respectively. The latter found a SMD inversely proportional to the ambient pressure.
More recently, the investigation of prefilming airblast atomization with a varying ambient pressure was conducted by 
\citet{batarseh2008spray} on an annular swirling injector with Phase Doppler Anemometry (PDA) located, at the closest, at 3 mm downstream of the nozzle exit. The author varied the ambient pressure from 1 to 10 bar and adapted the gas velocity to keep the mass flow rate constant, and observed that the SMD 
As shown in the present work, this non monotonic behaviour is explained by the fact that the author varied the ambient pressure and the velocity together.
\citet{bhayaraju2009planar} investigated prefilming airblast atomization at high pressure using a planar injector with PDA located, at the closest, 10 mm downstream the nozzle exit. The authors also reported a \textit{storage mechanims} of the liquid at the prefilmer tip.
\citet{gepperth2012ligament,gepperth2010pre,gepperth2013analysis} investigated a planar prefilming airblast nozzle (figure~\ref{fig_liquid_sheet_vs_accumulation} left) by means of shadowgraphy. Contrary to the previous investigations, they extracted the spray characteristics directly at the nozzle exit, and showed that the mean diameter of this primary spray scales with the atomizing edge thickness.
Within the operating conditions studied, they observed that (i) the liquid accumulates at the atomizing edge, (ii) the breakup events only occur at the liquid accumulation, and (iii) the liquid accumulation decouples the breakup events from the wave propagation of the film flow.
\citet{dejean2016experimental} investigated the prefilming airblast atomization of water for different film loadings, gas velocities and prefilming lengths by mounting a prefilming surface downstream a streamlined body. They highlighted the influence of the prefilming length, and proposed a regime map describing the surface of the film and the presence of a liquid accumulation depending of the prefilming length, the vorticity thickness and the momentum flux ratio.
They also observed that the transition from accumulation breakup to liquid sheet breakup is triggered by an increase of the film loading (figure~\ref{fig_intro_regime_transition}).\\
Prior to the publications of Gepperth and coworkers, the breakup regime which was associated to prefilming airblast atomization was the liquid sheet breakup only, which is mainly observed in non-prefilming airblast atomization (figure~\ref{fig_liquid_sheet_vs_accumulation} right). In this type of breakup, the liquid propagates downstream the nozzle in the form of a thin sheet, and is continuously fragmented by the air stream. This type of breakup can be broken down into different sub-regimes by increasing the momentum flux ratio $M$: cellular, stretched-ligament, torn-sheet and membrane-sheet breakups \citep{Stapper:1990,fernandez2011dynamic}. The distance to the atomizing edge where breakup takes place decreases with an increasing $M$.
In the contrary, in the regime observed by Gepperth and coworkers, the breakup always occurs directly at the liquid accumulation, \ie at the atomizing edge, and several sub-regimes of breakup are observed for the same operating conditions. Hence, this type of breakup is significantly different from the liquid sheet breakup, and the identification of the liquid accumulation was a major breakthrough for prefilming airblast atomization. This type of breakup is subsequently referred to as \textit{accumulation breakup}.
In this study, we restrict our analysis to accumulation breakup only.
This phenomenon, known as the teapot effect, was studied and quantified by \mbox{\citet{duez2010wetting}} for an inertial liquid flow. 

\begin{figure*}%
\centering
	\linespread{1.0}
	\def \svgwidth {0.67\textwidth}
	{\scriptsize
	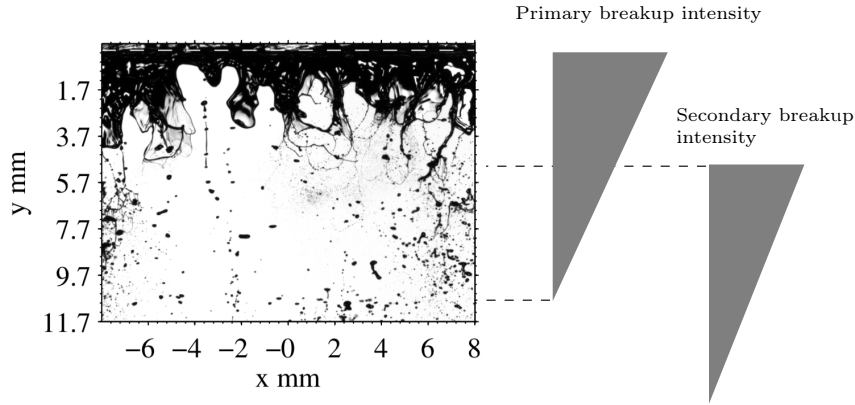
	}
	\caption{Snapshot of the spray generation in the vicinity of the atomizing edge (left) and corresponding illustration of the type of breakup (right)}
	\label{fig_intro_breakup}
\end{figure*}

At this point, it must be stated that the distance between the atomizing edge and droplet size measurement position is of primary importance. In the vicinity of the atomizing edge, the accumulation breakup produces
highly distorted blobs or ligaments. Further downstream, the liquid structures undergo a cascade fragmentation referred to as the secondary atomization, and lead to a cloud of spherical droplets. Between these zones a transition zone can be identified, where both regimes are observable, as schematically illustrated in figure~\ref{fig_intro_breakup}. Note that there is no dedicated quantitative criterion to distinguish between primary and secondary breakup.
These two types of breakup mechanisms feature different characteristics and therefore, they require different models. The most emblematic ones are the model of \mbox{\cite{reitz1987modeling}} for primary breakup, based on the linear stability analysis of a round jet, and the Taylor Analogy Breakup (TAB) model of \mbox{\cite{ORourke:1987}} for secondary breakup, based on the vibrational modes of a liquid droplet.
While secondary breakup is universal and independent of the atomizer type, primary breakup heavily depends on the injector design.
Therefore, a study focusing on primary breakup has to measure the spray quantities directly at the nozzle exit in order to minimize the impact of secondary breakup.
\\On the other hand, experimental diagnostic of the primary breakup is subject to limitations. First, in modern prefilming airblast atomizers the optical access is limited due to the presence of a diffusor downstream the atomizing edge \citep{shanmugadas2018characterization}. Second, the highly distorted liquid structures cannot be captured by PDA and lead to a signification deviation for LDT \citep{dumouchel2014laser}. This is the reason why, in most of the studies aforementioned, the measuring volume was placed further downstream in the zone where secondary breakup was already significant or even dominant. This limits the availability of experimental data which are suitable to develop a primary breakup model, and stresses the need to use alternative diagnostics for recording the spray characteristics in the vicinity of the atomizing edge.
\footnotetext{\footnotesize Reprinted from \textit{International Journal of Multiphase Flow 79, B. D{\'e}jean, P. Berthoumieu, P. Gajan, \href{https://doi.org/10.1016/j.ijmultiphaseflow.2015.09.001}{Experimental study on the influence of liquid and air boundary conditions on a planar air-blasted liquid sheet, part {II}: prefilming}, 214-224}, \textcopyright \ 2016, with permission from Elsevier.}

With respect to the modelling of a prefiming airblast atomizer under real engine condition, the presence of the flame front and zones of high temperature are to be considered. It is observed in experiments \citep{doll2017temperature} and numerical simulations \citep{Moin:2006,Boileau:2008a} of swirl-stabilized flames that the flame front is located approximately one diameter downstream the nozzle exit, but the temperature sharply increases directly at the prefilmer tip. This means that the evaporation starts at the prefilmer tip where primary breakup occurs. Hence, the droplet diameter is constantly reduced from the atomizing edge to the flame front. 
In this case, a numerical model calibrated on secondary atomization data, which would inject droplets at the prefilmer tip might not be satisfactory for a high-fidelity simulation of combustion. 
Moreover, the size of the droplets directly affects the flame position \citep{fiorina2016modeling}. 
This emphasizes the need to collect data on droplets and liquid blobs as soon as they are created, \ie at the atomizing edge.

The objectives of the present work is to analyze the influence of the ambient pressure and aerodynamic stress on the characteristics of the liquid accumulation and the primary spray with the aim to develop models for predicting primary spray characteristics at the atomizing edge. The second goal is to demonstrate the necessity to collect data directly at the atomizing edge, and therefore to rely on shadowgraphy technique, in order to capture large and non-spherical liquid blobs.
The experimental setup as well as the diagnostics are presented in Section~\ref{sec_expe_setup}.
PIV measurements of the air flow field are presented in Section~\ref{sec_gas_phase}
and a model to estimate the axial velocity at the center line is presented.
Thereafter, qualitative observations of the liquid accumulation are shown in Section~\ref{sec_qualit_liquid}, followed by a quantitative analysis in Section~\ref{sec_quantit_liquid}.
The influence of the aerodynamics stress is discussed in Section~\ref{sec_M_constant}.
Finally, the measurements of the SMD are compared to correlations from literature in Section~\ref{sec_compare_correlations}.

\section{Experimental setup and diagnostics \label{sec_expe_setup}}

\subsection{High pressure test-rig \label{ssec_highpress}}

\begin{figure*}%
\centering
\linespread{1.0}
	\def \svgwidth {\textwidth}
	{\scriptsize
	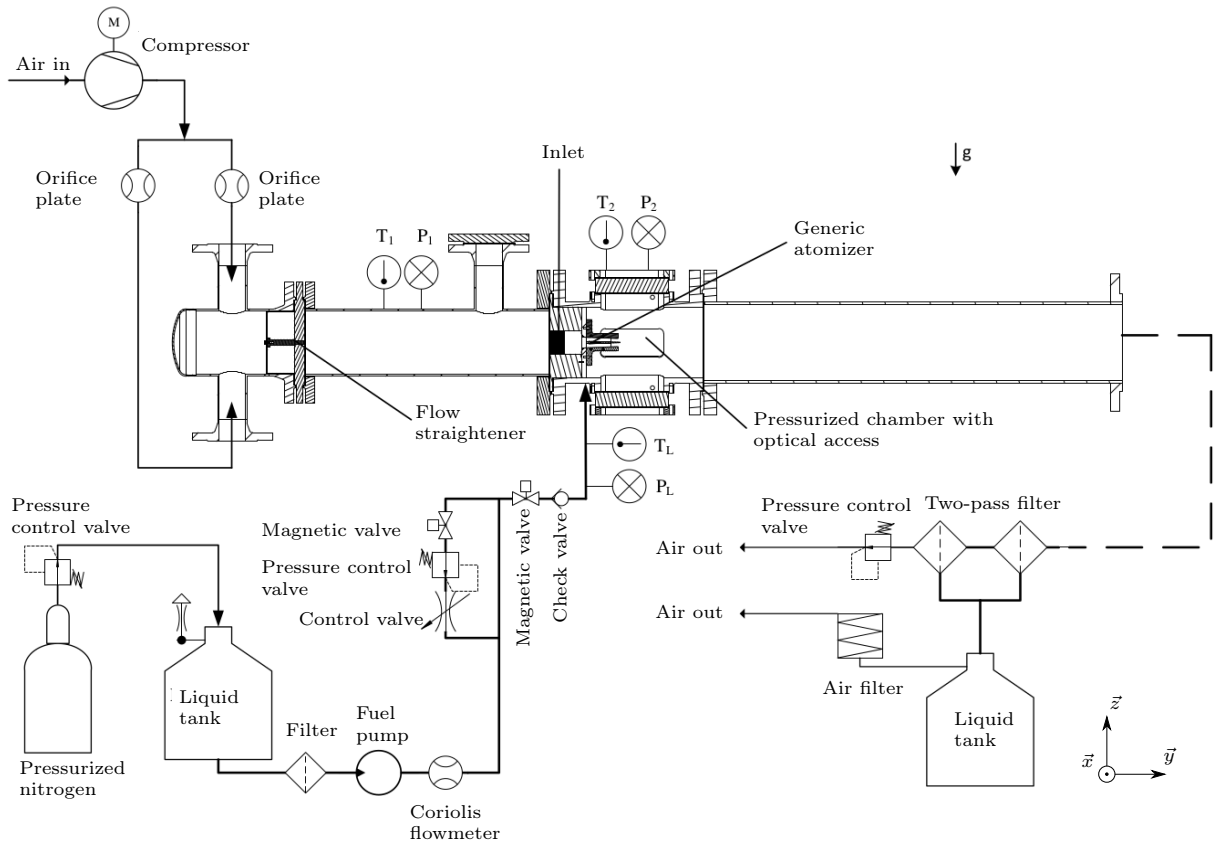
	}
	\caption{Schematic of the high pressure experimental setup, from \citet{gepperth2019experimentelle}.}
	\label{fig_schematic_exp_setup_hdt}
\end{figure*}

The test-rig is depicted in figure~\ref{fig_schematic_exp_setup_hdt}. It consists of a cylindrical pressurized duct featuring optical accesses.
The pressurized air is supplied by a compressor providing a maximum mass flow of 1.2 kg/s at 8~bar. A plenum is mounted upstream of the test section for homogenization of the air.
The large diameter of the plenum ensures that the air velocity as well as the turbulence intensity are low.
The air then enters the atomizer through a nozzle equipped with a flow straightener
to guarantee a smooth acceleration and to avoid the localised production of turbulence.
The liquid is supplied to the atomizer from a pressurized vessel that suppresses any flow rate fluctuations. The mass flow rate is controlled by a Coriolis mass flow meter. The atomizer discharges in an open and quiescent atmosphere. Downstream of the test-section, the liquid is separated from the air in a cyclone.

\subsection{Prefilmer geometry}

\begin{figure}%
	\begin{minipage}[b]{0.495\textwidth}
	\centering
	\linespread{1.0}
	\def \svgwidth {\textwidth}
	{\scriptsize
	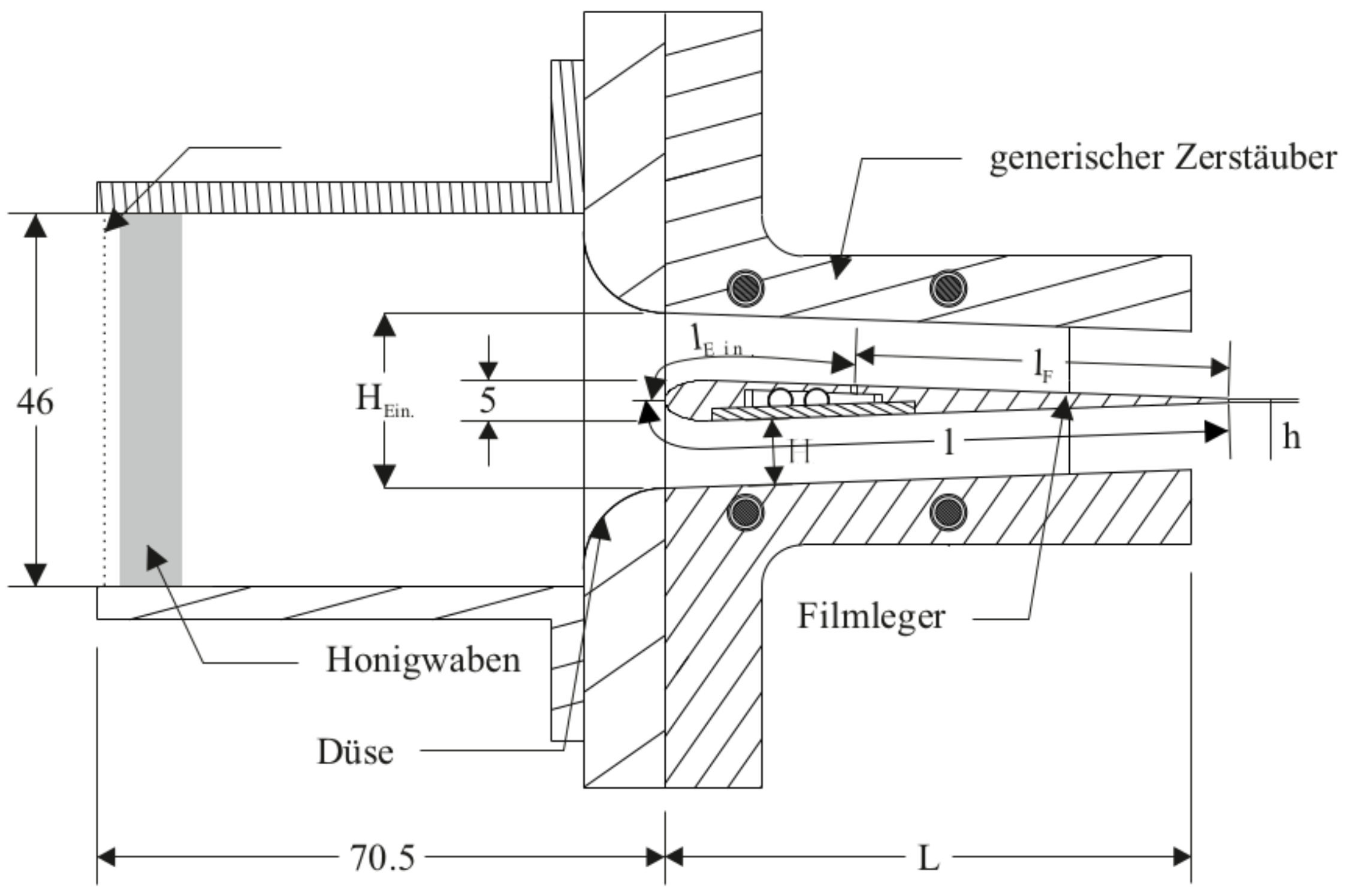
	}
	\end{minipage}
	\hspace{0.01\textwidth}
	\begin{minipage}[b]{0.495\textwidth}
	\centering
	\linespread{1.0}
	\def \svgwidth {\textwidth}
	{\scriptsize
	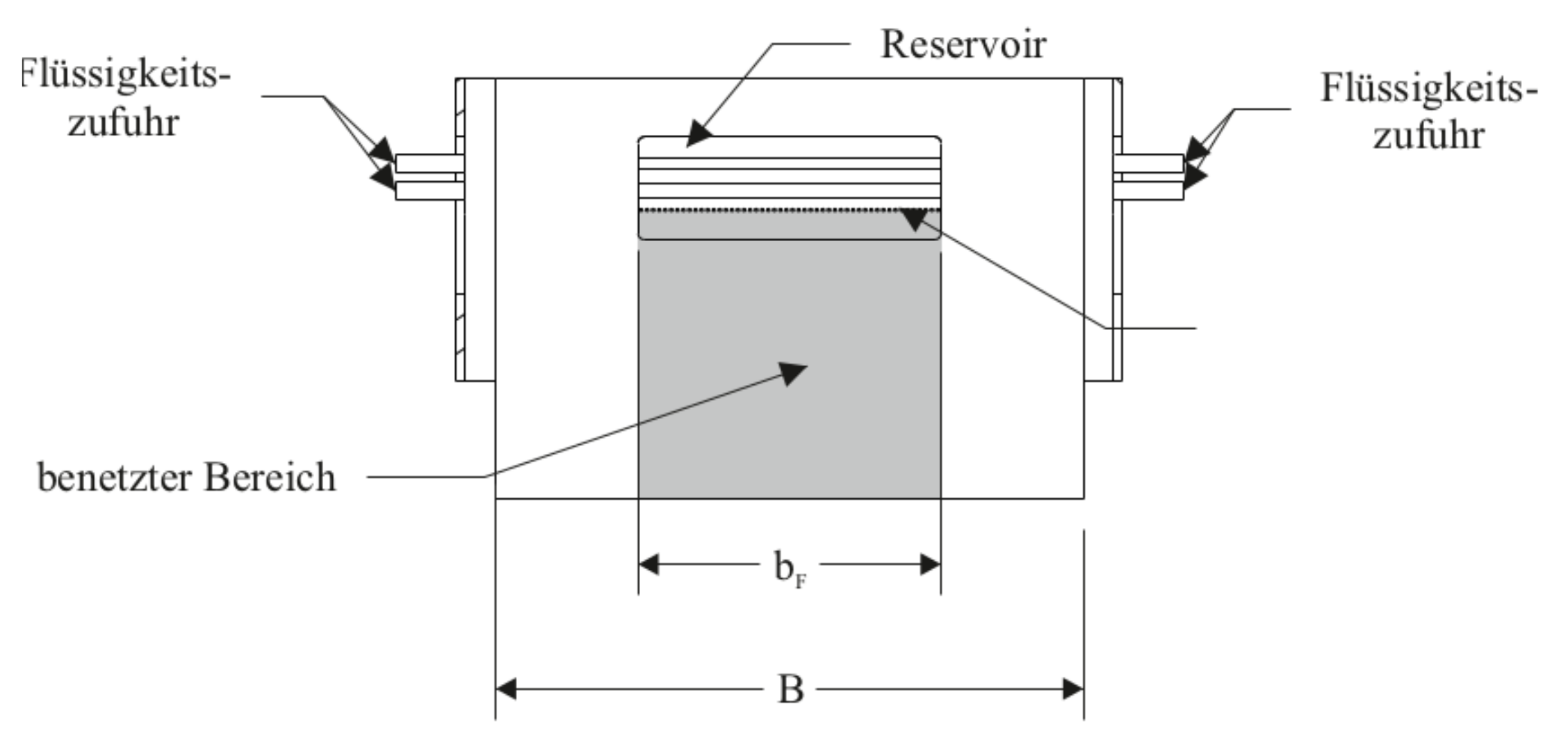
	}
	\end{minipage}
	\caption{Schematics of the planar prefilming airblast atomizer, from \citet{gepperth2019experimentelle}.}
	\label{fig_prefilmer_views}
\end{figure}

The investigated atomizer is depicted in figure~\ref{fig_prefilmer_views}. It is a planar prefilming airblast atomizer and can be considered as a 2D abstraction of a realistic annular atomizer. It has the advantage of featuring the same breakup mechanism like in annular atomizers \citep{gepperth2012ligament, holz2016comparison} in a deterministic plane that simplifies optical measurements. The geometry consists of a wing-shaped prefilmer located at the center of a rectangular channel. The liquid is injected in a small plenum inside the prefilmer through four ducts. It discharges from the plenum to the prefilmer surface through fifty holes equally distributed along the prefilmer width, allowing a homogeneous wetting. The thickness of the atomizing edge was measured using a technique based on foam filling. The geometrical characteristics of the prefilmer are recalled in Table~\ref{tab_geom_charac_prefilmer} in dimensional and non-dimensional forms.

\begin{table}%
	\small
	\centering
	\begin{tabular}{  l  l c  c c  c c  c c c c}
	 & Unit & $H_{in}$ & $l_{in}$ & $H$ & $L$ & $l$ & $l_F$ & $h$ & $B$ & $b_F$ \\
	Dimensional & [mm] & 21.6 & 23.3 & 8.11 & 65 & 70.9 & 47.6  & 0.64 & 96 & 50 \\
	Non-dimensional & [$-$] & 3.28 & 0.358 & 1 & 1 & 1.09 & 0.732  & 7.89 10\textsuperscript{-2} & 1 & 0.521 \\
	Normalizing parameter & [$-$] & H & L & H & L & L & L & H & B & B \\
	\end{tabular}
	\caption{Geometrical characteristics of the planar prefilmer, in dimensional and non-dimensional forms.}
	\label{tab_geom_charac_prefilmer}
\end{table} 

\subsection{Operating conditions and non-dimensional groups}

The operating conditions are summarized in Table~\ref{tab_operating_condition}. The bulk velocity and the ambient pressure were varied at constant temperature. The investigated liquid (Shellsol D70) possesses similar properties as usual kerosene for aeroengines.
The liquid mass flow rate $\dot{m}_l$ is expressed by the film loading $\Lambda_f=\dot{m}_l / b_F$ where $b_f$ is the prefilming width as shown in figure {\ref{fig_prefilmer_views}}.
\begin{table}%
		\small
		\centering
		\begin{tabular}{  l  c  r  l  }
		Mean air velocity & $U_g$& 40 $-$ 80  &m/s\\
		Air mass flow rate & $\dot{m}_g$ & 0.075 $-$ 1.20 &kg/s \\
		Air temperature & T & 298 &K \\
		Air density & $\rho_g$ &1.2 $-$ 9.6 &kg/m$^3$ \\
		Air pressure & $p_g$ &1 $-$ 8.2 &bar \\
		Air dynamic viscosity & $\mu_g$ & 1.8 $\cdot$ 10$^{-5}$ & kg/m.s \\
		Liquid density & $\rho_l$ & 770  &kg/m$^3$ \\
		Liquid dynamic viscosity & $\mu_l$ &1.56 & g/m.s \\
		Liquid surface tension & $\sigma$ &0.0275  & kg/s$^2$ \\
		Liquid volume flow rate & $\Lambda_f$ & 25 $-$ 50 & mm$^2$/s \\
		\end{tabular}
		\caption{Operating conditions and liquid properties}
		\label{tab_operating_condition}
\end{table}\\\\
The boundary layer that develops over the prefilmer is estimated at the location of the atomizing edge using the formula from \citet{white1991viscous} for a turbulent boundary layer on a flat plate: %
\begin{equation}
\delta = 0.16 \frac{l}{\Re_g^{1/7}}
\quad \text{with} \quad
\Re_g = \frac{\rho_g U_g l}{\mu_g}
\label{eq_BL_correl}
\end{equation}
with $\Re_g$ the gaseous Reynolds number based on the prefilmer length.
To estimate $\delta$ in case of adverse pressure gradient, the interested reader is advised to use another formulation.
The liquid Reynolds number is based on the mean liquid velocity $U_l$ and the film thickness $h_f$. These two quantities were not measured in the present configuration, but their product is equal to the film loading $\Lambda_f$. Hence:
\begin{equation}
\Re_l = \frac{\rho_l U_l h_f}{\mu_l} = \frac{\Lambda_f}{\nu_l}
\end{equation}
where $\nu_l$ is the kinematic viscosity of the liquid.
Two Weber numbers are derived in the following. First, the Weber number $\We_h$ is based on the atomizing edge thickness $h$
and the relative velocity between the gas and the Dimotakis velocity $U_D$:
\begin{equation}
U_D = \frac{U_g \sqrt{\rho_g}+U_l \sqrt{\rho_l}}{\sqrt{\rho_g}+\sqrt{\rho_l}}
\end{equation}
can be approximated as $r_{\rho} \, U_g$ where $r_{\rho} = \sqrt{\rho_l}/(\sqrt{\rho_g}+\sqrt{\rho_l})$ is called the density parameter.
It was shown in \citep{chaussonnet2016new} that the SMD scales
(i) with $1/\sqrt{\We_h}$ 
and (ii) with the transverse wavelength of the liquid accumulation instability measured in \mbox{\citep{mueller2004atomization}}.
The second Weber number $\We_{\delta}$ is based on the boundary layer thickness $\delta$. The two Weber numbers are expressed as:
\begin{equation}
\We_h= \frac{\rho_g h \left(r_{\rho} U_g \right)^2}{\sigma}
\quad \text{and} \quad
\We_{\delta} = \frac{\rho_g U_g^2 \delta}{\sigma}
\end{equation}
Finally, the aerodynamic stress $\tau_G=\rho_g \, u_g^2$ represents the amount of gaseous momentum available for disintegration of the liquid accumulation at the atomizing edge. As discussed in Section~\ref{sec_M_constant}, it is an important quantity governing primary atomization.
The extreme values of the non-dimensional numbers, the estimated boundary layer thickness at the atomizing edge and the aerodynamic stress are listed in Table~\ref{tab_non_dim_number}.

\begin{table}%
		\small
		\centering
		\begin{tabular}{  l  l  r  l }
		Boundary layer at the trailing edge &$\delta$ & 1.35 $-$ 2.00 &mm \\
		Aerodynamic stress &$\tau_G$ & 1220 $-$ 61440  &Pa\\
		Gaseous Reynolds number &$\Re_g$ & 190,000 $-$ 3,025,000 &$-$ \\
		Liquid Reynolds number &$\Re_l$ & 12.3 $-$ 24.7 &$-$  \\
		Edge Weber number &$\We_h$ & 41.35 $-$ 1157 &$-$  \\
		Boundary layer Weber number &$\We_{\delta}$ & 139.6 $-$ 3007 &$-$  \\
		Density parameter &$r_{\rho}$ & 0.90 $-$ 0.96 &$-$  \\
		\end{tabular}
		\caption{Derived quantities and important non-dimensional groups}
		\label{tab_non_dim_number}
\end{table}

\subsection{Diagnostics}

The gas velocity field was measured by means of PIV in the plane $(\vec{y},\vec{z})$ at the outlet of the atomizer.
The flow was seeded upstream the prefilmer by droplets of DEHS (Di-Ethyl-Hexyl-Sebacat) of $\approx$~1~\textmu m, resulting into a particle relaxation time of $\approx$ 2.5 \textmu s for the present conditions.
For a Stokes number lower than 0.1, the uncertainty of the PIV measurement is below 1\% \citep{tropea2007springer}. Therefore, single phase flow features of a frequency up to 40 kHz will be captured with an accuracy better than 1\%. 
The tracers were recorded in a plane illuminated by a 100 \textmu m thick laser sheet. The spatial correlation is computed over a square of 8x8 pixels of 26 \textmu m each, leading to a spatial resolution of 208 \textmu m. The statistics (mean and RMS) of the flow field are computed based on 200 double pictures. The overall uncertainty is $\approx$1\%.

The breakup process was investigated by means of shadowgraphy. Time series were recorded by a high-speed camera with a constantly illuminated background for qualitative analysis. 
For quantitative measurement, high-resolution double-frame images were recorded with a CCD camera at a frequency of 10~Hz. Hence, each double-frame picture is uncorrelated from the previous or next one. The double-frame image corresponds to an instantaneous sample of the primary breakup and the quantitative output can be treated in a statistical sense.
The field of view is 12 mm and 16 mm in the $y$ and $z$ direction, respectively, and the spatial resolution is 10 \textmu m. The Depth-of-Field (DoF) is 0.44 mm.
The background is illuminated with a dual cavity Nd:YAG pulse laser. The intensity of the expanded beam is homogenized by a diffuser disc. Then, the light passes a cuvette of laser dye (rhodamine) to be reemitted as a non-coherent light. This is a counter measure for eliminating spurious interferences.

\begin{figure}%
\centering
\linespread{1.0}
\subfloat[\label{fig_contour_detection_algo1}Image frame with normalized intensity]{\includegraphics[width=0.35\linewidth,keepaspectratio]{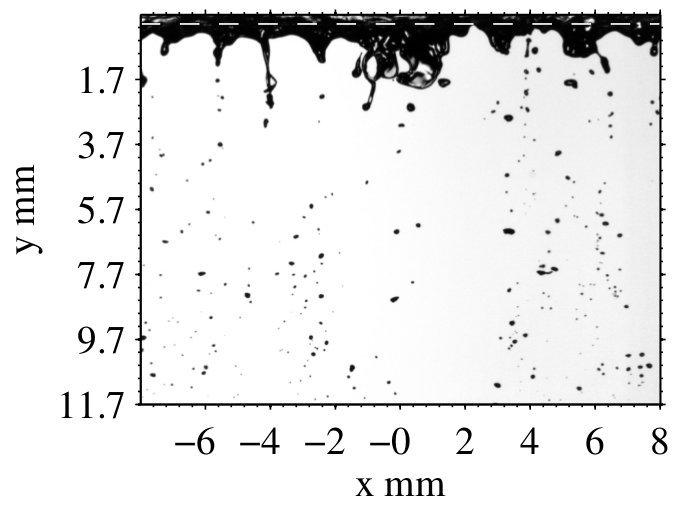}}
\hspace{3mm}
\subfloat[\label{fig_contour_detection_algo2}Contour of the liquid accumulation and position of the crests]%
{
\def \svgwidth {0.35\linewidth}
\scriptsize
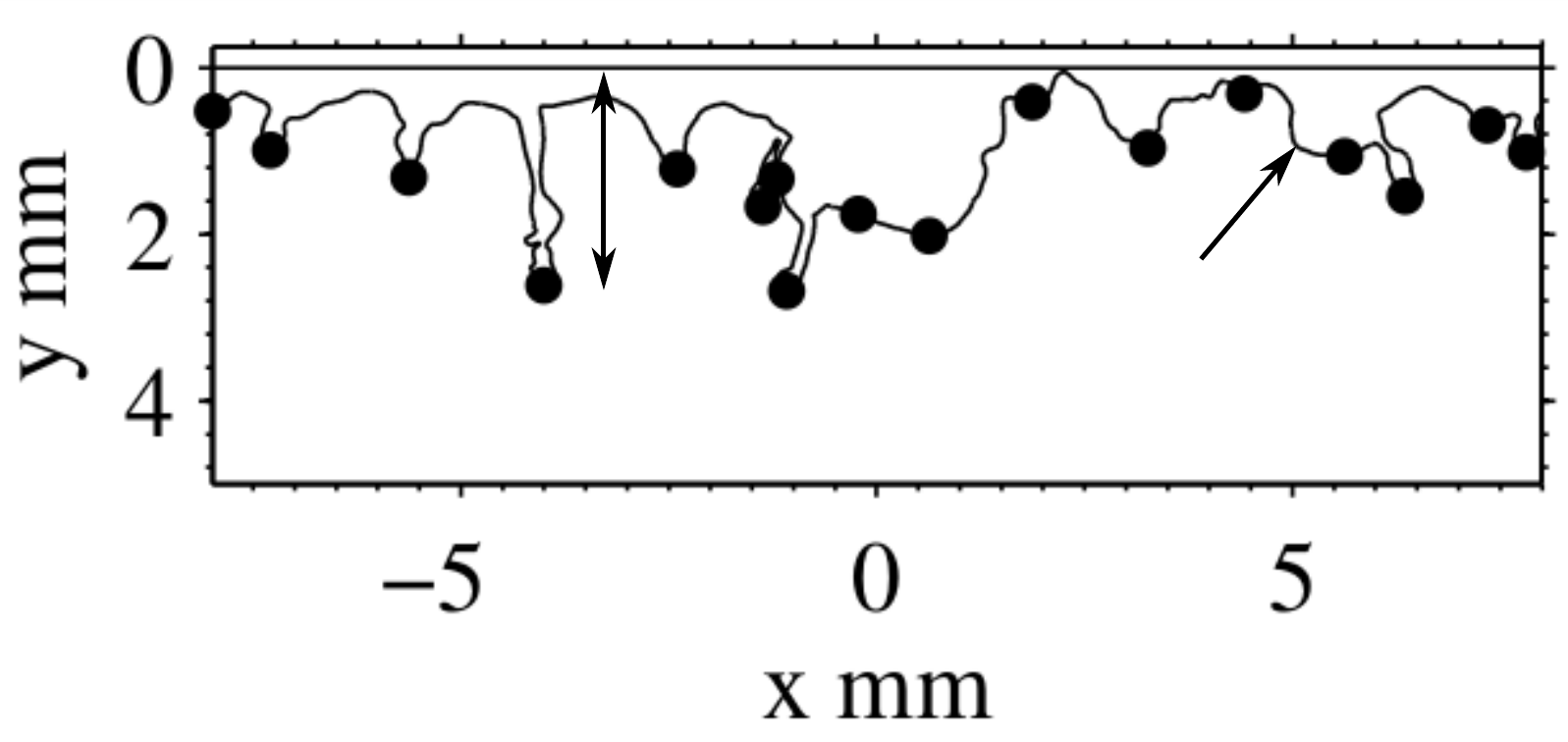
}
\\
\subfloat[\label{fig_contour_detection_algo3}Elongation velocity of ligaments]%
{
\def \svgwidth {0.35\linewidth}
\scriptsize
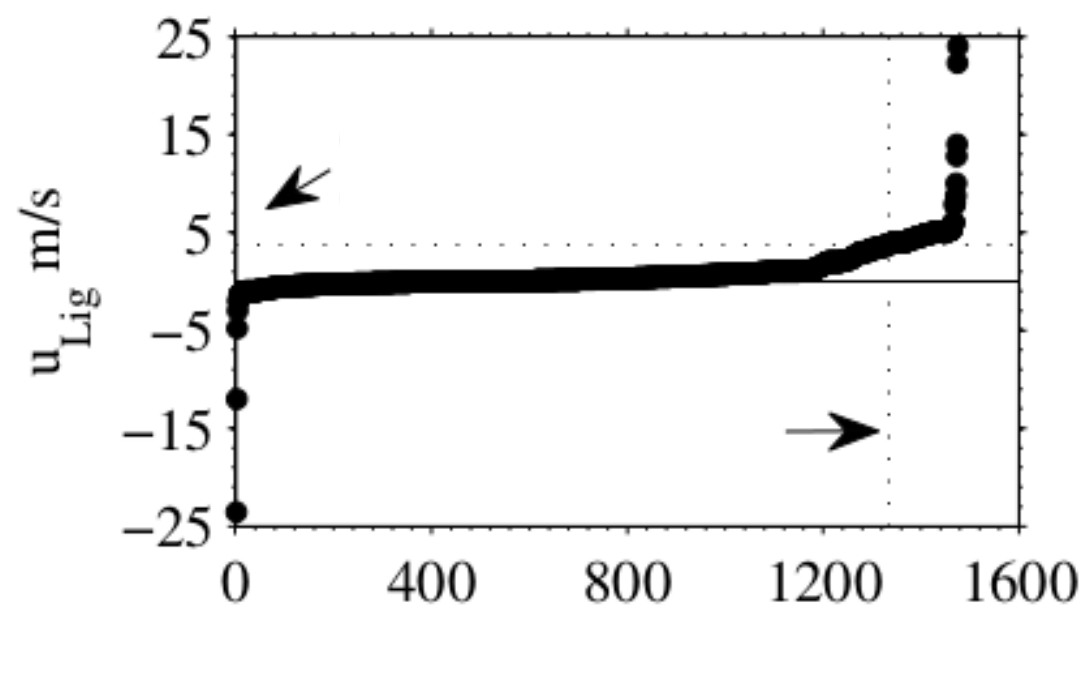
}
\hspace{3mm}
\subfloat[\label{fig_contour_detection_algo4}Projected surface of the liquid accumulation]{\includegraphics[width=0.35\linewidth,keepaspectratio]{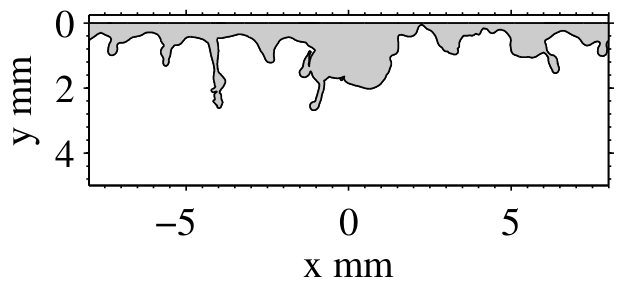}}
\caption{Details of the PLTV algorithm, from \citet{gepperth2019experimentelle}.}
\label{fig_contour_detection_algo}
\end{figure}

For post-processing a Particle and Ligament Tracking Velocimetry (PLTV) technique was applied to the double-frame pictures. This technique possesses the advantage to record the characteristics of the liquid accumulation and the generated droplets by the same image. In addition, non-spherical droplets are captured.
The algorithm was previously developed by \citet{kapulla2008droplet} and \citet{mueller2015experimentelle}. It consists of the following steps as illustrated in figure~\ref{fig_contour_detection_algo}. First, the position of the atomizing edge is manually set and the intensity of the background is normalized to enhance the contrasts between background and liquid (figure~\ref{fig_contour_detection_algo1}). Second, a contour detection algorithm based on gray scale threshold is performed to discriminate the liquid phase. The threshold value is set proportional to the median value of all the pixels. From this algorithm, the contour of the liquid accumulation and the position of the tip of each ligament are identified (figure~\ref{fig_contour_detection_algo2}).
Each position is determined as a local maximum with respect to its distance from the atomizing edge. It is considered as the ligament length and labeled $L_{Lig}$ in the following. The length $P_{Lig}$ of the contour corresponds to the perimeter of the liquid accumulation.
From the double frame, the axial velocity $U_{Lig}$ of the ligament tip is calculated.
A strong negative velocity can be recorded for receding ligaments after a bag breakup, whereas extreme high velocities are due to the uncertainty in the determination of the ligament tip when the surface is highly distorted \citep{mueller2015experimentelle}.
In order to eliminate unrealistic ligament velocities and to achieve an acceptable estimation of the ligament velocity close to breakup, $U_{Lig}$ is taken as the 90\textsuperscript{th}-percentile of the whole set for an operating point (figure~\ref{fig_contour_detection_algo3}).
Finally, the projected surface $A_{Lig}$ of the liquid accumulation is defined as the area delimited by the contour and the atomizing edge (figure~\ref{fig_contour_detection_algo3}). 

The liquid blobs, which are torn off from the accumulation, are detected as closed contours, and considered as droplets. Their equivalent diameter is estimated by two methods. First, the number of pixels enclosed within the contour is converted to a projected area. This area $A_{proj,1}$ is considered as the frontal area of a spherical droplet, leading to a diameter of $d_{meas.,1}=\sqrt{4 A_{proj,1} / \pi}$. Second, an ellipsoid is fitted to the shape of the closed contour, and the two major axis ($a$ and $b$) are determined. With the assumption that the short axis ($b$) is the extension of the droplet in the third dimension, the volume can be computed as $V=4 \pi a b^2 /3$, leading to a diameter $d_{meas.,2}=\sqrt[3]{a b^2}$.
The difference between the two equivalent diameters is an indicator of the sphericity of the droplet.
A Depth-of-Field (DoF) correction was applied to increase the accuracy of small droplets. The procedure is extensively explained by \citet{warncke2017experimental}.

In the following, the characteristic scales of the liquid accumulation are 
the mean ligament length $L_{Lig}$, the ligament tip velocity $U_{Lig}$, the breakup frequency $f_{BU}$ and the ratio $A_{Lig}/P_{Lig}$. The two formers were explained previously. The frequency $f_{BU}$ is expressed as $U_{Lig}/{L}_{Lig}$ and is referred to as the breakup frequency. Strictly speaking, it is the inverse of the time required for the ligament to be stretched over the distance ${L}_{Lig}$ if the velocity were constant.
It must be highlighted that ${L}_{Lig}$ is the mean length of all ligaments detected for one operating point. This includes the whole ligament, from formation to elongation and finally breakup. Consequently, it does not correspond to the ligament length at the moment of breakup. However, $f_{BU}$ provides a measure of the breakup time scale (or frequency) at the atomizing edge.
The ratio $A_{Lig}/P_{Lig}$ is a length scale that depends on the total volume of liquid inside the accumulation ($A_{Lig}$) over the distortion of the interface ($P_{Lig}$). Hence, it is a measure of the \textit{size-to-distortion} ratio of the liquid accumulation.

\section{Air flow\label{sec_gas_phase}}

\subsection{Flow structure downstream of the prefilmer in the literature}\label{ssec_theory_flow_structure}

In the case of a single-phase flow, the gas flow field downstream the prefilmer is similar to the turbulent wake downstream of a flat plate. Such a configuration was experimentally investigated by \citet{chevray1969turbulence}, \citet{pot1979measurements}, \citet{andreopoulos1980measurements} and \citet{ramaprian1982symmetric}. A theory was proposed by \citet{alber1980turbulent} and later completed by \citet{ramaprian1982symmetric}. All the authors agreed on the characteristics of the flow.
\begin{figure}%
	\begin{center}
	\linespread{1.0}
	{\scriptsize
	\def \svgwidth {0.7\textwidth}
	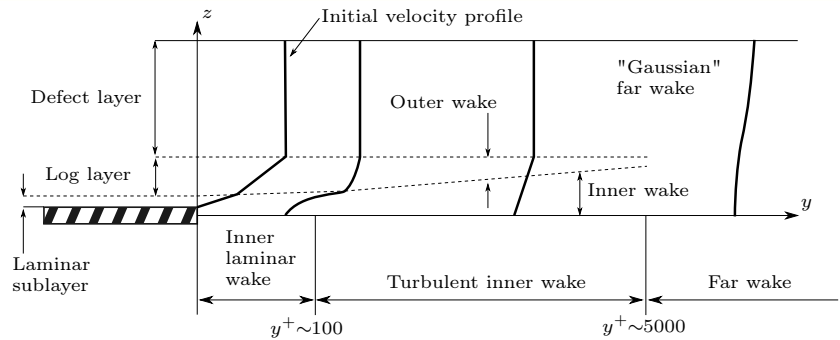
	}
	\caption{Near- and far-wake regions: turbulent wake of a flat plate, as sketched by \citet{alber1980turbulent}.}
	\label{fig_sketch_wake}
	\end{center}
\end{figure}

The structure of the flow is sketched in figure~\ref{fig_sketch_wake} and described as follows.
In the direction normal to the prefilmer surface, the wake is decomposed into (i) the outer wake, at the same level as the one of the boundary layers developing on the plate, and in (ii) the inner wake, on the same level of the plate, in the region of the centerline. The thickness of the outer wake tends to decrease in the streamwise direction downstream the atomizing edge, and eventually merges with the inner wake. As the liquid accumulation is located in the wake zone of the prefilmer, we will focus on the inner wake. In the streamwise direction, the inner wake can be decomposed into three zones. First, into the inner wake directly downstream the edge, where the flow is laminar, and the Goldstein's solution of the boundary layer is valid. In this zone, the wake "consumes" the viscous sublayer $\delta_{\nu}$ of the upstream boundary layer. According to the Goldstein's solution, the thickness of this layer and the centerline velocity both increase with $y^{1/3}$. The existence of this laminar zone is valid until its thickness reaches $\delta_{\nu}$, \ie until it consumes the upstream viscous sublayer completely. The transition to the second zone, namely the turbulent inner wake, occurs at $y^+\sim100$ or $y^*\sim25$, where the superscripts $^+$ and $^*$ refer to lengths normalized by $\delta_{\nu}$ and $\theta_0$, respectively:
\begin{equation}
y^+ = y \, u_{\tau} / \nu
\quad \text{and} \quad
y^* = y / \theta_0
\end{equation}
with $\theta_0$ the momentum thickness of the upstream boundary layer and $u_{\tau}$ the shear velocity of the upstream boundary layer, defined as $\tau_w = \rho_g \, u_{\tau}^2$.
When the viscous sublayer is totally consumed, the wake starts to influence the logarithmic region of the upstream boundary layer. \citet{alber1980turbulent} uses this idea to justify that the scales of the turbulence are similar to the ones in the logarithmic layer (\ie $\sqrt{u'^2} \sim O(u_{\tau})$ and $\mathscr{L} \sim O(z)$ for the velocity and length scale, respectively), and that the diffusion process is dominated by turbulence. Assuming a linear increase of the eddy viscosity $\epsilon = \kappa  u_{\tau} z$ in the normal direction and the existence of similarity, \citet{alber1980turbulent} derives, among other expressions, the centerline velocity $v_c(y)$:
\begin{equation}
\frac{v_c(y)} {u_{\tau}} = \frac{1}{\kappa} \left( \ln [g(y^+)] - \gamma \right) + B
\label{eq_v_centerline_final}
\end{equation}
where $\kappa$ is the von-Karmann constant equal to 0.41, B = 5.2 and $\gamma$ is the Euler constant equal to 0.5772157. The function $g(y^+)$ is representative of a length scale in the turbulent inner wake. It is defined by \citet{alber1980turbulent} as:
\begin{equation}
g(y^+) \left( \ln[g(y^+)] - 1 \right) = \kappa^2 y^+
\label{eq_g_turbulent_alber}
\end{equation}
As stated by \citet{alber1980turbulent}, there is some arbitrariness regarding Eq.~\ref{eq_g_turbulent_alber} in the origin of $y^+$. Therefore, $y_t^+ = (y-y_{0,t}) \, u_{\tau} / \nu$ will be substituted into $y^+$ in the following. This degree of freedom in the virtual origin of the turbulent inner wake will be used later to match the experimental profiles. 
The third zone is called the far-wake region and starts at $y^+>5000$ (or $y^* > 350$). In this zone, the flow admits self-similarity, and the wake half-width and the velocity defect scale with $y^{1/2}$ and $y^{-1/2}$, respectively.

\subsection{Raw measurements of the flow field \label{ssec_raw_PIV_maps}}

\begin{figure}[!htb]
	\centering
	\includegraphics[width=\textwidth,keepaspectratio]{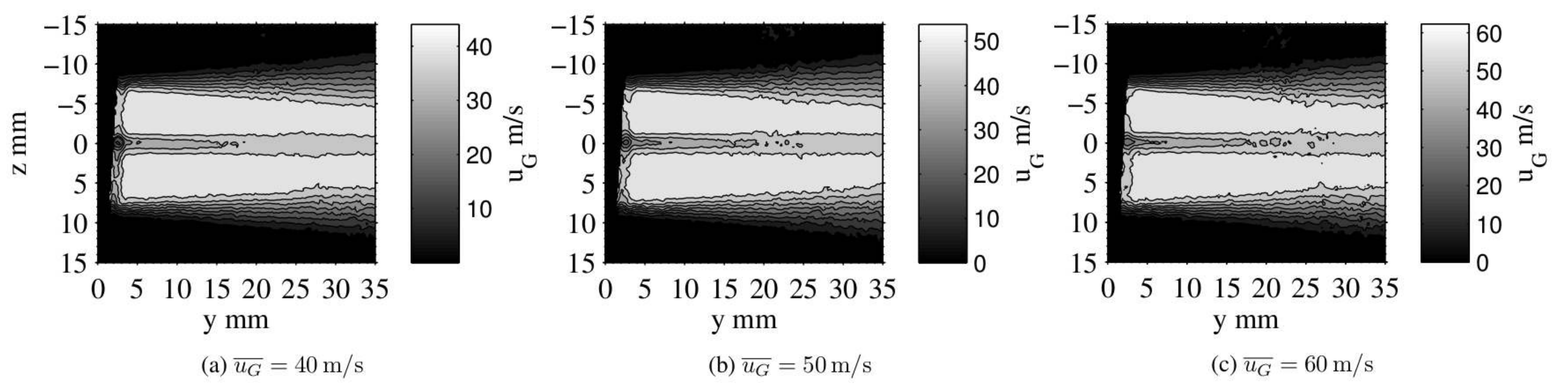}
	\caption{Air flow field at p=1~bar, measured by \citet{gepperth2019experimentelle}}
	\label{fig_PIV_gradient}
\end{figure}

Due to geometrical constraints, the lens of the PIV camera was not perpendicular to the optical access. Therefore, a geometrical transformation was necessary to recover the air flow field. 
The origin of the coordinate system is located on the atomizing edge.
Figure~\ref{fig_PIV_gradient} displays the mean velocity field for a bulk velocity $U_g$ = 40, 50 and 60 m/s at $p$ = 1~bar. The color map corresponds to an equidistant partitioning from 0 to $U_{max}$ and qualitatively shows the identical flow patterns.
All the typical features of shear layers are visible, such as the shear zones at the outside of the channel flow, the wake of the prefilmer and its subsequent inner shearing zones.

\begin{figure}[!htb]
		\centering
		\includegraphics[width=0.6\textwidth,keepaspectratio]{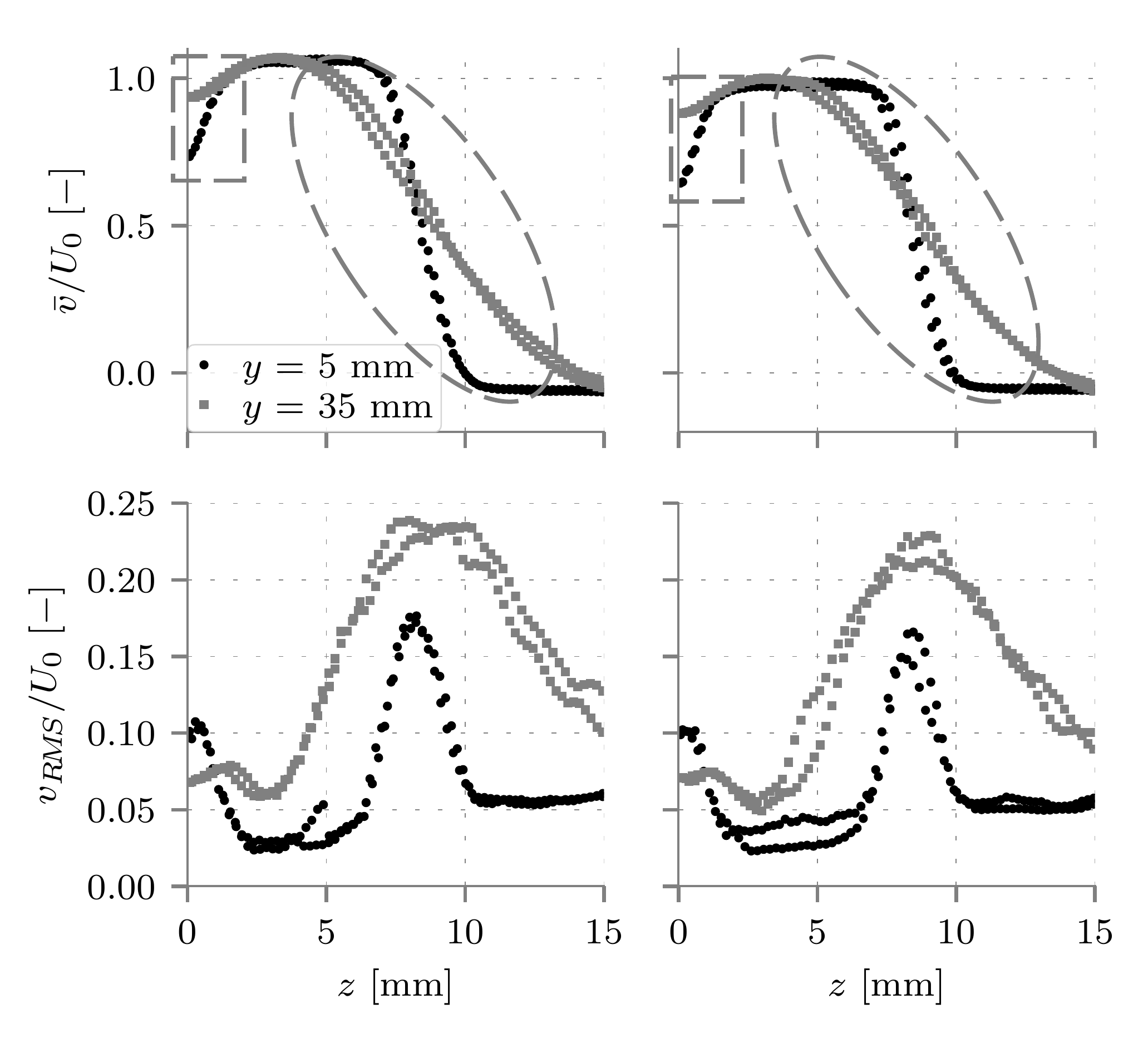}
		\caption{Mean (top) and RMS (bottom) normalized velocity profiles at p = 1~bar for $U_g$ = 40 m/s (left) and 60 m/s (right)}
		\label{fig_PIV_u_mean_RMS}
\end{figure}

\begin{figure}[!htb]
	\centering
	\includegraphics[width=0.495\textwidth,keepaspectratio]{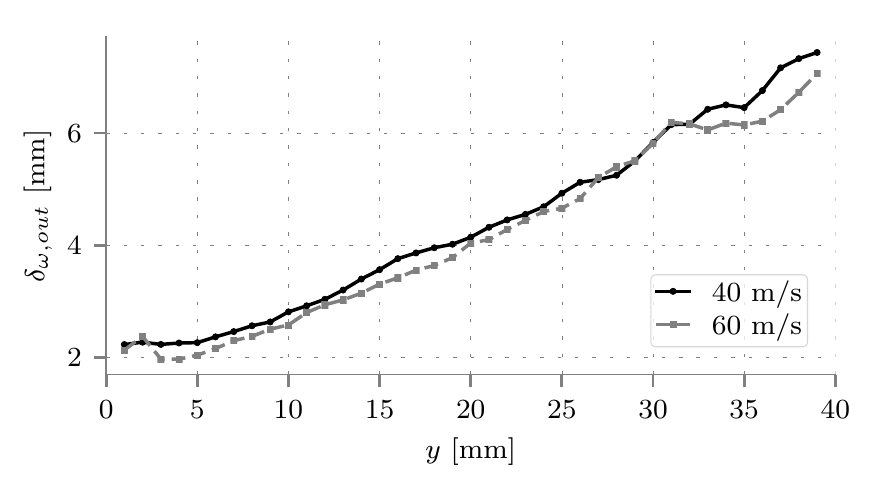}
	\includegraphics[width=0.495\textwidth,keepaspectratio]{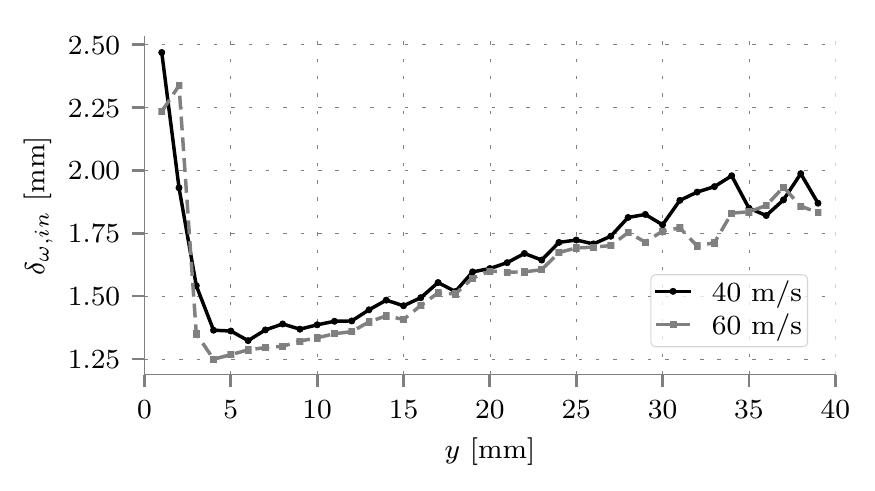}
	\caption{Evolution of the vorticity thickness $\delta_{\omega}$ of the outer (left) and the inner (right) shear layer in the longitudinal direction.}
	\label{fig_vorticity_thickness}
\end{figure}

The axial velocity profiles are plotted in figure~\ref{fig_PIV_u_mean_RMS}, in terms of mean (top) and RMS (bottom) values, at two different planes at $y$ = 5 and 35 mm for $U_g$ = 40 (left) and 60~m/s (right). 
The part of the profiles in negative $z$ direction are flipped to the positive direction in order to demonstrate the symmetry of the wake.
The mean velocity inside the wake strongly increases from $y$ = 0 mm to reach approximately 66\% of the bulk velocity at $y$ = 5 mm and finally more than 90\% at $y$~=~35 mm.
Recirculation zones are also observed at the outer part of the channel.
The RMS profiles show a typical peak at the outer shear zones (dashed line ellipse), which increases as the shear zone widens downstream. To the contrary, the peak located inside the wake zone (dashed line rectangle) decreases downstream, as a the two inner shear layers join and regularize the flow.
The profiles show a weak dependance on the bulk gas velocity, confirming the identical flow pattern as illustrated in figure~\ref{fig_PIV_gradient}.
\\The vorticity thickness $\delta_{\omega}$ is shown in figure~\ref{fig_vorticity_thickness} for the outer (left) and the inner (right) shear layers. It is defined as:
\begin{equation}
\delta_{\omega}(y) = \frac{U_{max}(y) - U_{min}(y)} {|\partial U / \partial z|_{y,max} }
\label{eq_vorti_thick}
\end{equation}
and thus gives a length scale of the momentum diffusion in the layer.
It was computed from the mean axial velocity profile at each axial position (\eg figure~\ref{fig_PIV_u_mean_RMS}) based on the PIV measurements.
The outer shear layers $\delta_{\omega,out}$ (figure~\ref{fig_vorticity_thickness} left) increases linearly with $y$, with a longitudinal gradient  $\mathrm d \delta_{\omega,out} / \mathrm dy$ of $\approx$~0.13, which is a typical value for a mixing layer \citep{Pope:2000}.
The value of $\delta_{\omega, out}$ at $y$=0 is the boundary layer thickness. It is slightly above 2 mm, which is consistent with the value given by Eq.~\ref{eq_BL_correl}. 
The value of the vorticity thickness of the inner layers (figure~\ref{fig_vorticity_thickness} right) undergoes a sharp drop inside the wake region of the prefilmer.
This is due to the fact that a recirculation zone does not exhibit the same feature as a canonical mixing layer. Therefore, $\delta_{\omega, in}$ for $y <$ 4~mm should not be considered as a vorticity thickness in a strict sense.
Outside the recirculation zone, the vorticity thickness features also a linear evolution with a longitudinal gradient  $\mathrm d \delta_{\omega,in} / \mathrm dy$ of $\approx$ 0.025, which is the half of the typical lower bound for regular mixing layers.

\subsection{Axial velocity profiles on the center line}

\begin{figure}[!htb]
		\centering
		\includegraphics[width=0.75\textwidth,keepaspectratio]{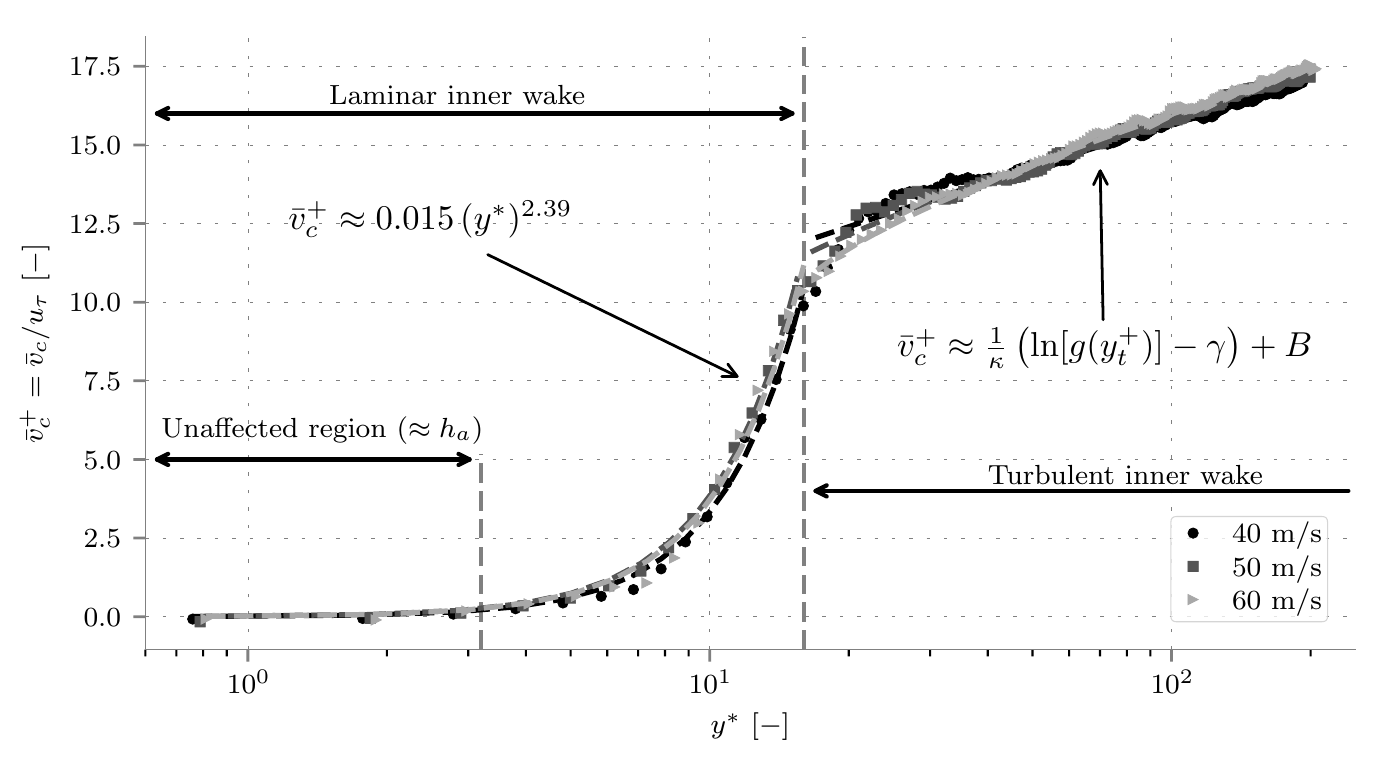}
		\caption{Mean normalized velocity profiles $\bar{v}_c/u_{\tau}$ at the center line at p = 1~bar for $U_g$ = 40, 50 and 60 m/s.}
		\label{fig_PIV_u_mean_centerline}
\end{figure}

In the following, we will compare the theory presented in Section~{\ref{ssec_theory_flow_structure}} to the present experimental results for a bulk velocity of 40, 50 and 60 m/s, for the laminar and turbulent inner wake region. We estimate the momentum thickness of the upstream boundary layer by the formula from \citet{Cousteix:1989}:
\begin{equation}
\frac{\theta_0} {l} = \frac{(A_1+1)A_2}{\Re_{\theta_0}^{A_1}}
\quad \text{with} \quad
\Re_{\theta_0} = \frac{U_g \theta_0}{\nu}
\quad \text{and} \quad
(A_1,A_2) = (0.2, \num{8.6e-3})
\label{eq_mom_thick_Cousteix}
\end{equation}
The shear velocity $u_{\tau}$ in the inner wake regions will be estimated by fitting Eq.~\ref{eq_v_centerline_final} on the experimental results in the turbulent inner wake region. The resulting $u_{\tau}$ will be compared to an estimation from \citet{Pope:2000} for a flat plate.
Note that in the aforementioned references the ratio of the momentum thickness to the splitter plate thickness ($\theta_0/h_a$) is in the range 10-20, whereas it is 0.32 in the present experiment. Hence, as it will be discussed in the following, the flow will be different inside the laminar inner wake region.\\
The mean axial velocity at the center line is shown in figure~\ref{fig_PIV_u_mean_centerline}. The axial distance is normalized by $\theta_0$, and the velocity is normalized by $u_{\tau}$. 
The three operating points match well, suggesting appropriate normalizing scales.
Different regions are observed. First, there exists a region where the velocity is almost zero.
The axial extent of this region is of the same order of magnitude as $h_a$ and corresponds to the zone unaffected by the laminar sublayer. This feature was not observed in previous experiments most presumably due to their large $\theta_0/h_a$ ratio. Then, for $3<y^*<16$, the velocity increase in a parabolic manner.
The transition between laminar and turbulent inner wake region is observed at $y^{*}\approx$16, somewhat closer to the atomizing edge than claimed in the literature ($y^*\approx$25). The turbulent inner wake extends to $y^*>250$. The zone of the far-wake ($y^*>350$) is behind the measurement volume.\\

\begin{table}%
	\small
	\centering
	\begin{tabular}{  l  l  c  c c }
	$U_g$ & [m/s] & 40 & 50 & 60 \\
	$\Re_g$ & [$- \times 1000$] & 189 & 236 & 284 \\
	\hline
	$u_{\tau}$ (Eq.~\ref{eq_v_centerline_final}) & [m/s] & 2.25 & 2.67 & 3.05 \\
	$\theta_0$ (Eq.~\ref{eq_mom_thick_Cousteix}) & [\textmu m] & 207 & 199 & 193 \\
	$y_{0,t}/\theta_0$ & [$-$] & 3.5 & 7.6 & 11 \\
	$u_{\tau}/u_{\tau,0}$ & [$-$] & 1.10 & 1.07 & 1.04 \\
	\end{tabular}
	\caption{Scales of the turbulent inner wake region and fitting constants}
	\label{tab_u_tau_gas}
\end{table} 

\begin{figure}[!htb]
		\centering
		\includegraphics[width=\textwidth,keepaspectratio]{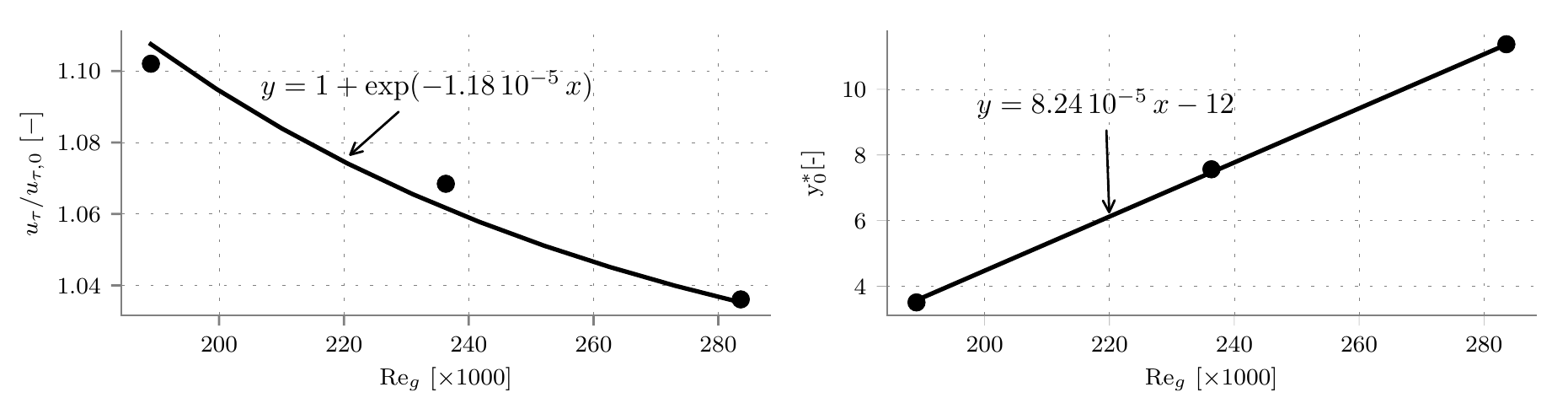}
		\caption{Evolution of $u_{\tau}/u_{\tau,0}$ (left) and $y_{0,t}/\theta_0$ (right) versus $\Re_g$.}
		\label{fig_correl_for_ratios_utau_y0star}
\end{figure}

Equation~\ref{eq_v_centerline_final} is fitted on the experiment for $y^*>16$. This leads to the determination of the shear velocity $u_{\tau}$ and the virtual origin $y_{0,t}/\theta_0$ of the logarithmic law (Eq.~\ref{eq_v_centerline_final}) for the three operating points. The quantities are listed in Table~\ref{tab_u_tau_gas}.
The shear velocity in the turbulent inner wake is compared to the shear velocity inside the boundary layer ($u_{\tau,0}$), as expressed by \citet{Pope:2000}:
\begin{equation}
\frac{U_g}{u_{\tau,0}} = \frac{1}{\kappa} \ln \left[ \Re_g \frac{u_{\tau,0}}{U_g} \right] + B +B_1
\end{equation}
where $B_1 \approx$ 0.7. The ratios $u_{\tau}/u_{\tau,0}$, as given in Table~\ref{tab_u_tau_gas}, are slightly larger than 1, suggesting that the velocity scale of the turbulence inside the turbulent inner wake region is somewhat larger than in the upstream boundary layer.
Even though the deviation is low ($<$10\%), it strongly influences the predictability of the mean velocity profile. Indeed, it was observed that the mean velocity profile is very sensitive to the value of $u_{\tau}$, and even 10\% deviation leads to significant discrepancy. Therefore, we propose empirical correlations for the ratios $u_{\tau}/u_{\tau,0}$ and $y_{0,t}/\theta_0$ versus the Reynolds number. They are shown in figure~\ref{fig_correl_for_ratios_utau_y0star}. The ratio $u_{\tau}/u_{\tau,0}$ is chosen such that its asymptotic behavior is 1, whereas the behavior on $y_{0,t}/\theta_0$ is assumed to be linear and valid only for $\Re_g$ > 190000. In this model we limit $y_0^* = y_{0,t}/\theta_0$ to 16.
\\The laminar inner wake is defined for $y^*<16$. In figure~\ref{fig_PIV_u_mean_centerline} the normalized velocities collapse remarkably well on a single curve for the three bulk velocities if $v_c^+ \approx 0.007 \, (y^*)^{2.58}$ is selected.
The discrepancy of the exponent between the Goldstein's solution (1/3) and the present results ($\sim$2.58) is striking. The most plausible cause is the different order of magnitude of the ratio $\theta_0/h_a$. As this ratio is large in the literature, the vertical distance between the center line and the plate surface is of the order of $\delta_{\nu}$. To the contrary, in this experiment, this distance is one order of magnitude larger.
This means that the fluid at the center line is affected by the upstream boundary layer at a large $z$. In figure~\ref{fig_PIV_u_mean_centerline}, the $y^*$ position where the mean velocity becomes significant (\ie $\bar{v}_c^+ \approx 1$) is found for $y^* \approx 3.2$.

\begin{figure}[!htb]
		\centering
		\includegraphics[width=0.75\textwidth,keepaspectratio]{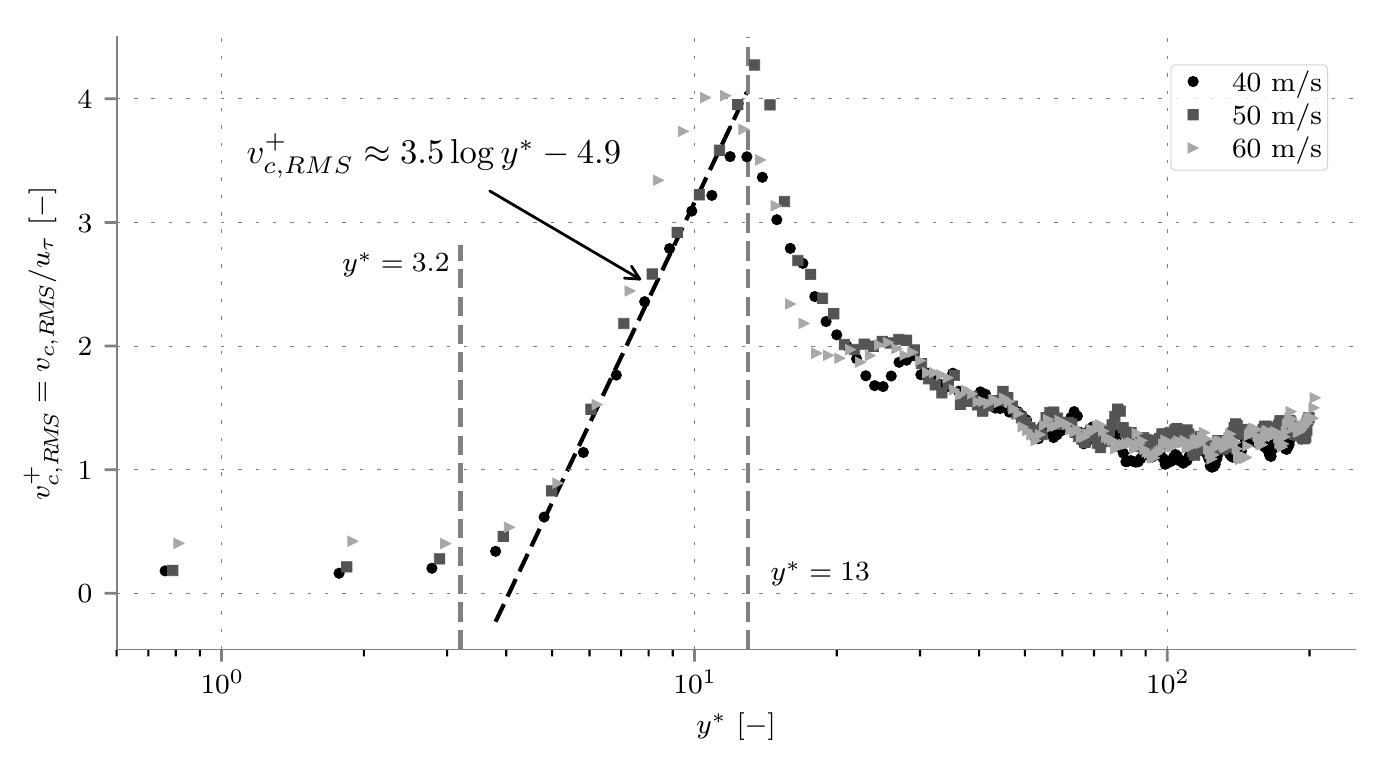}
		\caption{Profiles of the RMS of the normalized axial velocity $v_{c,RMS}/u_{\tau}$ at the center line at p = 1~bar for $U_g$ = 40, 50 and 60 m/s.}
		\label{fig_PIV_v_RMS_centerline}
\end{figure}

The RMS of the axial velocity is shown in figure~\ref{fig_PIV_v_RMS_centerline} with the same normalization as for the mean values. The data of the three operating points also collapse well together, confirming that $u_{\tau}$ and $\theta_0$ are the proper scales in the near wake flow.
The same three zones as for the mean velocity are distinguished. First, for $y^*< 3.2$ the fluctuations of the axial velocity $v_{RMS}^+$ is rather constant, and lower than $u_{\tau}$. This corresponds to the region where the upstream layer does not influence the flow at the center line. Second, for  $y^*$ in the range 3.2-13, $v_{c,RMS}^+$ follows a logarithmic growth versus the variable $y^*$. After reaching a peak at $y^* \approx$ 13, the axial fluctuations decrease to a constant value roughly equal to $u_{\tau}$. The location of the peak corresponds to the end of the laminar inner layer in figure~\ref{fig_PIV_u_mean_centerline}. 

\begin{figure}[!htb]
		\centering
		\includegraphics[width=0.75\textwidth,keepaspectratio]{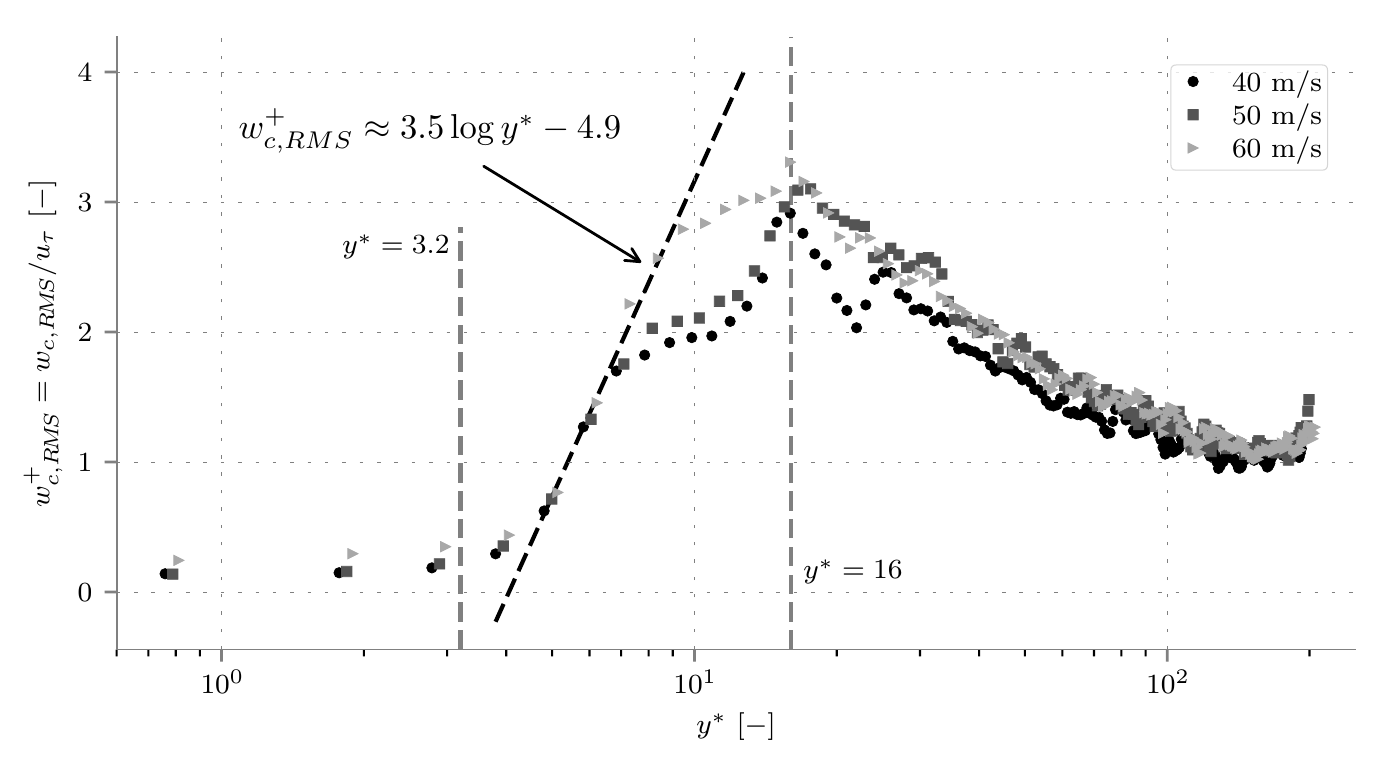}
		\caption{Profiles of the RMS of the normalized vertical velocity $w_{c,RMS}/u_{\tau}$ at the center line at p = 1~bar for $U_g$ = 40, 50 and 60 m/s.}
		\label{fig_PIV_w_RMS_centerline}
\end{figure}

The RMS of the vertical velocity (figure~\ref{fig_PIV_w_RMS_centerline}) feature also the three zones observed for $v_{c,RMS}^+$. 
The peak of $w_{c,RMS}^+$ is located $y^* \approx$ 16, which is somewhat larger than the location of the peak of $v_{c,RMS}^+$. In the zone of logarithmic growth (3.2$<y^*<$13), the same logarithmic law found for $v_{c,RMS}^+$ matches remarkably well with the data of $U_g$~=~60 m/s.
For $U_g$ = 40 and 50 m/s, there is a RMS velocity deficit at $y^*\approx$10. 
This is similar to the RMS of the axial velocity (figure~{\ref{fig_PIV_v_RMS_centerline}}) where $v_{c,RMS}/u_{\tau}$ is slightly higher for $U_g$~=~60 m/s in the range $y^*\approx$8-11. This could be attributed to a non-linear effect due to a larger turbulent mixing at higher Reynolds number.
Apart from this discrepancy between the different velocities, the data points collapse well together.

The interesting feature of the velocity profiles plotted in figures~\ref{fig_PIV_u_mean_centerline}-\ref{fig_PIV_w_RMS_centerline} is that they solely depend on the scales $u_{\tau}$ and $\theta_0$. The shear velocity inside the inner wake is predicted by a classical correlation of $u_{\tau,0}$ and corrected by the behavior shown in figure~\ref{fig_correl_for_ratios_utau_y0star}, while the momentum thickness is estimated by Eq.~\ref{eq_mom_thick_Cousteix}. 
Since these expressions depend mainly on the Reynolds number, the air density
is taken into account for the velocities in the inner wake.
This will be useful to estimate the mean and the fluctuations of the gas velocity in the region where the liquid is fragmented at different ambient pressures. It is to be expected that the gas flow will be disturbed by the presence of the liquid, nevertheless, it gives a good and simple approximation of the flow field in the vicinity of the atomizing edge.

\section{Qualitative observations of the liquid phase\label{sec_qualit_liquid}}

\begin{figure}%
\centering
		\includegraphics[width=0.495\textwidth,keepaspectratio]{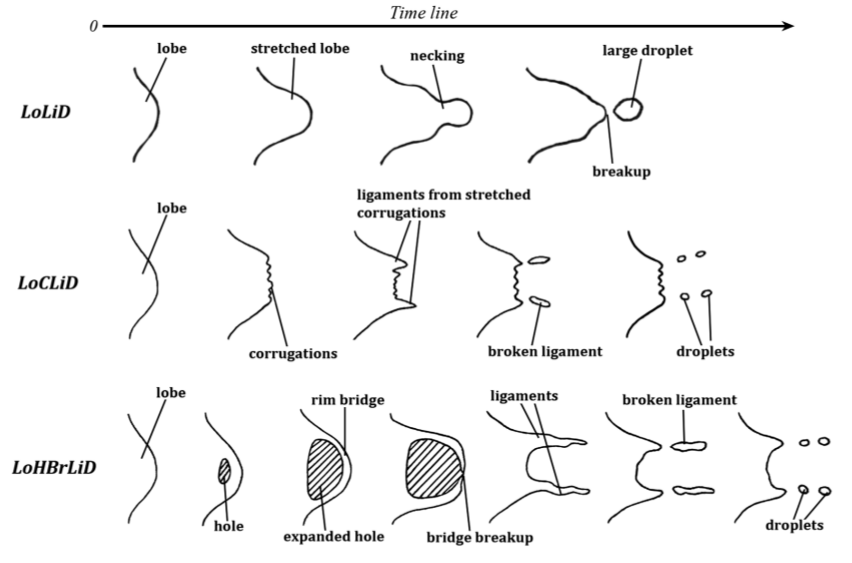}
	\caption{Nomenclature of different breakup sequences.\protect\footnotemark}
	\label{fig_nomenclature_zandian}
\end{figure}

In this section the breakup process at the tip of the prefilmer will be qualitatively discussed.
Recently, \citet{zandian2017planar} performed a DNS of the breakup of a liquid sheet immersed into a turbulent flow by means of a VoF-Level-Set method. They identified several breakup sequences and proposed a nomenclature, as compiled in figure~\ref{fig_nomenclature_zandian}.
\footnotetext{\footnotesize Reproduced from \textit{A. Zandian, W. Sirignano, F. Hussain, \href{https://aip.scitation.org/doi/abs/10.1063/1.4986790}{Planar liquid jet: Early deformation and atomization cascades}, Physics of Fluids 29 (6) (2017)}, with the permission of AIP Publishing.}
The LoLiD sequence refers to as Lo=Lobe, Li=Ligament and D=Droplet, and occurs when the initial liquid lobe is stretched axially and later is disrupted into droplets by capillary forces. This regime is triggered at low Reynlods and low Weber numbers and strongly depends on the viscosity of the liquid. 
The LoHBrLiD sequence (H=Hole, Br=Bridge) occurs at large Weber numbers.
In this case, the dynamic pressure of the gas at the center of the lobe is large enough to thin it.
This creates a membrane which expands and generate a bag. The outer sides of the lobe remain thick and constitute a so-called bridge (or rim) that encloses the bag.
The bag is eventually disrupted while the rim collapses due to capillary forces. The remnants of the bag rims are elongated and undergo a ligament breakup.
Finally, the LoCLiD sequence (C=Corrugation) occurs at larger Reynlods and lower Ohnesorge number. In this case, no membrane is generated and thinner ligaments appears on the sides of the lobe.\\
Although their configuration is different from the present one, the sequence and the characteristics of the individual breakup events are similar to the ones observed here. As it will be seen later, the flow conditions leading to the appearance of these breakup sequences differ between the two configurations. 
First, the fundamental mechanism of the breakup occurring at the atomizing edge is different from the one occurring on a liquid sheet. Second, the complexity of the flow field at the atomizing edge leads to a large variety of Reynolds and Weber number, and thus different breakups are involved simultaneously.
For the second possibility, it means that in the case of accumulation breakup, the location of the breakup event is primordial to determine its type of sequence. Indeed, the velocity seen by the ligament will be very different if the breakup occurs in the wake of the prefilmer or in the high velocity flow. Since the shear imposed by the aerodynamic stress is proportional to $U_g^2$, it varies on a range larger than the one of the velocity, and thus might lead to different breakup types.
Recently, \mbox{\cite{ling2019two}} investigated the interaction between gas turbulence and interfacial instabilities with a Direct Numerical Simulation (DNS) of a liquid layer sheared by a turbulent flow. The authors showed that the waves on the liquid interface, due to a blockage effect on the mean flow, enhance the production of turbulence, which in turn is expected to promote the atomization downstream. This effect should to be active here. However, this topic is out of the scope of the present paper.

\subsection{Time series of single breakup processes}

\begin{figure}[!htb]
\centering
\subfloat[\label{fig_time_serie_LoHBrLiD1}0]{\includegraphics[width=\timeseriewidth,keepaspectratio]{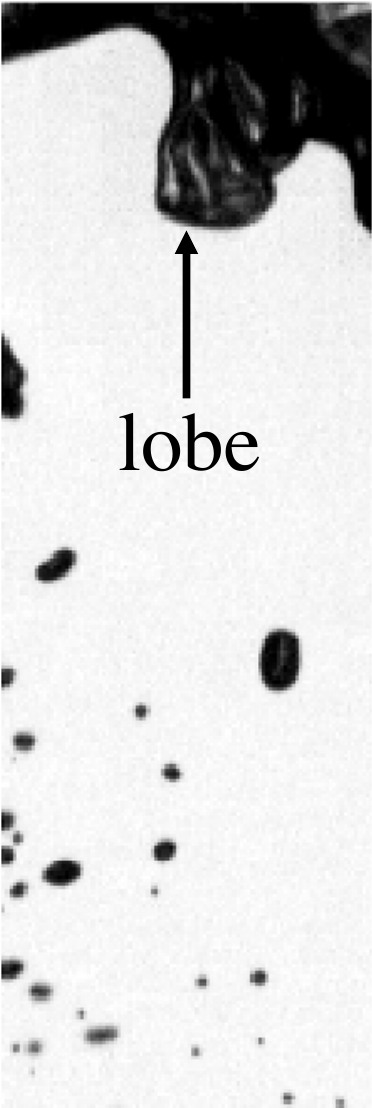}}
\hspace{2mm}
\subfloat[\label{fig_time_serie_LoHBrLiD2}71]{\includegraphics[width=\timeseriewidth,keepaspectratio]{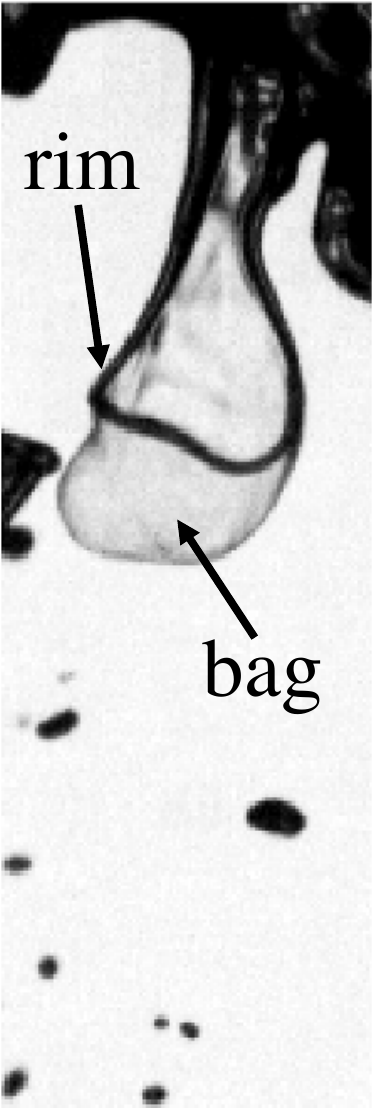}}
\hspace{2mm}
\subfloat[\label{fig_time_serie_LoHBrLiD3}143]{\includegraphics[width=\timeseriewidth,keepaspectratio]{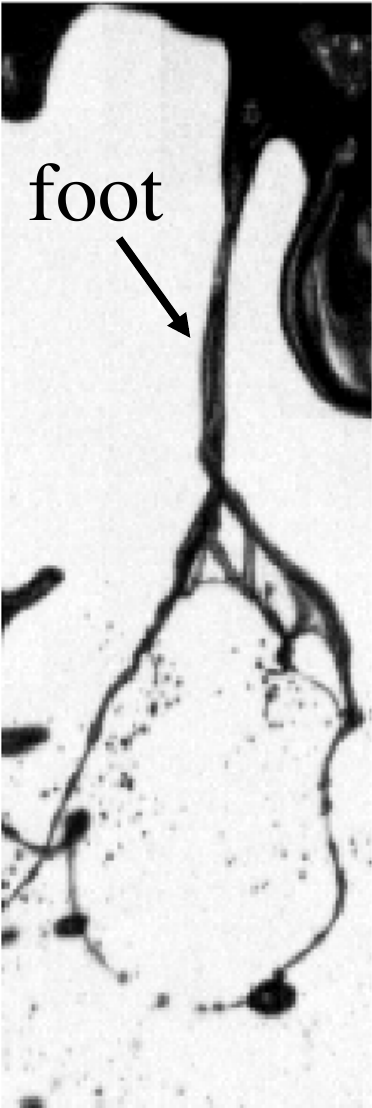}}
\hspace{2mm}
\subfloat[\label{fig_time_serie_LoHBrLiD4}214]{\includegraphics[width=\timeseriewidth,keepaspectratio]{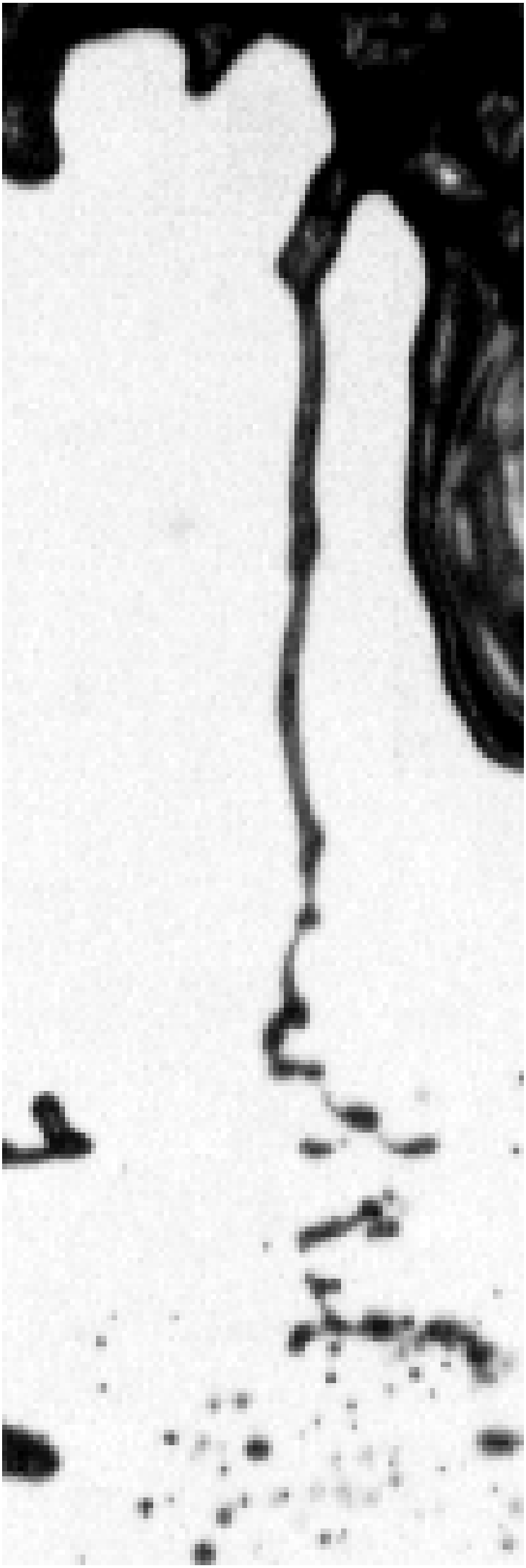}}
\hspace{2mm}
\subfloat[\label{fig_time_serie_LoHBrLiD5}286]{\includegraphics[width=\timeseriewidth,keepaspectratio]{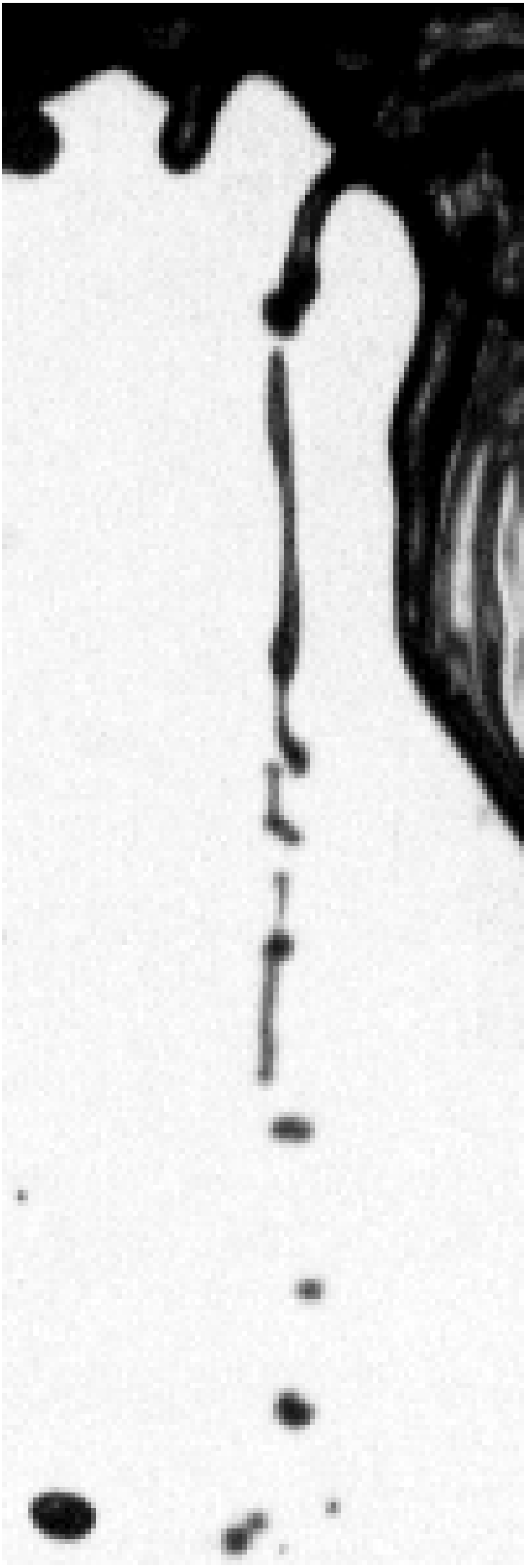}}
\hspace{2mm}
\subfloat[\label{fig_time_serie_LoHBrLiD6}357]{\includegraphics[width=\timeseriewidth,keepaspectratio]{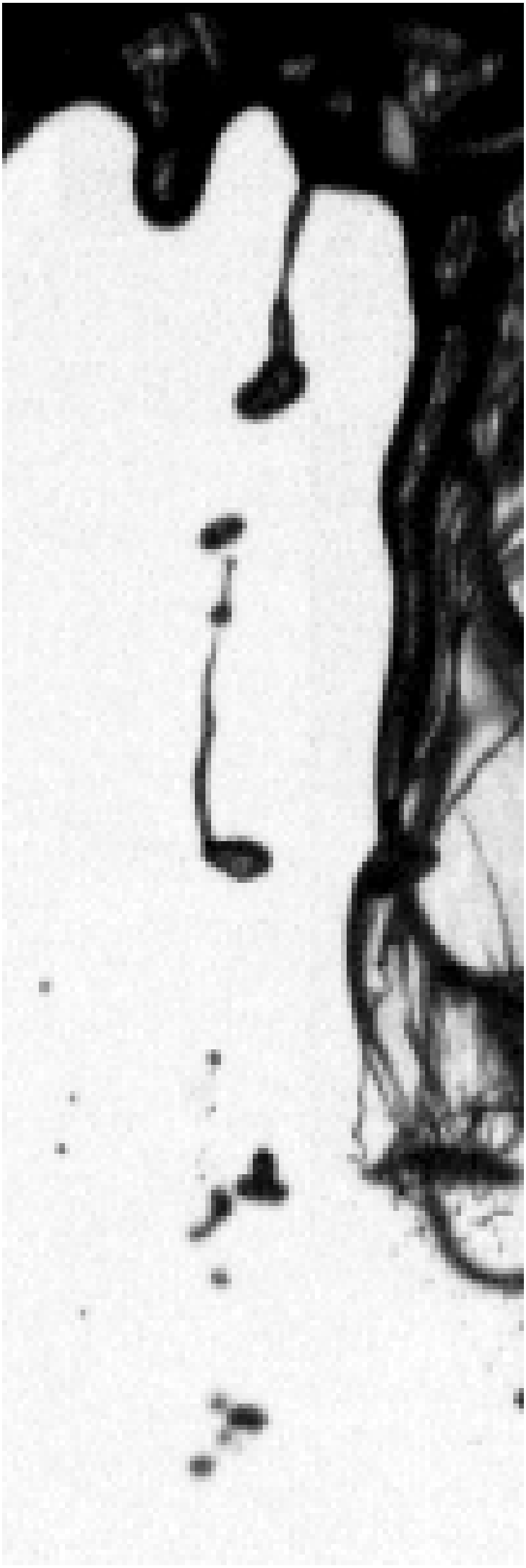}}
\hspace{2mm}
\subfloat[\label{fig_time_serie_LoHBrLiD7}429]{\includegraphics[width=\timeseriewidth,keepaspectratio]{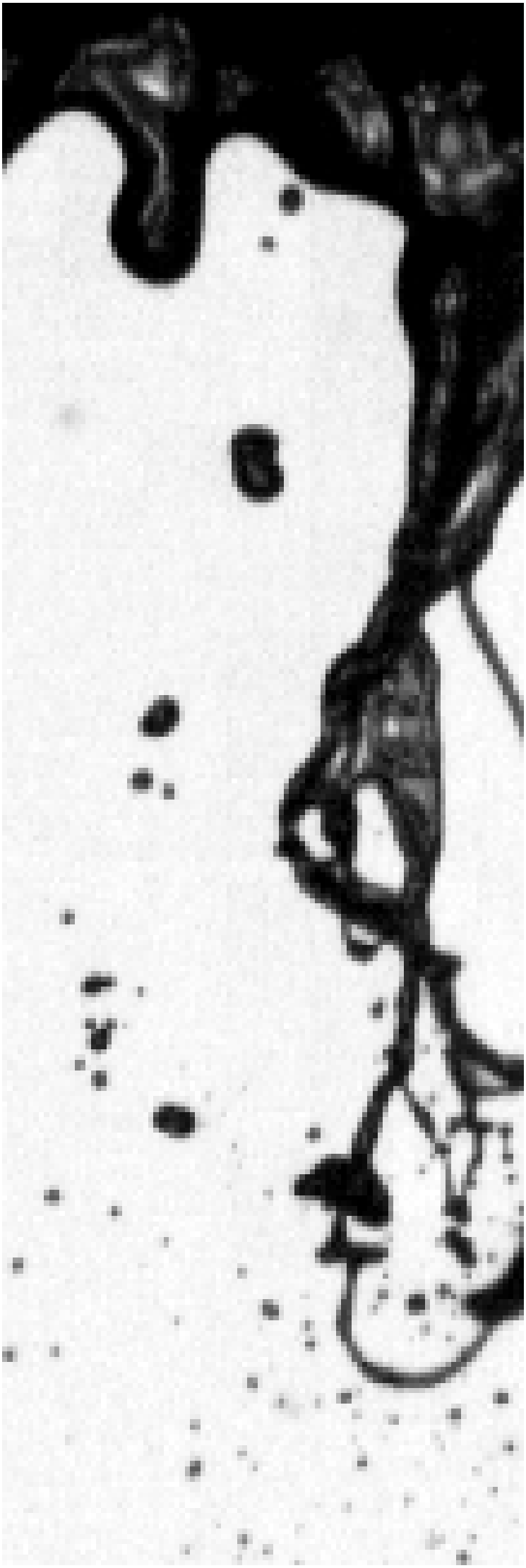}}
\caption{Time series of a single breakup event similar to LoHBrLiD for p = 3~bar and $U_g$ = 40 m/s. The time scale under each image is in \textmu s.}
\label{fig_time_serie_LoHBrLiD}
\end{figure}

In this section the different types of breakup sequence will be described. First, in figure~\ref{fig_time_serie_LoHBrLiD} the most identifiable type is depicted, which shows strong similarities to LoHBrLiD. 
The process starts with a sufficiently large amount of liquid accumulated at the prefilmer tip.
The mass of liquid is stretched in the streamwise direction in the shape of a tongue (figure~\ref{fig_time_serie_LoHBrLiD1}).
Due to the large dynamic pressure that acts on the lobe surface, the center of the tongue is stretched out like in a bag-breakup phenomenon, while the sides of the tongue form the rim (figure~\ref{fig_time_serie_LoHBrLiD2}). During the formation of this structure, it is continuously stretched in the streamwise direction.
The membrane of the bag as well as the top of the rim are quickly fragmented (figure~\ref{fig_time_serie_LoHBrLiD3}).
Out of the membrane very fine droplets are formed whereas from the rim significantly larger droplets are generated.
The remaining part of the structure, \ie the rest of the rim and the stretched foot contract back to reform into a ligament (figure~\ref{fig_time_serie_LoHBrLiD4}), which is further stretched and eventually breaks up (figure~\ref{fig_time_serie_LoHBrLiD5}) into droplets of larger size (figure~\ref{fig_time_serie_LoHBrLiD6} and \ref{fig_time_serie_LoHBrLiD7}).
Several deviations from this canonical behavior can occur. For instance, if the foot of the structure remaining after the first breakup
contains a sufficient amount of liquid, the same process may be reinitiated.

\begin{figure}[!htb]
\centering
\subfloat[\label{fig_time_serie_LoLiD1}0]{\includegraphics[width=\timeseriewidth,keepaspectratio]{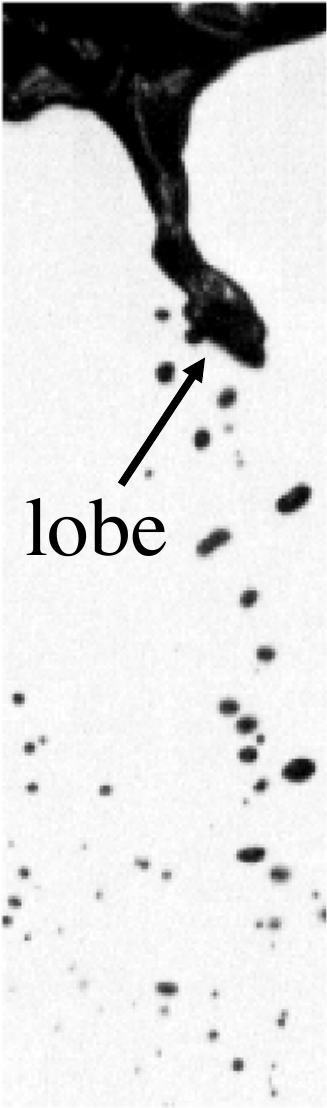}}
\hspace{2mm}
\subfloat[\label{fig_time_serie_LoLiD2}71]{\includegraphics[width=\timeseriewidth,keepaspectratio]{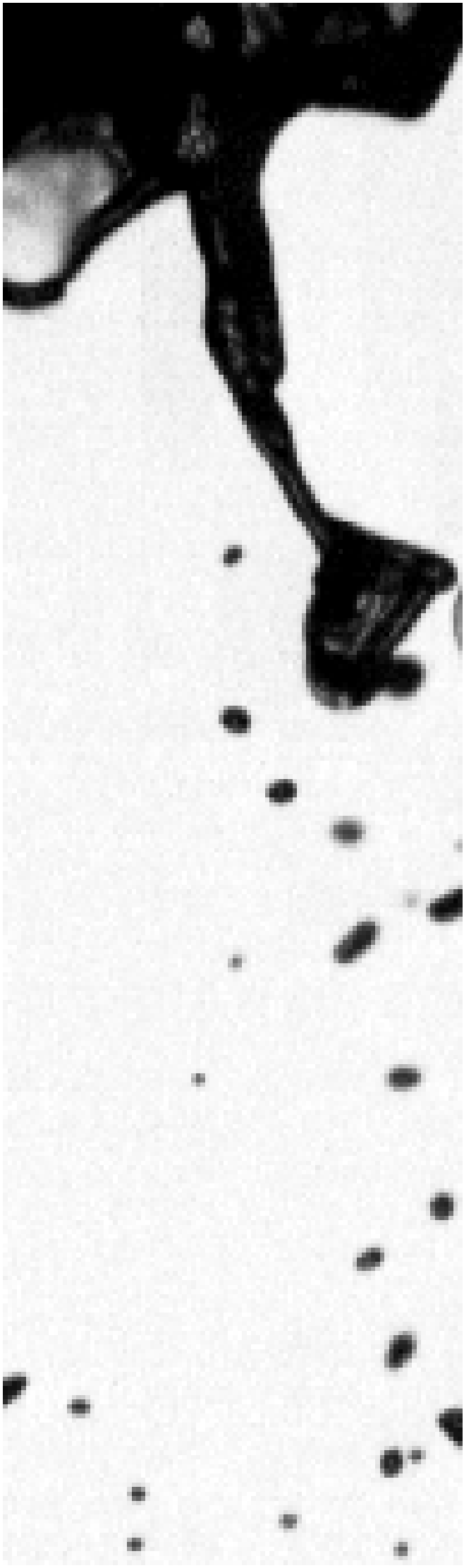}}
\hspace{2mm}
\subfloat[\label{fig_time_serie_LoLiD3}143]{\includegraphics[width=\timeseriewidth,keepaspectratio]{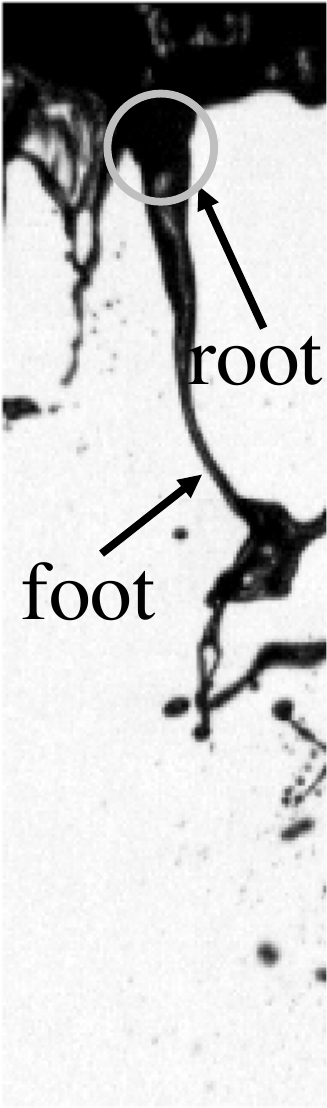}}
\hspace{2mm}
\subfloat[\label{fig_time_serie_LoLiD4}214]{\includegraphics[width=\timeseriewidth,keepaspectratio]{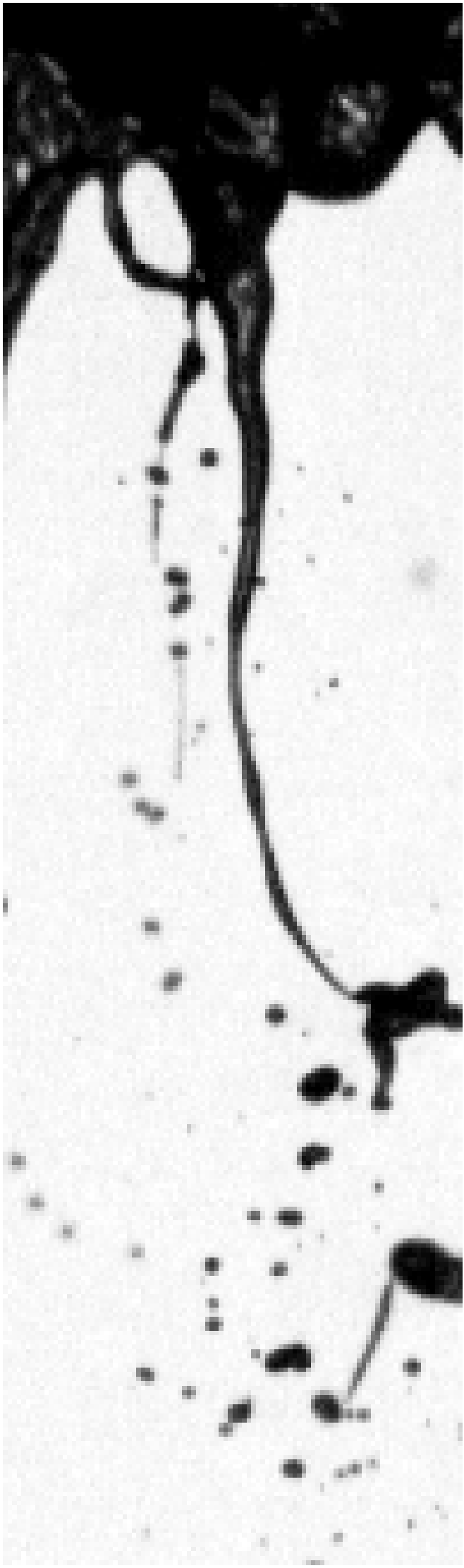}}
\hspace{2mm}
\subfloat[\label{fig_time_serie_LoLiD5}286]{\includegraphics[width=\timeseriewidth,keepaspectratio]{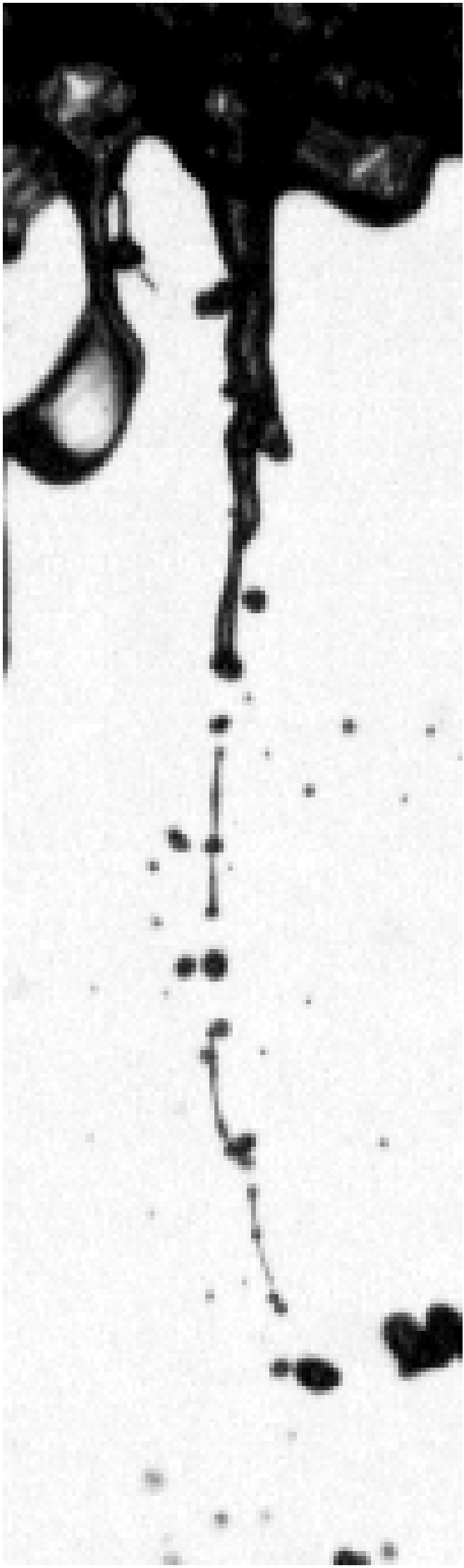}}
\caption{Time series of a single breakup event similar LoLiD for p = 3~bar and $U_g$=40 m/s. The time scale under each image is in \textmu s.}
\label{fig_time_serie_LoLiD}
\end{figure}

The ligament breakup sequence, also referred to as LoLiD \citep{zandian2017planar} is depicted in figure~\ref{fig_time_serie_LoLiD}. The starting point of this sequence may be a lobe emerging from the breakup accumulation, or it may also be the remnant of 
a previous bag breakup
as it is the case in figure~\ref{fig_time_serie_LoLiD1}. This initial lobe exhibits a surface prependicular to the streamwise direction, but, contrary to LoHBrLiD, the dynamic pressure of the gas is not large enough with respect to the frontal area to pinch the surface of the lobe. Hence, it is stretched in the streamwise direction (figure~\ref{fig_time_serie_LoLiD2}).
The lobe is extended faster than liquid is provided by the accumulation at the ligament root. As consequence, the lobe becomes thinner (figure~\ref{fig_time_serie_LoLiD3}) and eventually forms a ligament (figure~\ref{fig_time_serie_LoLiD4}).
Finally, this ligament is disrupted by the so-called ligament breakup, as described in \citet{Marmottant:2004}, producing droplets larger than the ligament diameter at the moment of breakup (figure~\ref{fig_time_serie_LoLiD5}).

\begin{figure}[!htb]
\centering
\subfloat[\label{fig_time_serie_hybrid1}0]{\includegraphics[width=\timeseriewidth,keepaspectratio]{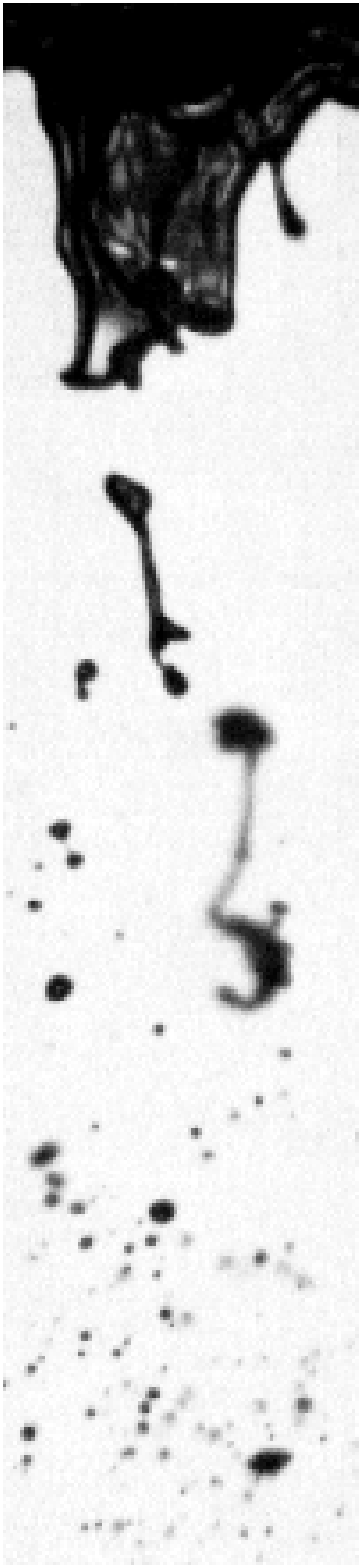}}
\hspace{2mm}
\subfloat[\label{fig_time_serie_hybrid2}71]{\includegraphics[width=\timeseriewidth,keepaspectratio]{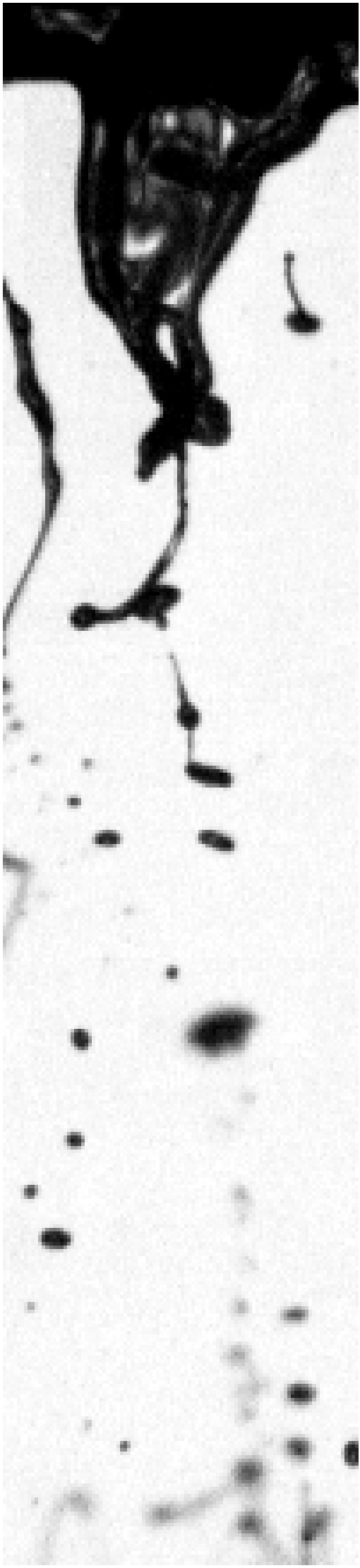}}
\hspace{2mm}
\subfloat[\label{fig_time_serie_hybrid3}143]{\includegraphics[width=\timeseriewidth,keepaspectratio]{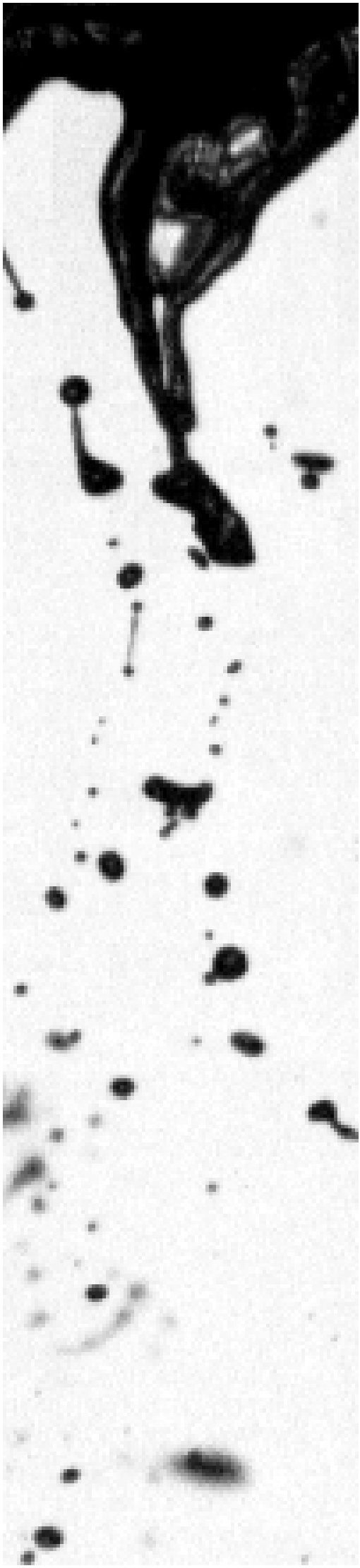}}
\hspace{2mm}
\subfloat[\label{fig_time_serie_hybrid4}214]{\includegraphics[width=\timeseriewidth,keepaspectratio]{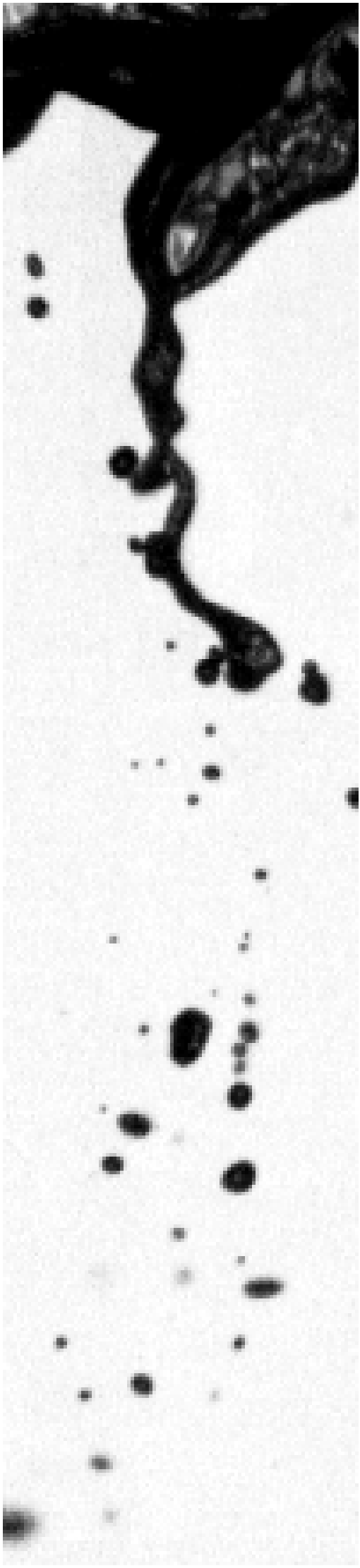}}
\hspace{2mm}
\subfloat[\label{fig_time_serie_hybrid5}286]{\includegraphics[width=\timeseriewidth,keepaspectratio]{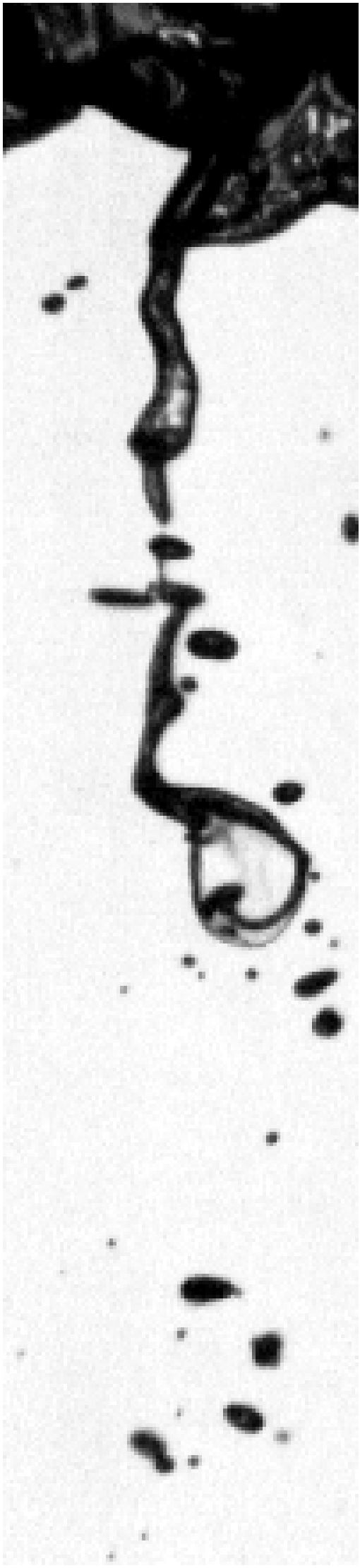}}
\hspace{2mm}
\subfloat[\label{fig_time_serie_hybrid6}357]{\includegraphics[width=\timeseriewidth,keepaspectratio]{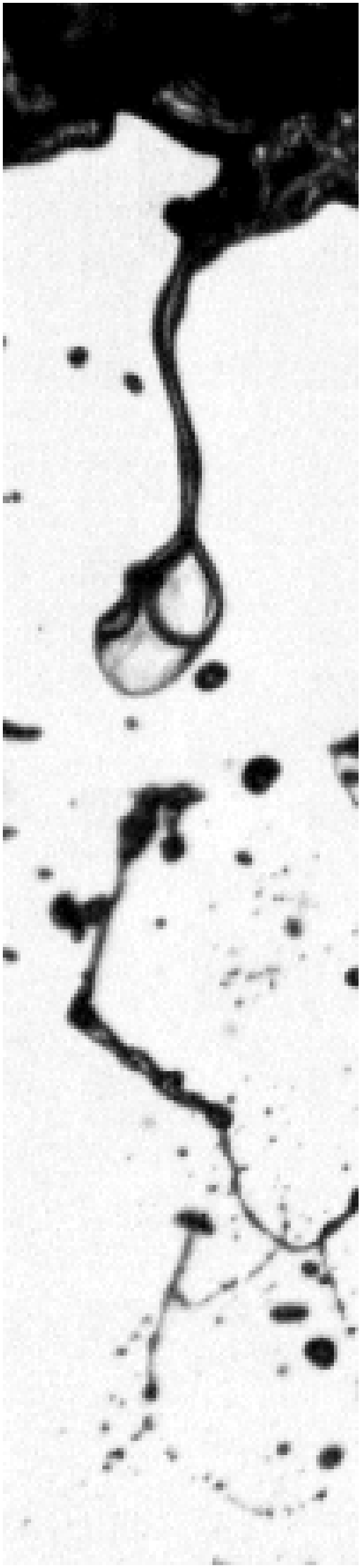}}
\hspace{2mm}
\subfloat[\label{fig_time_serie_hybrid7}429]{\includegraphics[width=\timeseriewidth,keepaspectratio]{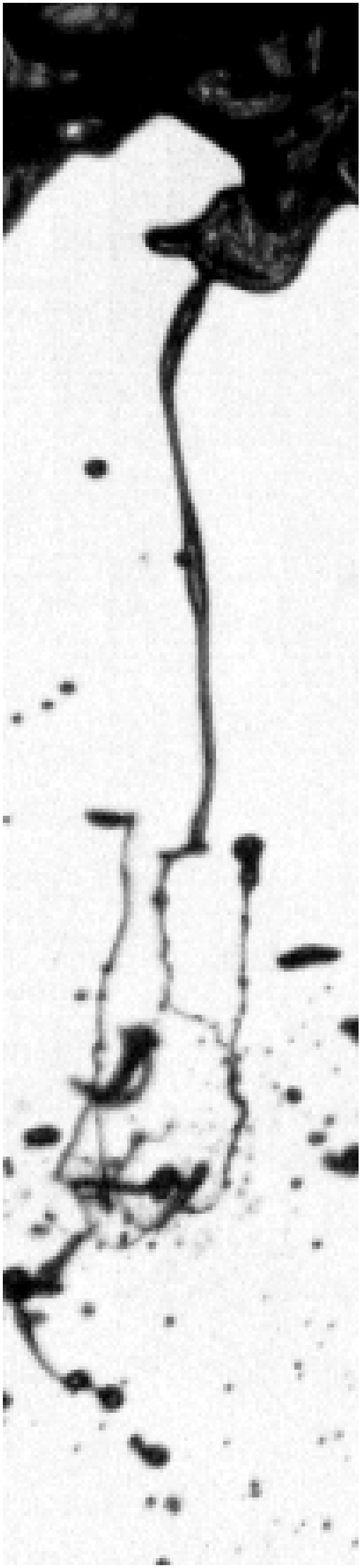}}
\hspace{2mm}
\subfloat[\label{fig_time_serie_hybrid8}500]{\includegraphics[width=\timeseriewidth,keepaspectratio]{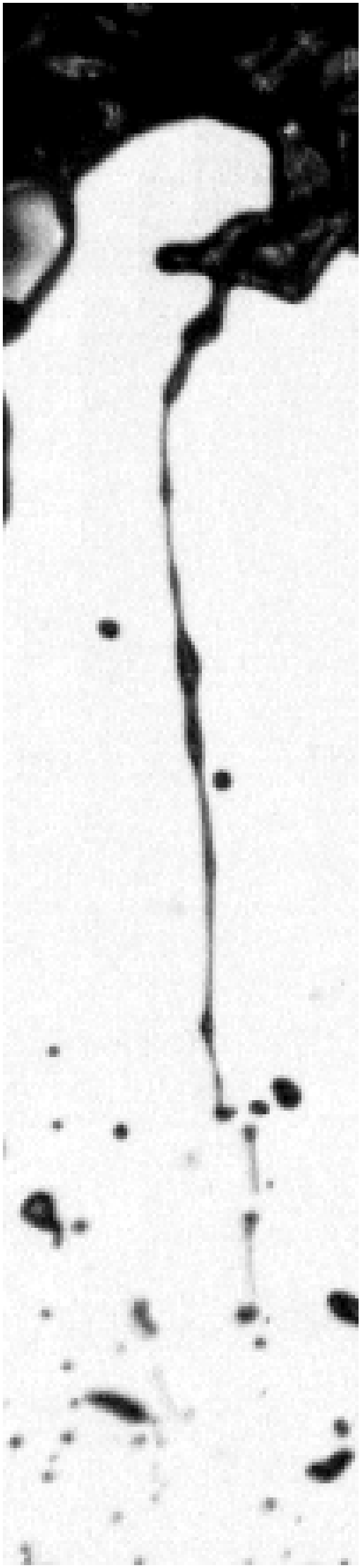}}
\hspace{2mm}
\subfloat[\label{fig_time_serie_hybrid9}571]{\includegraphics[width=\timeseriewidth,keepaspectratio]{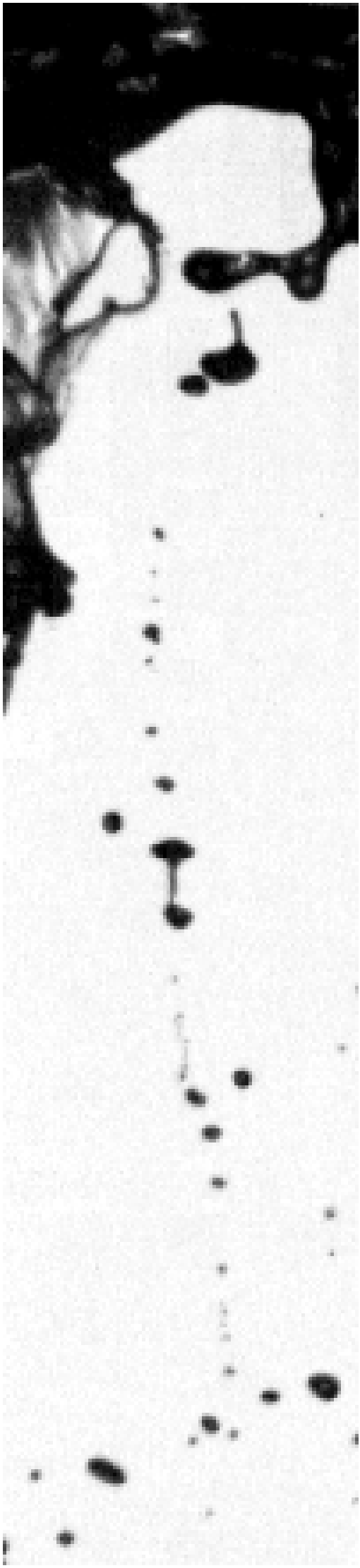}}
\caption{Time series of a single hybrid breakup event for p = 3~bar and $U_g$=40 m/s. The time scale under each image is in \textmu s.}
\label{fig_time_serie_hybrid}
\end{figure}

The sequence with the corrugation of the lobe (LoCLiD) was rarely observed in the present configuration. Hence, we were not able to isolate such an event. Consequently, it is not described in detail here.
It was frequently observed that the breakup sequence was a combination of LoHBrLiD and LoLiD. This hybrid breakup sequence is depicted in figure~\ref{fig_time_serie_hybrid}. First, a large lobe is stretched and appears to undergo a LoHBrLiD sequence (figure~\ref{fig_time_serie_hybrid1}). It is further stretched and a proto-bag is discernible in figure~\ref{fig_time_serie_hybrid2} and \ref{fig_time_serie_hybrid3}. However, the two sides of the rim come closer, so that the bag does not inflate, and the ligament is further stretched (figure~\ref{fig_time_serie_hybrid4}). As the tip of the ligament contains a significant amount of liquid, a LoHBrLiD occurs at the tip of the ligament (figure~\ref{fig_time_serie_hybrid5}), which is detached itself from the root of the initial ligament. The LoHBrLiD sequence further proceeds on the detached ligament, which is advected downstream, while another LoHBrLiD starts at the tip of the remaining part (figure~\ref{fig_time_serie_hybrid6}). Then the second LoHBrLiD is quickly dragged away by the gas, leading to a quickly elongated ligament (figure~\ref{fig_time_serie_hybrid7} and \ref{fig_time_serie_hybrid8}) which eventually breaks up (figure~\ref{fig_time_serie_hybrid9}).

\subsection{Influence of the ambient pressure and the gas velocity on the length scales}

\begin{figure*}%
	\begin{center}
	\linespread{1.0}
	{\scriptsize
	\def \svgwidth {\textwidth}
	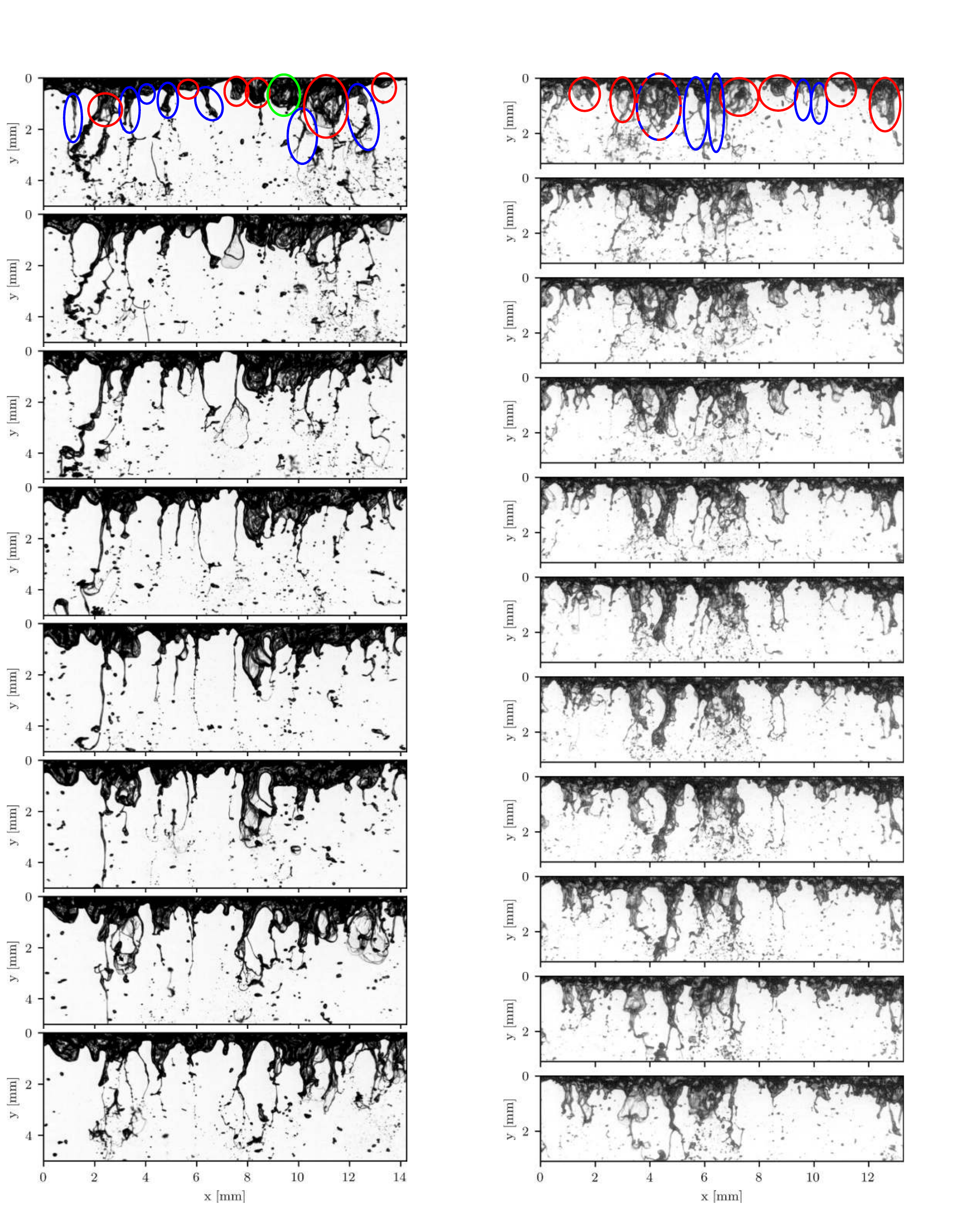
	}
	\caption{Time series of the liquid accumulation breakup for an ambient pressure of 3 (left) and 7 (right)~bar at 40 m/s.}
	\label{fig_time_serie_p_influence}
	\end{center}
\end{figure*}

The previous time series clearly shows the role of the liquid accumulation during the breakup process. Indeed, the liquid accumulation provides a liquid tank which feeds the ligaments during the stretching and breakup phase. During the whole process, no influence of the film flow is observed. The breakup process occurs continuously, independently of the incoming film waves. This independence is due to the moderate film loading, in comparison to \citet{dejean2016experimental}.\\
As mentioned earlier, although \citet{zandian2017planar} observed that one given operating point corresponds to one given type of sequence, it is not the case here. To illustrate this statement, two time series are displayed in figure~\ref{fig_time_serie_p_influence} for an ambient pressure of 3 (left) and 7 (right)~bar at 40 m/s. At the beginning of each series, different lobes are identified depending of their subsequent breakup sequence. It is observed that (i) different breakup sequences can occur at the same time for the same operating point and (ii) the pressure does not influence the type of sequence. It must be emphasized that the surface of the liquid accumulation is very distorted, which makes it difficult to analyze the morphological structure of accumulation breakup in general. Any automated shape recognition technique results in poor results or lack of representativeness.

\begin{figure}%
	\centering
	\begin{tabular}{ c  c  c  c}
      &
	p = 3~bar & p = 5~bar & p = 7~bar \\ 
	{\small \rotatebox{90}{\ \ \ \ \ \ \ \ \ $U_g$ = \SI[per-mode=symbol]{40}{\meter \per \second}}} &
	\includegraphics[width=\gommettes,keepaspectratio]{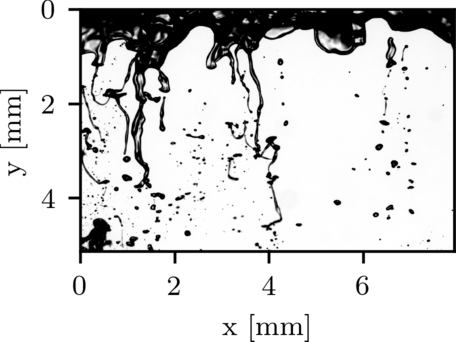}&
	\includegraphics[width=\gommettes,keepaspectratio]{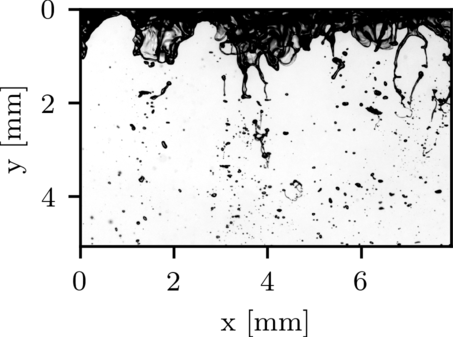}&
	\includegraphics[width=\gommettes,keepaspectratio]{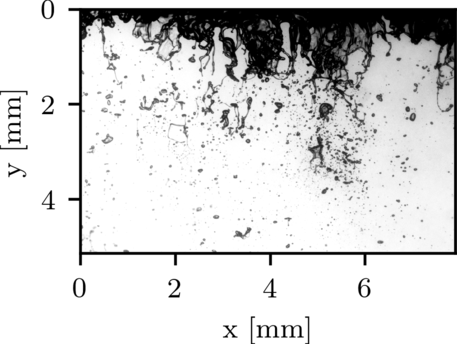}\\
	{\small \rotatebox{90}{\ \ \ \ \ \ \ \ \ $U_g$ = \SI[per-mode=symbol]{80}{\meter \per \second}}} &
	\includegraphics[width=\gommettes,keepaspectratio]{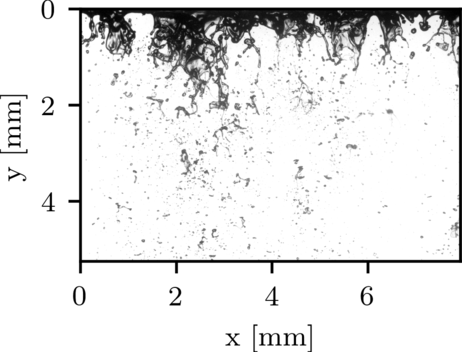}&
	\includegraphics[width=\gommettes,keepaspectratio]{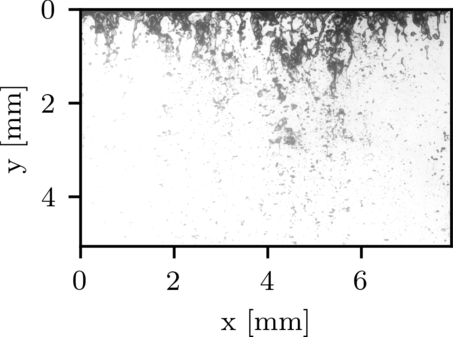}&
	\includegraphics[width=\gommettes,keepaspectratio]{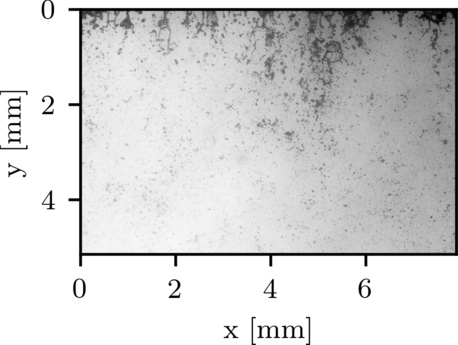}\\

	\end{tabular}\\
	\caption{Snapshots of liquid accumulation breakup for different bulk velocities and ambient pressures.}
	\label{fig_p_and_u_influence}
\end{figure}

The effect of the ambient pressure and bulk velocity is shown in figure~\ref{fig_p_and_u_influence}.
The structure of the liquid accumulation and the ligament is left unchanged with the same occurrences of LoCLiD and LoHBrLiD for all operating points. This suggests that the same mechanism occurs independently of the velocity and the pressure for the ranges investigated here.
However, a clear change in length scale is observed. With an increasing pressure and/or velocity, the global length scale of the bags, ligaments and generated droplets decrease. Obviously, the velocity has a stronger influence than ambient pressure.

\section{Quantitative observations\label{sec_quantit_liquid}}

\subsection{Influence of the gas velocity}

\begin{figure}%
\centering
		\includegraphics[width=0.495\textwidth,keepaspectratio]{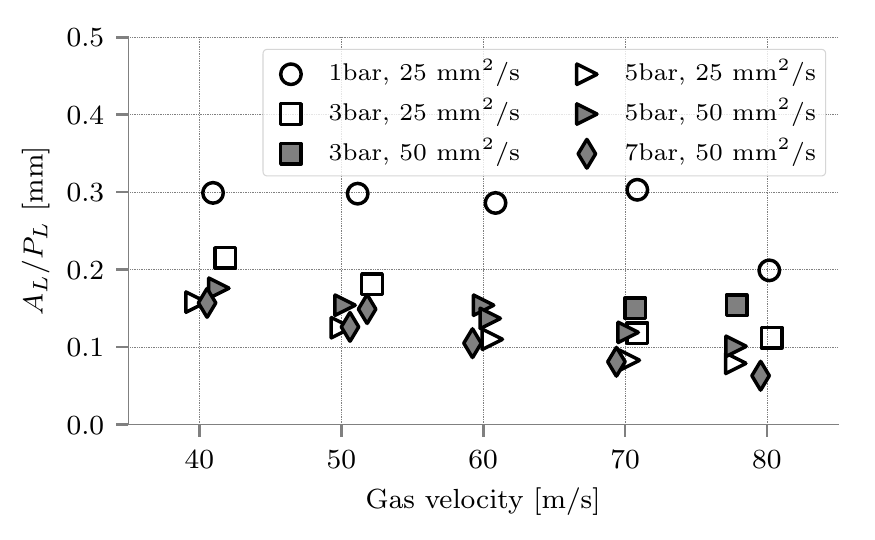}
		\includegraphics[width=0.495\textwidth,keepaspectratio]{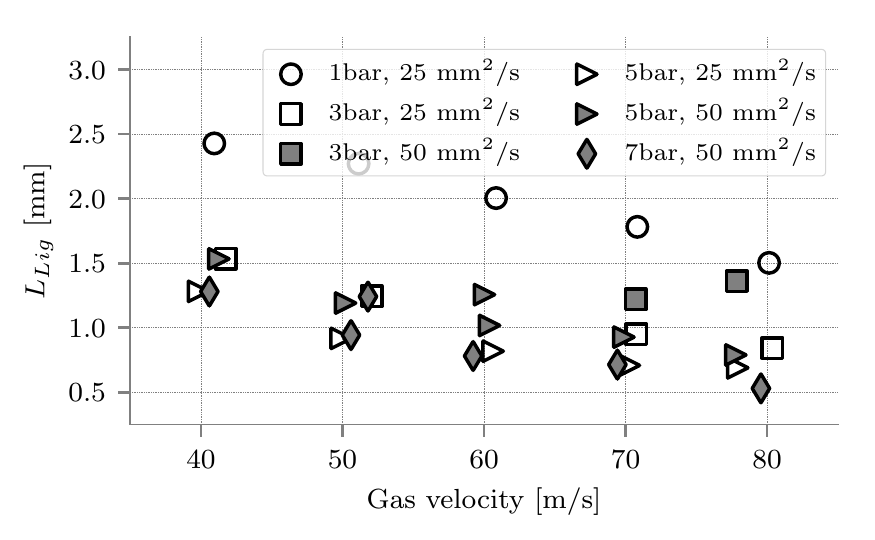}
	\caption{Evolution of the Area-to-Perimeter ratio (left) and Ligament longitudinal extent (right) versus the mean gas velocity for different ambient pressures and film loadings.}
	\label{fig_p_influence_on_liq_accu_1}
\end{figure}

\begin{figure}%
\centering
		\includegraphics[width=0.495\textwidth,keepaspectratio]{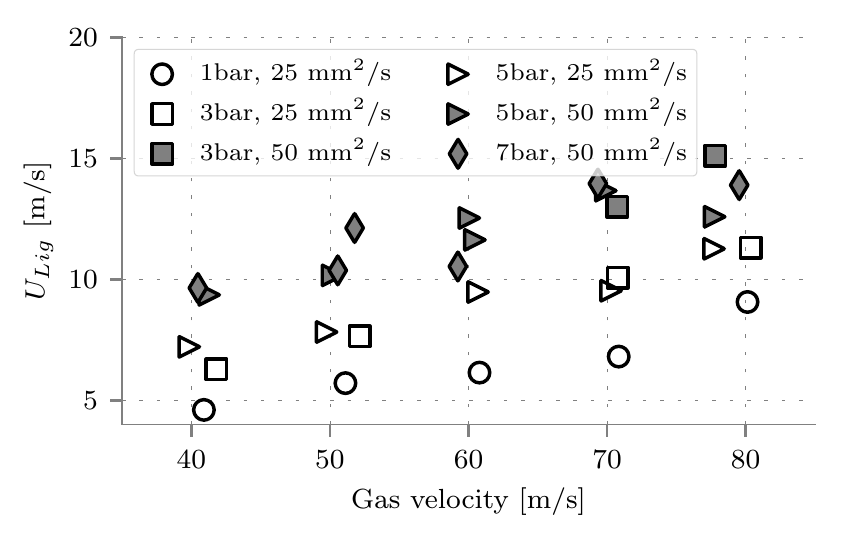}
		\includegraphics[width=0.495\textwidth,keepaspectratio]{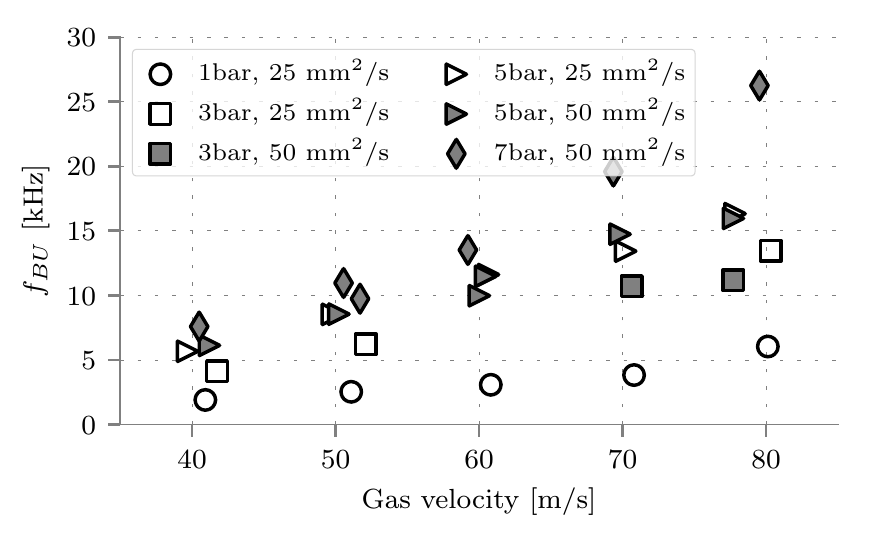}
	\caption{Evolution of the ligament velocity (left) and breakup frequency (right) versus the mean gas velocity for different ambient pressures and film loadings.}
	\label{fig_p_influence_on_liq_accu_2}
\end{figure}

Typical dimensions of the liquid accumulation are presented in figures~\ref{fig_p_influence_on_liq_accu_1} and \ref{fig_p_influence_on_liq_accu_2}. At $p$~=~1~bar, the ratio $A_{Lig}/P_{Lig}$ is almost constant versus the gas velocity, but decreases with the gas velocity for $p >$ 1~bar. The slope becomes slightly steeper with an increase of ambient pressure. $A_{Lig}/P_{Lig}$ is smaller for lower film loading, which can be explained as follows.

The liquid accumulation is fragmented due to the turbulent high speed airflow outside the recirculation zone. Therefore, the surface of the liquid accumulation is highly distorted by the aerodynamic stress, which controls the \textit{tearing rate}, \ie the mass flow rate to be detached from the liquid accumulation. This means that the perimeter of the liquid accumulation is mostly determined by the gas flow.
When the film loading is increased, it increases in turn the feeding rate of the liquid accumulation, but the tearing rate remains constant.
This increases the total volume of the liquid accumulation, which finally leads to a larger area. Hence, the ratio $A_{Lig}/P_{Lig}$ increases for an increasing film loading.
The mean longitudinal extent of the ligaments $L_{Lig}$ is depicted in figure~\ref{fig_p_influence_on_liq_accu_1} (right). 
The global trend is also a decrease of $L_{Lig}$ with an increase of the aerodynamic stress, \ie with an increase of (i) the gas velocity and/or an increase of (ii) the ambient pressure. 
This is because the ligaments are disrupted faster, so that they are less stretched in axial direction.
This observation also confirms the aerodynamic stress as the driving phenomenon, both in terms of velocity and density.
Like for $A_{Lig}/P_{Lig}$, the increase of film loading slightly increases $L_{Lig}$.

The characteristic extension velocity of the ligament is shown in figure~\ref{fig_p_influence_on_liq_accu_2} (left).
The gas velocity acts also a driving force, but contrary to the two previous quantities, the film loading has a stronger effect than the ambient pressure.
For $p >$ 1~bar,
all the points collapse on two different lines corresponding to two different film loadings.
Finally the breakup frequency is shown in figure~\ref{fig_p_influence_on_liq_accu_2} (right) where an increasing trend is clearly observable with the aerodynamic stress. The influence of the film loading cannot be clearly identified.

\begin{figure}%
\centering
		\includegraphics[width=0.495\textwidth,keepaspectratio]{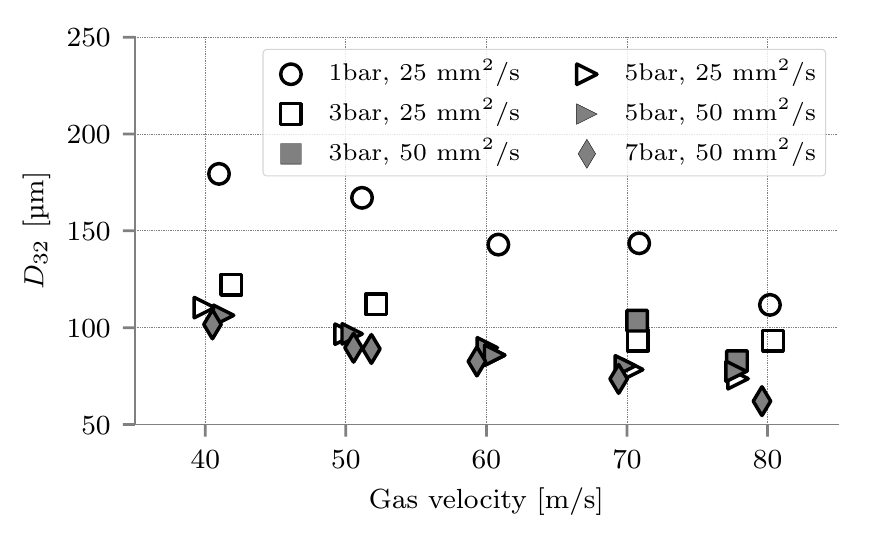}
		\includegraphics[width=0.495\textwidth,keepaspectratio]{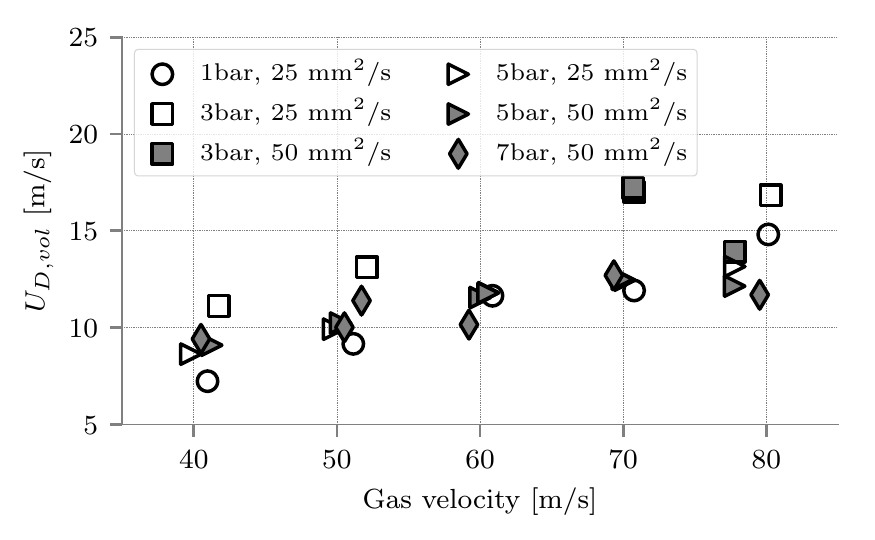}
	\caption{Evolution of the primary spray characteristics versus the mean gas velocity for different ambient pressures and film loadings.}
	\label{fig_quant_spray_charac}
\end{figure}

The Sauter Mean Diameter (SMD) of the spray is depicted in figure~\ref{fig_quant_spray_charac} (left). As expected, it decreases monotonically with an increasing gas velocity.
With an increasing ambient pressure, the SMD is reduced but with a decreasing rate: the relative change of SMD is smaller as the pressure is larger. This trend confirms the SMD $\propto p^{\alpha}$ where $\alpha$ is negative, as initially observed by \citet{rizkalla1975influence}.\\
The volume-based mean velocity $u_{D,vol}$ is shown in figure~\ref{fig_quant_spray_charac} (right), where a linear dependency on the gas velocity is observed. However, the influence of the pressure and the film loading is complex, and no simple conclusion can be drawn at this point.

\subsection{Influence of the ambient pressure}

\begin{figure}%
\centering
		\includegraphics[width=\textwidth,keepaspectratio]{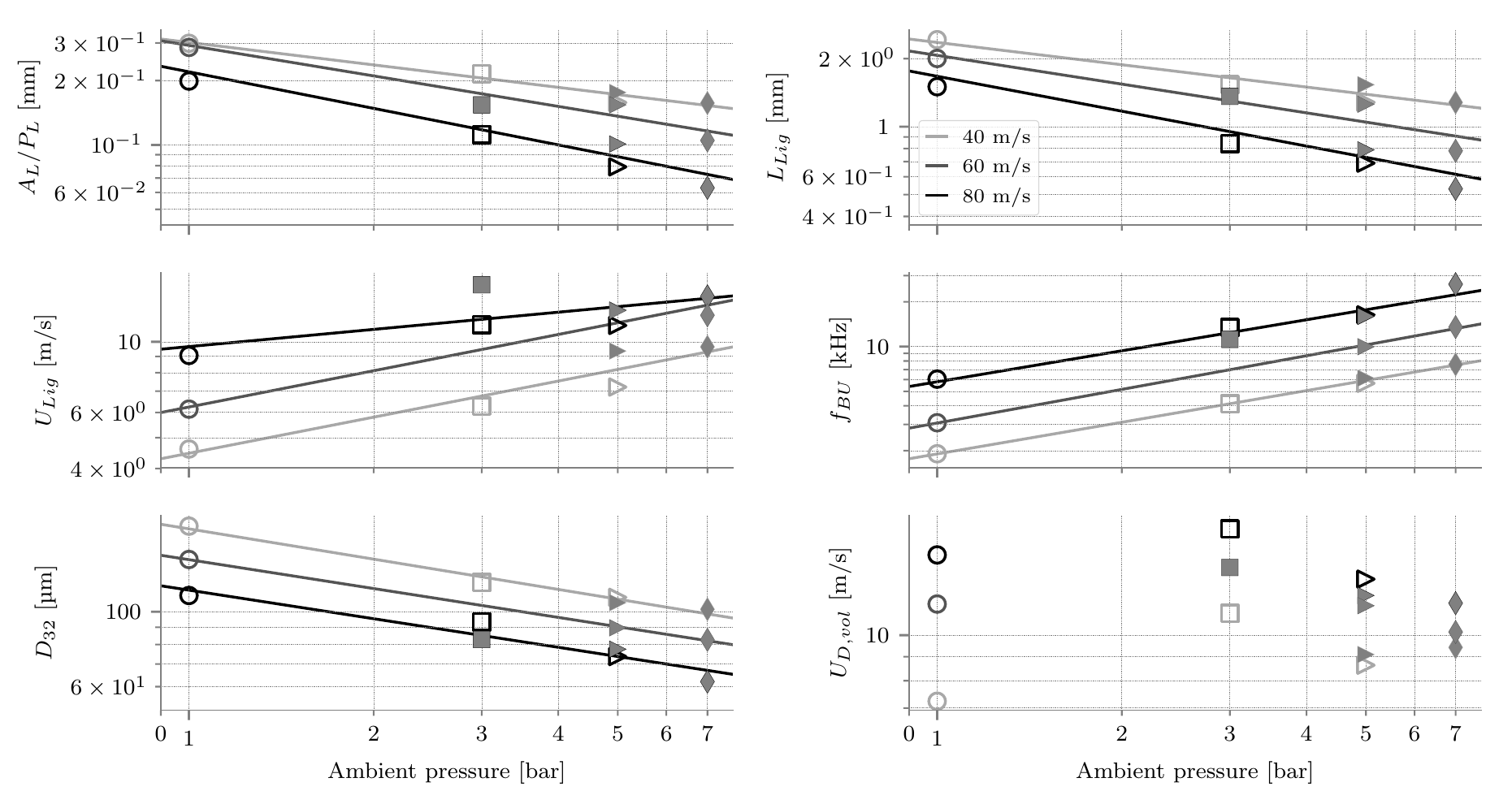}
	\caption{Evolution of all the accumulation quantities versus the ambient pressure for $U_g$~=~40 (light gray), 60 (gray) and 80 m/s (black) superimposed with the lines of equation $y=a \, x^b$. Values of $a$ and $b$ are recalled in Table~\ref{tab_fitting}.}
	\label{fig_all_qtity_vs_pressure}
\end{figure}

All the quantities are plotted versus the ambient pressure for $U_g$ = 40, 60 and 80~m/s in figure~\ref{fig_all_qtity_vs_pressure}. 
Each variable is fitted with the line of equation $y=a \, x^b$ for a single velocity and different film loadings. 
This is to identify the accumulation characteristics whose pressure dependency is weakly influenced by the film loading.
The coefficients of the fitting are listed in Table~\ref{tab_fitting}.
\begin{table}%
\footnotesize
\begin{tabular}{c | ccc | ccc | ccc | ccc| ccc}
& \multicolumn{3}{c|}{ y = $A_L/P_L$ [-] } & \multicolumn{3}{c|}{ y = $L_{Lig}$ [mm] } & \multicolumn{3}{c|}{ y = $U_{Lig}$ [m/s] } & \multicolumn{3}{c|}{ y = $f_{BU}$ [kHz] } & \multicolumn{3}{c}{ y = SMD [\textmu m] } \\
& a & b & R & a & b & R & a & b & R & a & b & R & a & b & R \\
40 m/s & 0.30 & -0.35 & 10.5 & 2.36 & -0.33 & 22.7 & 4.47 & 0.38 & 41.6 & 1.90 & 0.71 & 3.24 & 176 & -0.30 & 3.53 \\
60 m/s & 0.29 & -0.47 & 25.7 & 2.07 & -0.42 & 56.9 & 6.24 & 0.38 & 13.3 & 3.06 & 0.75 & 1.33 & 143 & -0.28 & 0.10 \\
80 m/s & 0.22 & -0.57 & 136 & 1.67 & -0.51 & 186.5 & 9.66 & 0.18 & 87.3 & 5.80 & 0.69 & 62.0 & 116 & -0.28 & 19.0 \\
Mean & 0.27 & -0.46 & 57.3 & 2.03 & -0.42 & 88.7 & 6.79 & 0.31 & 47.4 & 3.59 & 0.72 & 22.2 & 145 & -0.29 & 7.53 \\
\end{tabular}
	\caption{Fitting parameters for $y=a\,p^b$, from figure~\ref{fig_all_qtity_vs_pressure}. R, the residual of the fitting, is multiplied by 1000.}
	\label{tab_fitting}
\end{table}
\\In figure~\ref{fig_all_qtity_vs_pressure}, the curves of the SMD and the breakup frequency show a regular parallel shift with a constant slope. This means that the exponent $b$ is rather constant and that the prefactor $a$ varies monotonically. Hence, the dependence of the SMD and $f_{BU}$ on the ambient pressure is rather constant, and it is not affected by the gas velocity and of the film loading. 
By averaging the exponent $b$ of the fitting function for the three velocities, the global trend can be expressed as:
\begin{equation}
D_{32} \propto p^{-0.29} 
\quad \text{,} \quad
f_{BU} \propto p^{0.72}
\end{equation}
The quantities related to the liquid accumulation $A_L/P_L$, $L_{Lig}$ and $U_{Lig}$ show a monotonic dependency on the ambient pressure with a slight influence of the gas velocity and the film loading. To obtain a global trend on pressure dependence, one can average the exponent $b$ for the three velocities:
\begin{equation}
A_L/P_L \propto p^{-0.463}
\quad \text{,} \quad
L_{Lig} \propto p^{-0.422}
\quad \text{,} \quad
U_{Lig} \propto p^{0.312}
\end{equation}
The similar scaling of $A_L/P_L$ and $L_{Lig}$ with the pressure is to be highlighted, as it could be a suitable candidate for characterizing the influence of pressure on the geometrical characteristic of the liquid accumulation.
The complex dependency of the primary droplets velocity $U_{D,vol}$ on the ambient pressure is confirmed by figure~\ref{fig_all_qtity_vs_pressure}, with a maximum between $p$ = 1 and 5~bar.
This maximum can be qualitatively explained by the influence of the ambient pressure on the drag exerted on primary droplet. 
The gas density, the droplet diameter and gas velocity at the tip of the ligament contribute to the droplet acceleration due to the drag. These three variables depend on pressure.
At lower pressure (p~=~1~bar), the ligament length $L_{Lig}$ is large, which means that the droplets are created relatively far from the atomizing edge, where the gas velocity is already large. On the other hand, at p~=~1~bar, the droplet diameter is large and the gas density is low. Considering together all these effects of gas density, droplet diameter and local velocity, the droplets will be moderately accelerated. In the contrary, at high pressure (\ie high air density), the droplets diameter will be smaller. As $L_{Lig}$ is smaller, the droplets are created closer to the atomizing edge, resulting in a smaller local gas velocity. Considering all terms also leads to a moderate acceleration. Therefore, in between these two extreme cases, there exists a pressure at which the acceleration of the droplet is at a maximum.

Finally, the values of the SMD are plotted versus $L_{Lig}$ and $A_L/P_L$. The black dashed line is the best linear fit and the grey dashed lines are the $\pm$10\% boundaries.
The SMD correlates well with the characteristic lengths of the liquid accumulations.
This observation strengthens the assumptions that the scale of the liquid accumulation determines the scale of the spray characteristics.

\begin{figure}%
	\centering
	\includegraphics[width=0.495\textwidth,keepaspectratio]{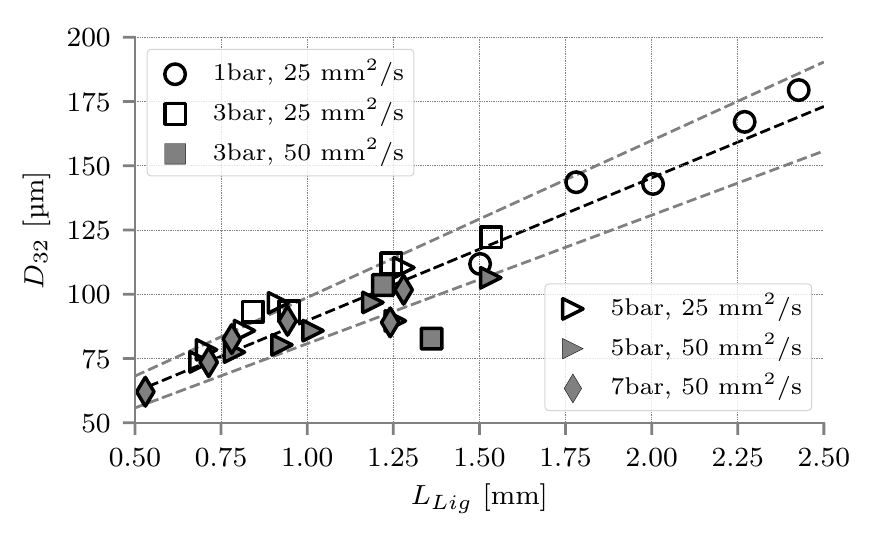}
	\includegraphics[width=0.495\textwidth,keepaspectratio]{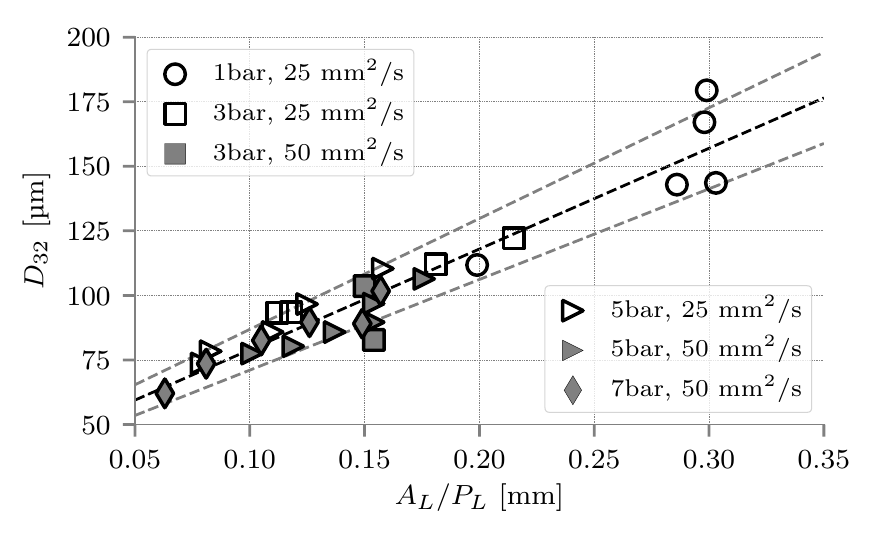}
	\caption{Linear correlation between the SMD and $L_{Lig}$ (left) and $A_L/P_L$ (right). Dashed gray lines depict the $\pm$ 10\% limits.}
	\label{fig_SMD_vs_L_Liq}
\end{figure}

\subsection{Probability Functions}

\begin{figure}%
\centering
		\includegraphics[width=0.495\textwidth,keepaspectratio]{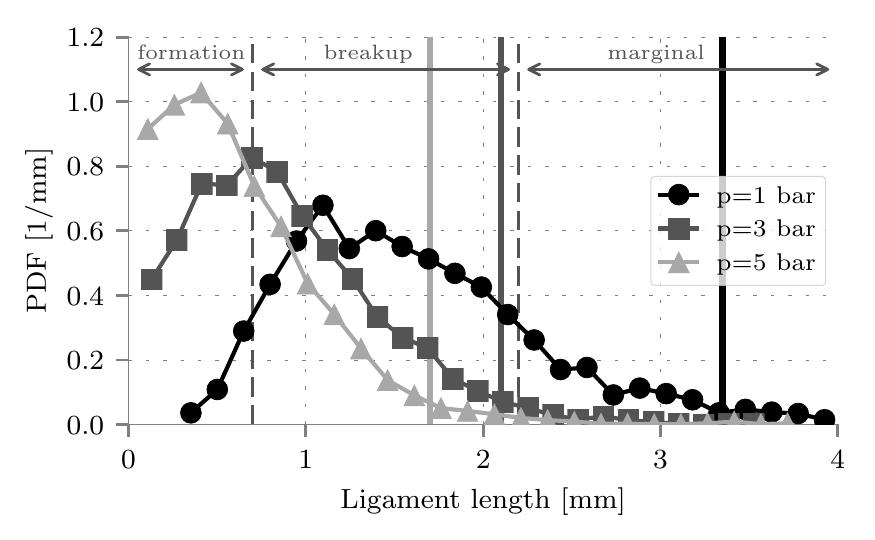}
		\includegraphics[width=0.495\textwidth,keepaspectratio]{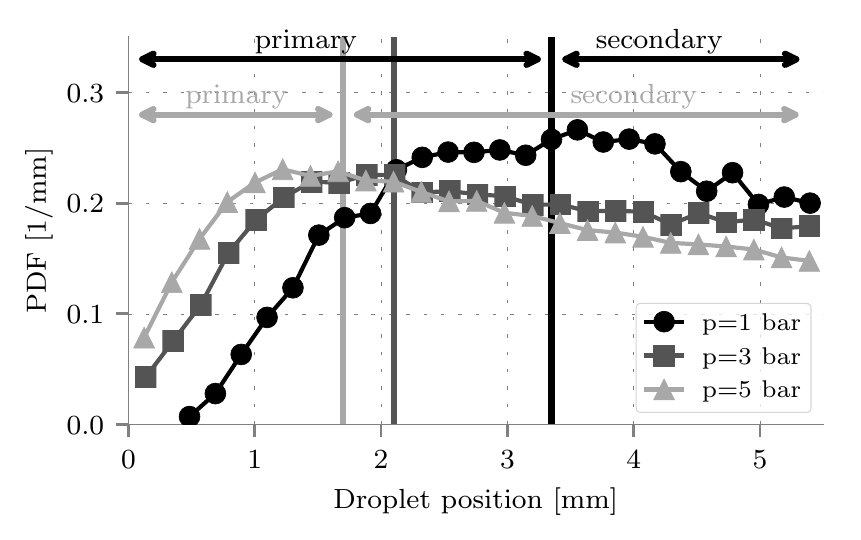}
	\caption{Influence of the ambient pressure on the ligament length (left) and droplet axial position (right) distributions at $U_g$=70~m/s and $\Lambda_f$=25~mm\textsuperscript{2}/s.
The vertical dashed lines (left) correspond to the limit between the three zones of the ligament breakup for p = 3 bar.}
	\label{fig_pdf_Lliga_drop_position}
\end{figure}

In this section the liquid accumulation and the primary droplets are investigated in terms of probability density function (PDF).
The link between the length of the ligament and the location where the primary droplets are created is illustrated in figure~\ref{fig_pdf_Lliga_drop_position} for different ambient pressures.
The distribution of $L_{Liga}$ (left) can be subdivided into three linear regimes. They are depicted on figure~\ref{fig_pdf_Lliga_drop_position} (left) for p = 3 bar. First, the linear increase corresponds to ligaments mostly in the process of formation, \ie stretched liquid streaks in a stable form. It is referred to as the \textit{ligament formation zone} in the following.
Second, a linear decrease depicts the final step of ligament stretching, during which ligament are fragmented, and it is labeled \textit{ligament breakup zone}.
The last linear part is the slow decreasing tail of the PDF and corresponds to marginal long ligaments.
Globally, figure~\ref{fig_pdf_Lliga_drop_position} (left) shows (i) the significant dispersion of the ligament length and (ii) the influence of the ambient pressure on the scale and the shape of the PDF. The meaning of the vertical solid lines in figure~\ref{fig_pdf_Lliga_drop_position} (left) will be explained later.
On the right of figure~\ref{fig_pdf_Lliga_drop_position} the distribution of the axial position of the droplets is depicted. It can be regarded as the evolution of the droplet concentration (or liquid fraction) in the axial direction.
Two zones can clearly be discriminated. First, the PDF increases linearly, representing droplet production. This corresponds to the zone where most of the ligaments and bags are fragmented. In this zone the breakup is dominant over the droplet transport. In the second zone, the droplet concentration decreases because of the acceleration by the gas. As the droplet velocity increases with the axial position, the cloud of droplets is expanded in the axial direction. In virtue of the mass conservation of the liquid, $\dot{m}_l = \alpha_l \, \rho_l \, A \, u_l$, a larger droplet velocity leads to a lower volume fraction $\alpha_l$. In this zone, the breakup phenomenon still occurs, but it is dominated by the effect of droplet acceleration. The identification of these zones is an objective distinction between (i) primary breakup where the liquid volume fraction due to the droplets increases, \ie the droplet creation is dominant over the cloud expansion, and (ii) secondary breakup where the droplet creation is dominated by the acceleration due to the gas. The limit between the two zones is defined at the location where droplet number PDF over droplet position starts to follow a linear decrease. It is shown as vertical solid line in figure~\ref{fig_pdf_Lliga_drop_position} (right) and coincides well with the end of the \textit{ligament breakup zone} of the $L_{Lig}$ distribution (vertical lines in figure~\ref{fig_pdf_Lliga_drop_position} left). Indeed, when most of the ligaments are fragmented at the end of primary breakup zone, most of droplets are created, and the cloud starts to expand.
As expected, the increase of ambient pressure leads to shorter ligaments. Hence, droplets are created closer to the atomizing edge, and the distribution is shifted towards the atomizing edge.
\begin{figure}%
\centering
		\includegraphics[width=0.495\textwidth,keepaspectratio]{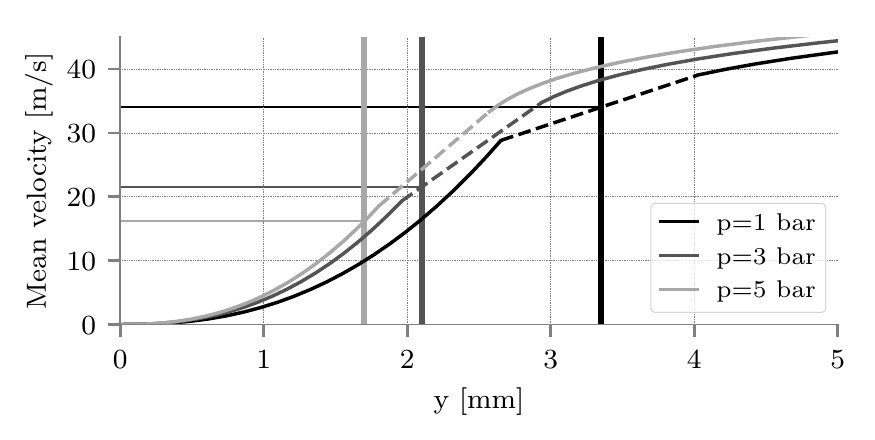}
		\includegraphics[width=0.495\textwidth,keepaspectratio]{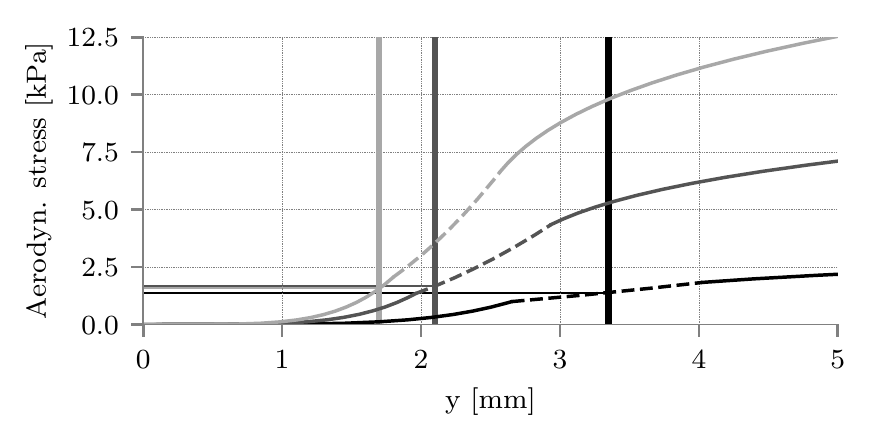}
	\caption{Left: Mean axial velocity from the model presented in Section~\ref{sec_gas_phase} with $U_g$=70~m/s. Right: Corresponding aerodynamic stress. Vertical solid lines corresponds to the transition between primary and secondary breakup as in figure~\ref{fig_pdf_Lliga_drop_position}.}
	\label{fig_umean_aerostress_profile}
\end{figure}
\\The profile of the mean axial velocity in the vicinity of the prefilmer is extrapolated from the model derived in Section~\ref{sec_gas_phase}, and its corresponding local aerodynamic stress is shown in figure~\ref{fig_umean_aerostress_profile}. The vertical lines correspond to the transition between primary and secondary breakup like in figure~\ref{fig_pdf_Lliga_drop_position}. The dashed portion of the curves corresponds to a buffer layer between the laminar and the turbulent near wake.  At the interface of the two zones, the velocity profiles did not match, \ie $u_{lam}(y^*=16)>u_{turb}(y^*=16)$. Therefore a linear profile, depicted as the dashed line, is applied in this buffer zone.
From figure~\ref{fig_umean_aerostress_profile} (left), it is obvious that the gas velocity at the transition between primary and secondary breakup significantly depends on the ambient pressure. 
However, when the profile of the local aerodynamic stress $\rho_g u_g^2$ is plotted versus the axial position, the primary/secondary breakup transition is found for a similar value of the stress.
This result emphasizes the hypothesis that the aerodynamic stress is one of the most important parameters to characterize primary breakup. Also, note that the primary/secondary breakup transition does not always occur at the same near-wake region. In figure~\ref{fig_umean_aerostress_profile}, it occurs in the laminar (5 bar) or in the buffer region (1 and 3 bar).

\begin{figure}%
\centering
		\includegraphics[width=0.495\textwidth,keepaspectratio]{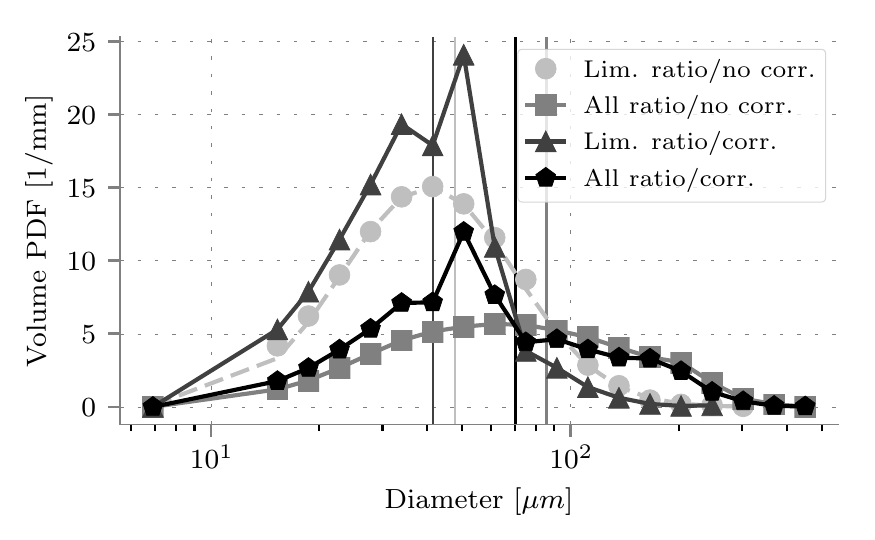}
	\caption{Influence of spherical droplet filtering and depth-of-field correction on the VPDF of droplets for $U_g$=80 m/s, p=5~bar and $\Lambda_f$=50~mm\textsuperscript{2}/s. For the curve denoted "Lim. ratio/no corr", the dashed line corresponds to the best fitting of the log-normal function.}
	\label{fig_pdf_dropsize_q3_features_ratio_small_drops}
\end{figure}

\begin{figure}%
\centering
		\includegraphics[width=0.495\textwidth,keepaspectratio]{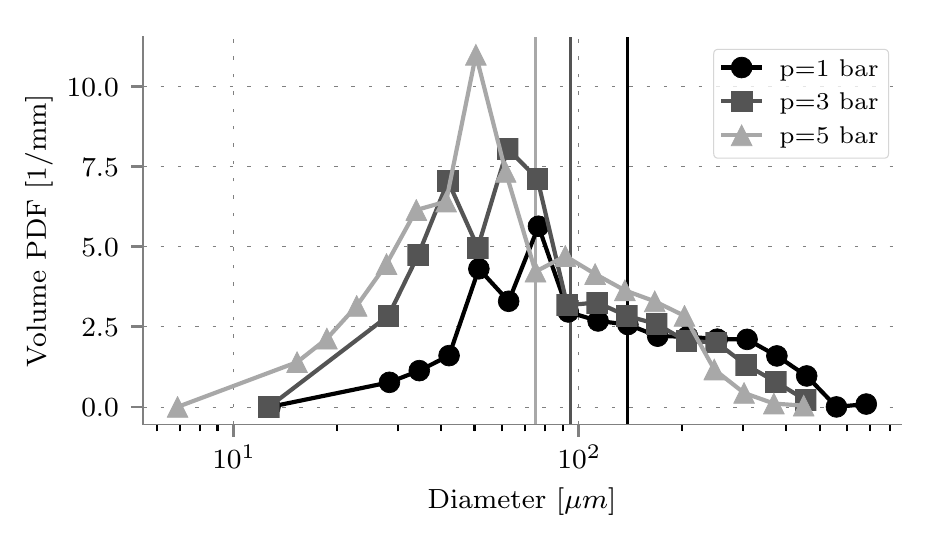}
		\includegraphics[width=0.495\textwidth,keepaspectratio]{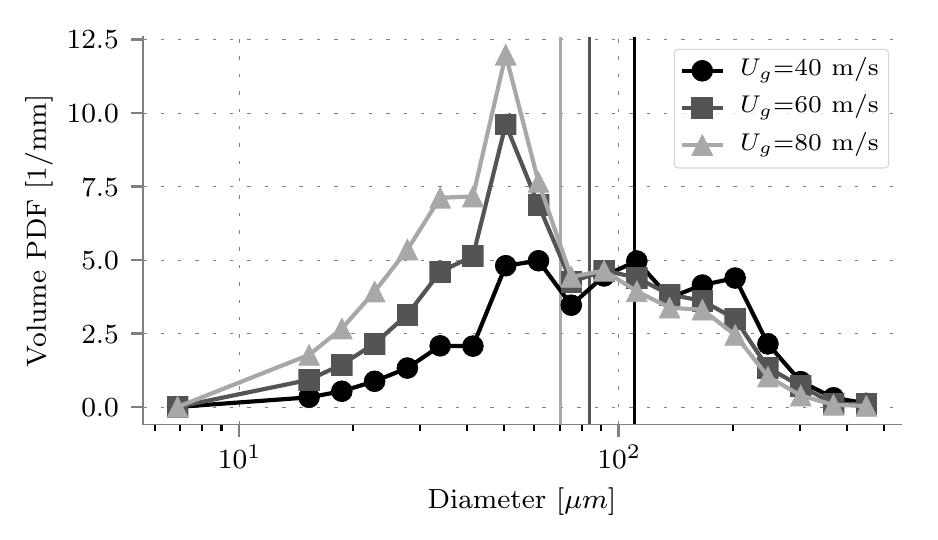}
	\caption{Left: Influence of the ambient pressure on the volume drop size distributions at $U_g$=70~m/s and $\Lambda_f$=25~mm\textsuperscript{2}/s. Right: Influence of the gas velocity on the volume drop size distributions at $p$ = 5~bar and $\Lambda_f$=50~mm\textsuperscript{2}/s}
	\label{fig_pdf_dropsize_q3}
\end{figure}

\begin{figure}%
\centering
		\includegraphics[width=0.495\textwidth,keepaspectratio]{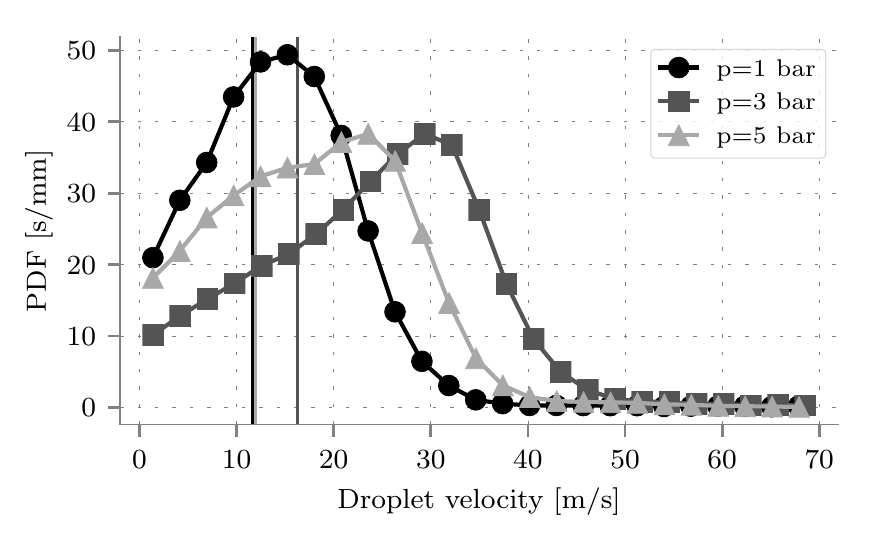}
		\includegraphics[width=0.495\textwidth,keepaspectratio]{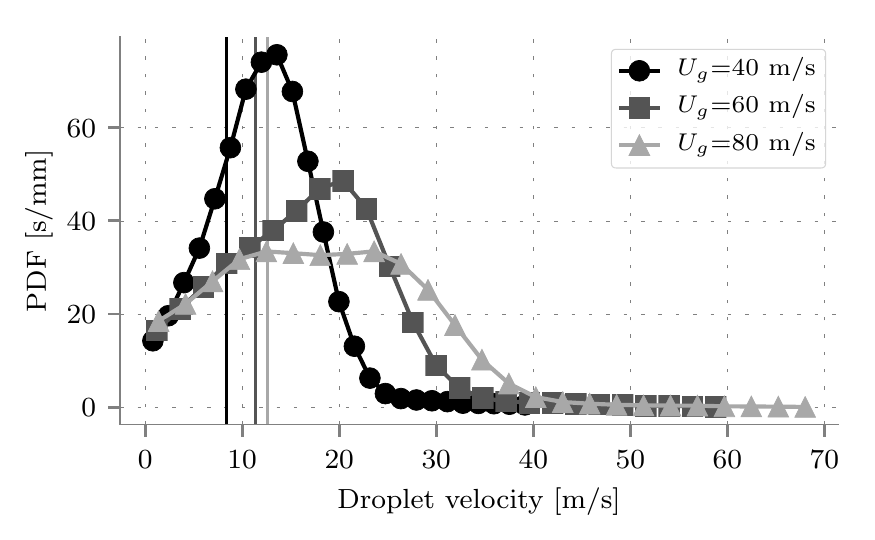}
	\caption{Left: Influence of the ambient pressure on the droplet velocity distributions at $U_g$=70~m/s and $\Lambda_f$=25~mm\textsuperscript{2}/s. Right: Influence of the gas velocity on the droplet velocity distributions at p=5~bar and $\Lambda_f$=50~mm\textsuperscript{2}/s}
	\label{fig_pdf_dropvelocity_q3}
\end{figure}

The Volume Probability Density Function (VPDF) of the droplet size and the PDF of the droplet velocity are now discussed. First, the influence of the post-processing is explained. The present post-processing methodology features two major differences with regards to traditional methodologies for spray characterization such as LDA and LDT. First, the present methodology captures non-spherical droplets, whereas PDA requires spherical or nearly spherical droplets, and LDT leads to a deviation in case of non spherical droplets \citep{dumouchel2014laser}. Second, our methodology includes a correction for small droplets located out of the focal plane, based on calibration \citep{warncke2017experimental}.
These two features extend the resolution for large and small diameter. While the small droplets are better detected with the calibration, non spherical droplets are usually in the range of large diameters. The influence of these features are depicted in figure~\ref{fig_pdf_dropsize_q3_features_ratio_small_drops}. The round symbols show the VPDF with close-to-spherical droplets only (Lim. ratio) and without small diameter correction. 
The overall shape is similar to the typical shape, such as the Rosin-Rammler, the Gamma, or the log-normal function. For instance, the data has been fitted by a log-normal function (dashed line in figure~\ref{fig_pdf_dropsize_q3_features_ratio_small_drops}).
Taking non-spherical droplets into account (square symbol) increases the PDF at the tail of the distribution (large droplets) and flattens the peak. 
On the other hand, correcting the diameter of small droplets (triangle symbols) accentuates the peak of the distribution and hence decreases the density of the tail.
Applying both corrections (pentagon symbols) leads to a VPDF made of two parts (i) a concave tail, due to larger droplets and (ii) a sharp, almost triangular, peak due to the small droplets. In this case, the resulting SMD is 70~\textmu m, significantly larger than without corrections (48~\textmu m). The two corrections are always applied in the rest of this work. Note that all distributions are normalized, and the apparent difference of the area is due to the logarithmic scale of the $x$-axis.
\\The VPDF of the droplet size is presented in figure~\ref{fig_pdf_dropsize_q3} for different pressures (left) and different gas velocities (right). The value of the SMD is represented by a vertical line for each case. 
As explained previously, the overall shape of the distributions is somewhat different from the typical shape. Nevertheless, the influence of pressure is visible with a global shift of the whole distribution towards smaller diameters at higher pressures. 
On the contrary, the increase of the velocity has a weak influence on the location of the peak. It promotes creation of small droplets (15 - 60 \textmu m) and leaves the tail almost unchanged.
This highlights two different contributions from the pressure and the velocity. The pressure affects the fragmentation on all scales of droplets, whereas the velocity affects only the smaller scales (left of the peak in figure~\ref{fig_pdf_dropsize_q3} right) which corresponds to droplets created in a high velocity air flow.
This is due to the particular geometry of prefilming airblast atomization. In the wake zone of the prefilmer, the gaseous velocity is low, almost independent of the bulk velocity of the free flow. Therefore, the breakup process occurring in this zone will result in large droplets with a weak dependence on the free flow velocity.
In contrary, when the ligaments are immersed in the high-speed air flow, they \textit{feel} the total of the bulk velocity. Hence, at this location, the droplets are small and are fully influenced by the bulk velocity of the flow.
As for figure~{\ref{fig_pdf_dropsize_q3_features_ratio_small_drops}}, figure~{\ref{fig_pdf_dropsize_q3}} is a semi-log plot so that the areas below the curve appear different whereas the distributions are normalized.
\\The influence of ambient pressure and gas velocity on the droplet velocity is shown in figure~\ref{fig_pdf_dropvelocity_q3}. The mean value is depicted by the vertical lines. The non-monotonic influence of the ambient pressure is well visible with a shift of the peak. At $p$ = 5~bar, the dispersion of velocities normalized by the mean value is larger, which could be the consequence of the primary droplets generated closer to the recirculation zone.
When the velocity increases (figure~\ref{fig_pdf_dropvelocity_q3} right), the mean value of the droplet velocity increases monotonically. The same type of dispersion as for the influence of pressure is also observed.

\section{Investigation at constant aerodynamic stress \label{sec_M_constant}}

\subsection{Using the liquid momentum flux for characterization of prefilming airblast atomization}

\begin{figure}%
\centering
\linespread{1.0}
	\def \svgwidth {0.9\textwidth}
	{\scriptsize
	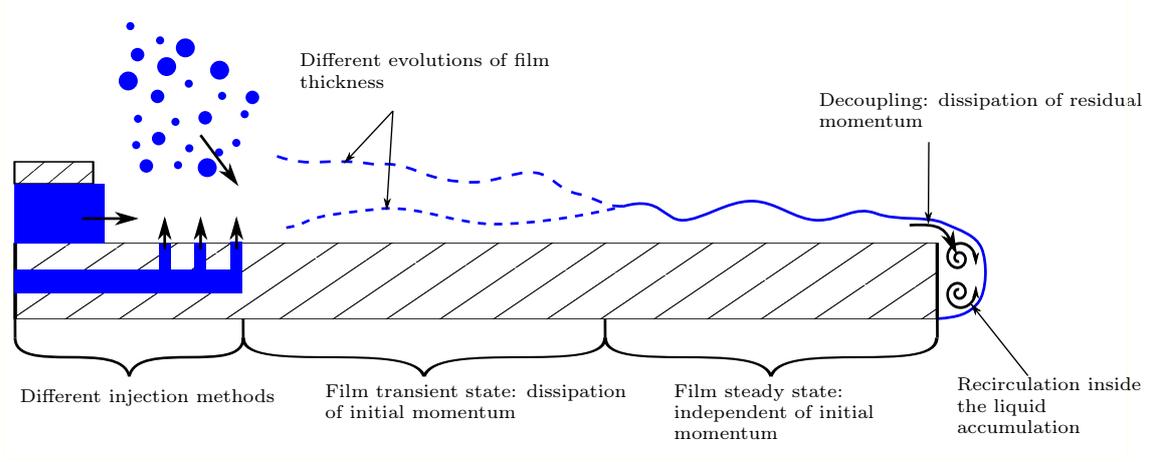
	}
	\caption{Successive phenomena dissipating the initial momentum between injection and breakup.}
	\label{fig_sketch_mom_flux}
\end{figure}

In the literature, the momentum flux ratio $M = \rho_G \, u_G^2 / \rho_L \, u_L^2$ was identified as an important parameter governing airblast atomization.
It reflects the rate of gaseous momentum (\ie the aerodynamic stress) available to
disintegrate a given amount of liquid continuously injected by the nozzle. Therefore, the aerodynamic stress is normalized by the liquid momentum flux at the location of breakup.
In the case of prefilming airblast atomization, the question of liquid momentum flux at the location of breakup is not straightforward, because the liquid can be injected in different manners, and the
initial momentum is dissipated in several ways, as illustrated in figure~\ref{fig_sketch_mom_flux}.
Downstream of the injection, the liquid forms a thin film, where it reaches its steady state after a period of $t_{st} = h_f^2/\nu$ \citep{chaussonnet2014Modeling}.
In this case, the state of the film is independent of the initial conditions, and the information of the initial momentum flux is lost when the film reaches the atomizing edge.
This means that for a prefilmer long enough to allow the film to reach the steady-state,
the liquid momentum flux at the injection is not representative of the momentum flux when the film reaches the atomizing edge.
In addition, since the liquid accumulation operates a decoupling between the film flow and the breakup process, the momentum of the film when it reaches the atomizing edge is significantly different from the momentum of the liquid accumulation, which is stationary.
As a final consequence, the momentum of the liquid accumulation is very low and usually negligible compared to the momentum of the gas. For these reasons, the momentum flux ratio is not pertinent to characterize prefilming atomization, and only the term $\tau_G$ will be investigated in the following. 

\subsection{Mean values of the liquid accumulation and the primary spray}

\begin{figure*}[!htb]
\centering
		\includegraphics[width=0.5\textwidth,keepaspectratio]{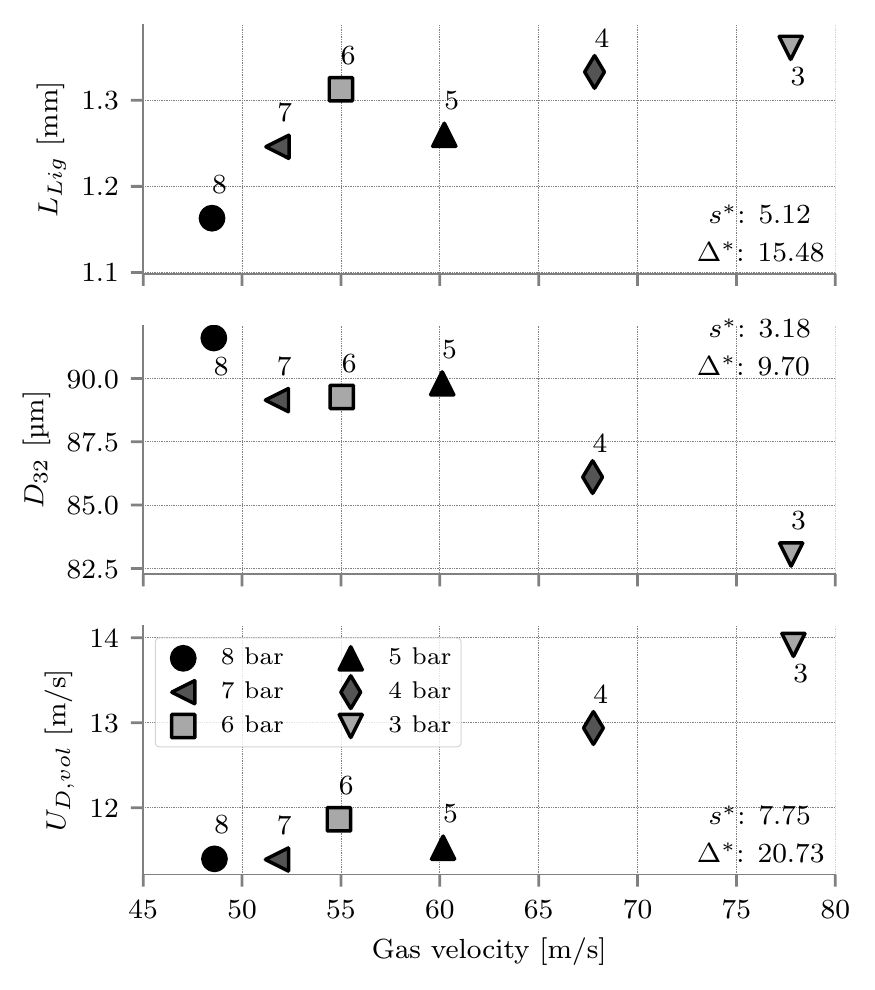}
	\caption{Evolution of the longitudinal ligament extent $L_{Lig}$, the SMD and the mean droplet velocity $U_{D,vol}$ versus the mean gas velocity for the same aerodynamics stress. $s^*$ and $\Delta^*$ are the standard deviation and the amplitude, respectively, normalized by their mean and expressed in \%. Numbers close to the symbols (3-8) is the pressure.}
	\label{fig_p_influence_M_constant}
\end{figure*}

In Section~\ref{sec_quantit_liquid}, the density and the velocity of the gas were independently varied. 
This led to comparisons with different aerodynamic stresses $\tau_G = \rho_G \, u_G^2$. 
In this section, $\tau_G$ is kept constant to $\approx$~22,000~Pa, and the two parameters are varied simultaneously.
The operating conditions are listed in Table~\ref{tab_pressure_ug_constant_aer_stress}.
\begin{table}%
	\small
	\centering
	\begin{tabular}{ c c  c c  c c  c c}
	Pressure & [bar] & 3 & 4 & 5 & 6 & 7 & 8\\
	Velocity & [m/s] & 78 & 68 & 60 & 55 & 52 & 48\\
	Reynolds number & [$\times$\num{e6}] & 1.11 & 1.29 & 1.42 & 1.56 & 1.72 & 1.82
	\end{tabular}
	\caption{Corresponding values of ambient pressure, gas velocity and Reynolds number for a constant aerodynamic stress $\approx$ 22,000 Pa.}
	\label{tab_pressure_ug_constant_aer_stress}
\end{table} 
Figure~\ref{fig_p_influence_M_constant} presents the evolution of
of the longitudinal ligament extent $L_{Lig}$, the SMD and the mean droplet velocity $U_{D,vol}$
versus the gas velocity. The corresponding ambient pressure is annotated at each symbol.
The quantities $s^*$ and $\Delta^*$ are the standard deviation and the maximum amplitude, respectively.
They are used to quantify the statistical deviation of the six quantities for the different operating points investigated in the present section.
They are both normalized by their mean value and expressed in \%.
The evolution of the longitudinal ligament extent $L_{Lig}$ suggests a regime close to saturation for lower ambient pressures.
With the smallest $s^*$ and $\Delta^*$, the SMD of the spray is globally kept unchanged with a slight decrease at high velocity.
$U_{D,vol}$ exhibits a linear evolution for p between 3 and 5~bar, and a constant value for p larger than 5~bar. It closely follows the trend of the breakup frequency.
For all quantities, the normalized standard deviation is below 10\% and, apart from the velocity ($U_{D,vol}$), the maximum amplitude of the variation is lower than 10\%.
These relative variations of the accumulation and spray quantities are rather small in comparison to the large range of velocity and density. This clearly demonstrates the stronger influence of the aerodynamic stress over the gas velocity or the ambient pressure taken individually.

\subsection{Densities of probability}

\begin{figure}[!htb]
\centering
		\includegraphics[width=0.495\textwidth,keepaspectratio]{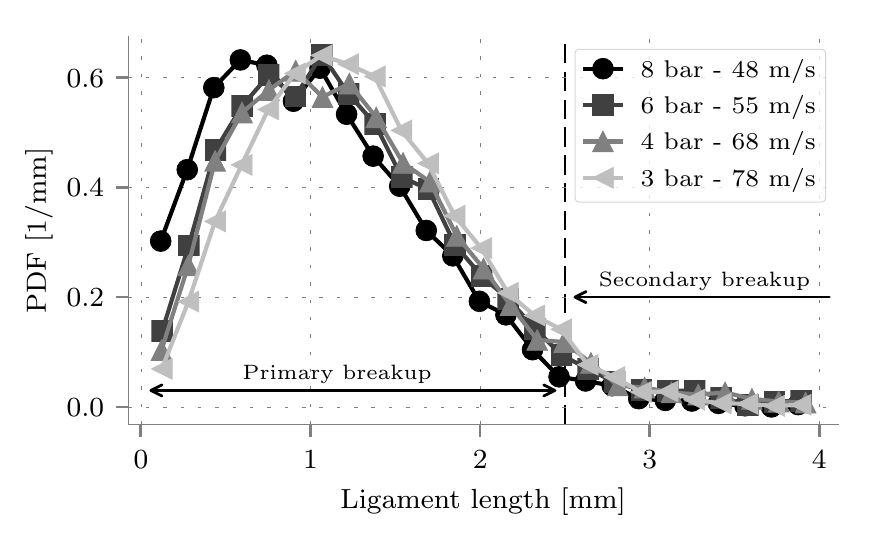}
		\includegraphics[width=0.495\textwidth,keepaspectratio]{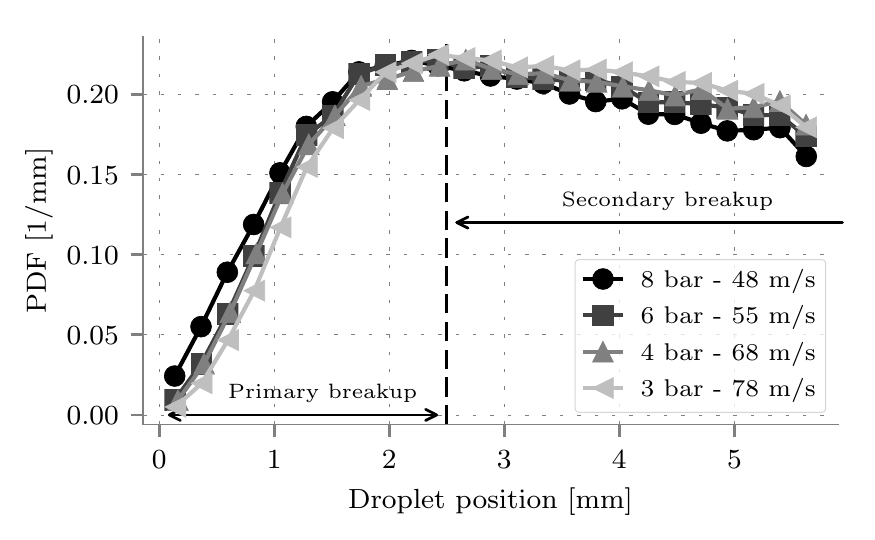}
	\caption{Influence of the aerodynamic stress on the ligament length (left) and droplet axial position (right) distributions.}
	\label{fig_pdf_Lliga_drop_position_M_cst}
\end{figure}

\begin{figure}[!htb]
\centering
		\includegraphics[width=0.495\textwidth,keepaspectratio]{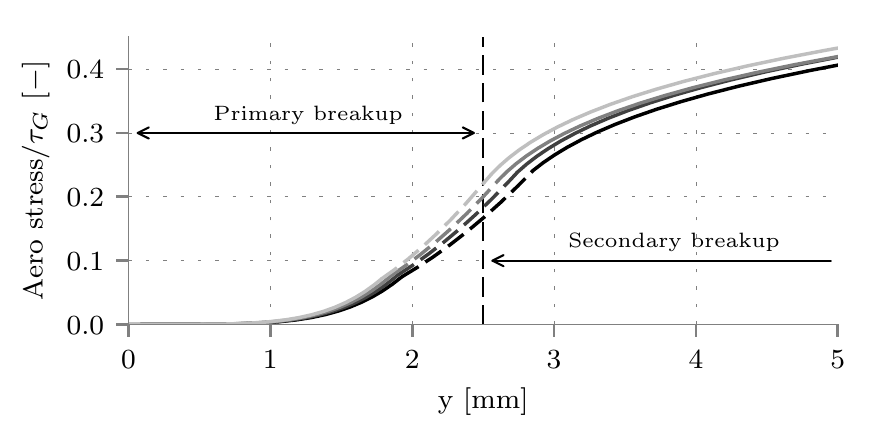}
	\caption{Profile of the aerodynamic stress normalized by $\rho_g u_g^2$.}
	\label{fig_stress_profile_M_cst}
\end{figure}

The probability density function of the ligament length and of the droplet position is given in figure~\ref{fig_pdf_Lliga_drop_position_M_cst}, (left) and (right), respectively. The similarity between all curves is striking. 
Once again, it demonstrates the superiority of the aerodynamic stress versus the gas velocity or the ambient pressure to describe the liquid accumulation and the primary spray location in prefilming airblast atomization.
In addition, the limit between the primary and secondary breakup ($y \approx$ 2.5 mm) in figure~\ref{fig_pdf_Lliga_drop_position_M_cst} (right) coincides with the end of the \textit{ligament breakup zone} (figure~\ref{fig_pdf_Lliga_drop_position_M_cst} left), as also observed in figure~\ref{fig_pdf_Lliga_drop_position}.
The profile of the local aerodynamic stress is shown in figure~\ref{fig_stress_profile_M_cst}. It is estimated from the model presented in Section~\ref{sec_gas_phase} and normalized by $\tau_G$. The different profiles match remarkably well. Not only this confirms the tight link between liquid breakup and shear stress imposed by the gas flow, but also it suggests that the gas flow model is valid over a larger range of Reynolds numbers. For these operating points, the primary/secondary breakup transition occurs in the second half of the buffer region of the velocity profile.

\begin{figure}[!htb]
\centering
		\includegraphics[width=0.495\textwidth,keepaspectratio]{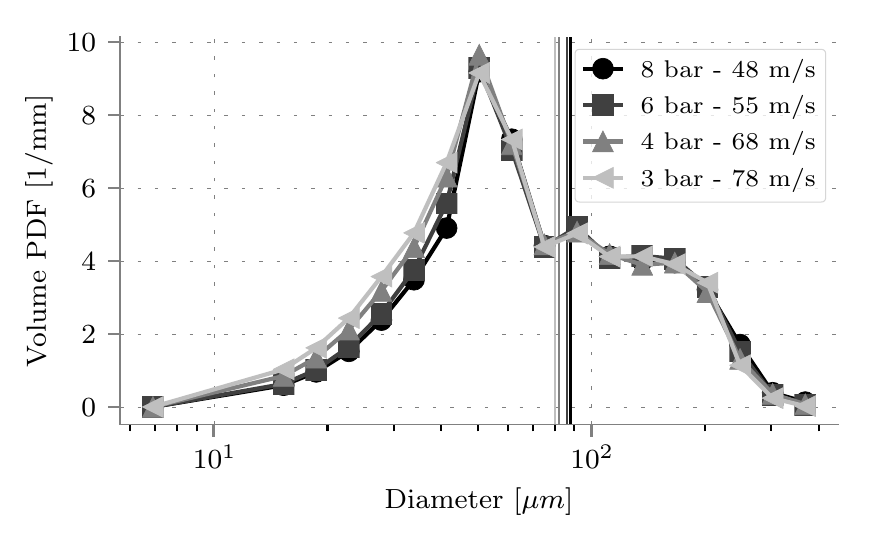}
		\includegraphics[width=0.495\textwidth,keepaspectratio]{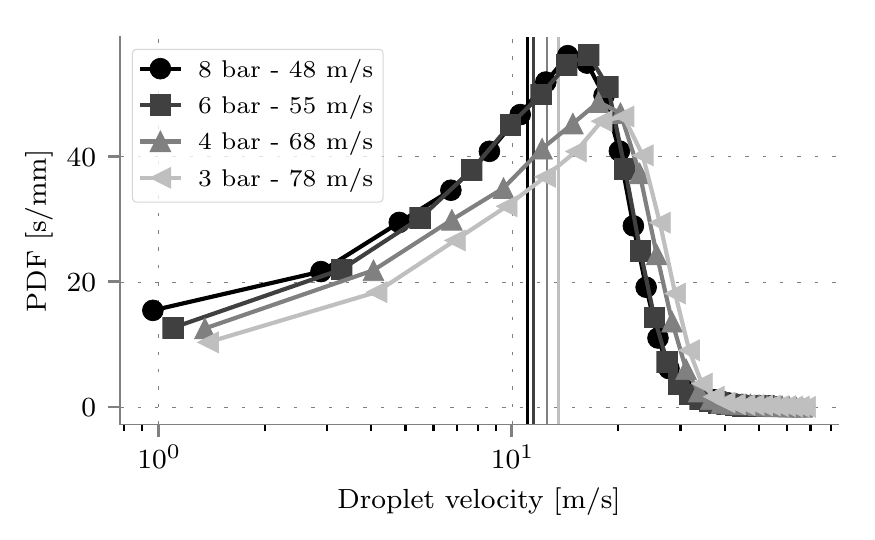}
	\caption{Volume drop size (left) and droplet velocity (right) distributions at different pressure and same aerodynamic stress.}
	\label{fig_p_influence_M_constant_VPDF}
\end{figure}

The distribution of the droplet diameter and velocity is shown in figure~\ref{fig_p_influence_M_constant_VPDF} (left). As expected, the VPDF of the spray droplets is almost unchanged for a constant aerodynamic stress. The only difference comes from the larger production of small droplets at larger velocities. As mentioned earlier, this is related to the fact that the gas velocity has a strong influence outside the recirculation zone whereas the effect of pressure is applied everywhere. The influence of the gas velocity on $U_{D,vol}$ (figure~\ref{fig_p_influence_M_constant_VPDF} right) leads to a shift of the peak towards larger velocity for larger gas velocities. This is because larger gas velocity generate more smaller droplets which are accelerated faster.

As a conclusion of this investigation of the aerodynamic stress, it can be stated that the geometrical output ($A_{Lig}/P_{Lig}$, $L_{Lig}$ and the SMD) and the frequency $f_{BU}$ of the breakup phenomenon is rather constant for a constant aerodynamic stress $\tau_G$, whereas the velocities of the ligament and the droplet slightly increase with the gas velocity.
The striking observation is that the probability density functions are also very similar. %
This means, if the aerodynamic stress is kept constant, not only the mean values but also the fluctuating components remain the same.

\section{Comparison with other correlations\label{sec_compare_correlations}}

In this section the SMDs are compared with correlations from the literature. The objective of this section is (i) to assess the dependence on ambient pressure given by different correlations and (ii) to stress the need of using shadowgraphy to calibrate primary breakup models.

\subsection{Literature survey}

In addition to the papers cited in the introduction \citep{rizkalla1975influence, el1980airblast, aigner1988swirl, gepperth2013analysis}, correlations of \citet{inamura2012spray}, \citet{eckel2013semiempirical}, \citet{chaussonnet2016new} and \citet{shanmugadas2017canonical} have been published. 
They are all listed in Table~\ref{tab_correlations_SMD}. 
A brief summary of the operating parameters for each correlation cited in the introduction is given by \citet{gepperth2013analysis}, and more details are provided by \citet{shanmugadas2018characterization}.
In the correlation of \citet{rizkalla1975influence}, $L_c$ represents a characteristic length, which was later attributed to a film thickness, even though the authors did not measure the film thickness in the original experiment. In the present work, we considered $L_c$ as the hydraulic diameter of the gas duct where the film is present. This corresponds to twice the channel height ($2H$).\\
The correlation by \citet{el1980airblast} was derived from experiment at an annular atomizer, and the term $D_p$ refers to the nozzle diameter at the throat, which corresponds also to the hydraulic diameter. Therefore $D_p$ was set to $2H$.\\
In the correlation by \citet{aigner1986charakterisierung}, the term $\delta$ refers to the boundary layer thickness over the prefilmer at the atomizing edge. It is estimated from Eq.~\ref{eq_BL_correl}. In its original form, the correlation does not contain any proportionality constant. It is set arbitrarily here to 20 to match approximately the present results.
The correlation from \citet{inamura2012spray} is derived from an advanced model that was calibrated using an experiment similar to the present one. It was investigated at atmospheric pressure with water, and the droplets were collected 50 mm downstream with LDT. To summarize, the model supposes two different instabilities on the film surface, characterized by two different wavelengths, which define a volume and a mean diameter. This mean diameter constitutes the first parameter of a Gamma function to describe the droplet size distribution of the primary spray. Then, the authors use a Taylor Analogy Breakup (TAB) model to mimic the secondary breakup that occurs between the atomizing edge and the measurement distance. The interested reader is referred to \citep{inamura2012spray} for further details, and to \citep{chaussonnet2017timeGTP} for a concise explanation of how to implement the model.\\
The correlation from \citet{eckel2013semiempirical} is a model that predicts a bimodal drop size distribution downstream the atomizing edge. The two modes correspond to the bag and ligament breakups.
It is calibrated using the experiment of \citet{gepperth2012ligament}. It also supposes two different instabilities on the film surface. Further details can be found in \citep{eckel2013semiempirical}, and a condensed description of implementation can be found in \citep{chaussonnet2017timeGTP}.\\
The correlation from \citet{gepperth2013analysis} was derived from the same experimental setup as in this work, based on a previous campaign \citep{gepperth2012ligament} with a different atomizing edge. When the correlation was developed, the nominal atomizing edge thickness was set to 1 mm. However, later measurements showed that the actual thickness of the atomizing edge was 0.64 mm. Therefore, the correlation from \citet{gepperth2013analysis} was recalibrated by considering the actual $h_a$.
The correlation from \citet{chaussonnet2016new} is derived from 
the idea that the liquid accumulation is accelerated in axial direction, leading to a Rayleigh-Taylor instability. It is calibrated using the experiment of \citet{gepperth2010pre}. For this correlation, the correction of the atomizing edge thickness $h_a$ is also taken into account. Therefore $C_{12}$ is set to 1.69, and not to 1.40 as stated in the original paper.
In the correlation by \citet{shanmugadas2017canonical}, $t_{rim}$ corresponds to the thickness of the liquid accumulation at the atomizing edge. It is approximated here as $h_a$.
In the perspective of the present work, and because of the significance of the aerodynamic stress, it is worth to note that almost all the correlations depend on the term $\rho_g^{\alpha} U_g^{\beta}$, with the ratio $\beta / \alpha$ either explicitly equal to 2 or close to 2. On the contrary, the momentum flux of the liquid is not found in these correlations.

\begin{table}
	\renewcommand*{\arraystretch}{2.2}
	\scriptsize
	\centering
	\begin{tabular}{  l  l  l  }
	1 & \citet{rizkalla1975influence} &
SMD = \num{3.33d-3}
$\ddfrac{\sqrt{\rho_l \sigma L_c}}{\rho_g U_g^2} \left(1 + \ddfrac{\dot{m}_l} {\dot{m}_g} \right) $\\
& & \ \ \ \ \ \ \ \ + \num{13d-3} $ \left(  \ddfrac{\mu_l^2}{\sigma \rho_l} \right)^{0.425} L_c^{0.575} \left(1 + \ddfrac{\dot{m}_l} {\dot{m}_g} \right)^2$  \\

	2 & \citet{el1980airblast} &
SMD = $\left[ 0.073 \left( \ddfrac{\sigma}{\rho_g U_g^2}  \right)^{0.6} \left( \ddfrac{\rho_l}{\rho_g}  \right)^{0.1} D_p^{0.4} + 0.0015 \sqrt{ \ddfrac{\mu_l^2 D_p}{\sigma \rho_l} } \right] \left(1 + \ddfrac{\dot{m}_l}{\dot{m}_g} \right)$ \\

	3 & \citet{aigner1986charakterisierung} &
SMD $\propto \sigma^{0.5} \rho_g^{-0.4} U_g^{-1.05} \delta^{0.3} \left( \ddfrac{\dot{m}_l}{\rho_l} \right)^{0.15} \mu_l^{0.15}$ \\

	4 & \citet{inamura2012spray} &
SMD = $\ddfrac{m}{q} \ddfrac{\Gamma(q+3)}{\Gamma(q+2)}$
\quad with $m$ and $q$ functions of the model \\

	5 & \citet{eckel2013semiempirical} &
SMD extracted from a global Volume Probability Density Function\\

	6 & \citet{gepperth2013analysis} &
$\ddfrac{\textrm {SMD}}{\delta} = 12.08 \left( \ddfrac{\rho_g U_g \delta}{\mu_g} \right)^{-0.28} \left( \ddfrac{\rho_g U_g^2 \delta}{\sigma} \right)^{-0.31} \left( \ddfrac{\rho_l}{\rho_g} \right)^{-0.037} \left( \ddfrac{h_a}{\delta} \right)^{0.27}$ \\

	7 & \citet{chaussonnet2016new} &
SMD = $\ddfrac{C_{12}}{0.7 U_g} \sqrt{ \ddfrac{h_a \sigma}{\rho_l \rho_g} } \left(\sqrt{\rho_l}+\sqrt{\rho_g} \right)$ \\

	8 & \citet{shanmugadas2017canonical} &
$\ddfrac{\textrm {SMD}}{\delta} = 0.07 \left( \ddfrac{\rho_g U_g^2 \delta}{\sigma} \right)^{-0.33} \left( 1 + \ddfrac{\dot{m}_g}{\dot{m}_l} \right)^{-4.772} \left( \ddfrac{t_{rim}}{\delta} \right)^{0.221}$ \\
	\end{tabular}
	\caption{Correlations on the SMD  from the literature, compared in figures~\ref{fig_correlations_expe_comparison_single} and \ref{fig_correlations_expe_comparison_all}}
	\label{tab_correlations_SMD}
\end{table} 

\subsection{Discussion}

\begin{figure}%
\centering
	\includegraphics[width=\textwidth,keepaspectratio]{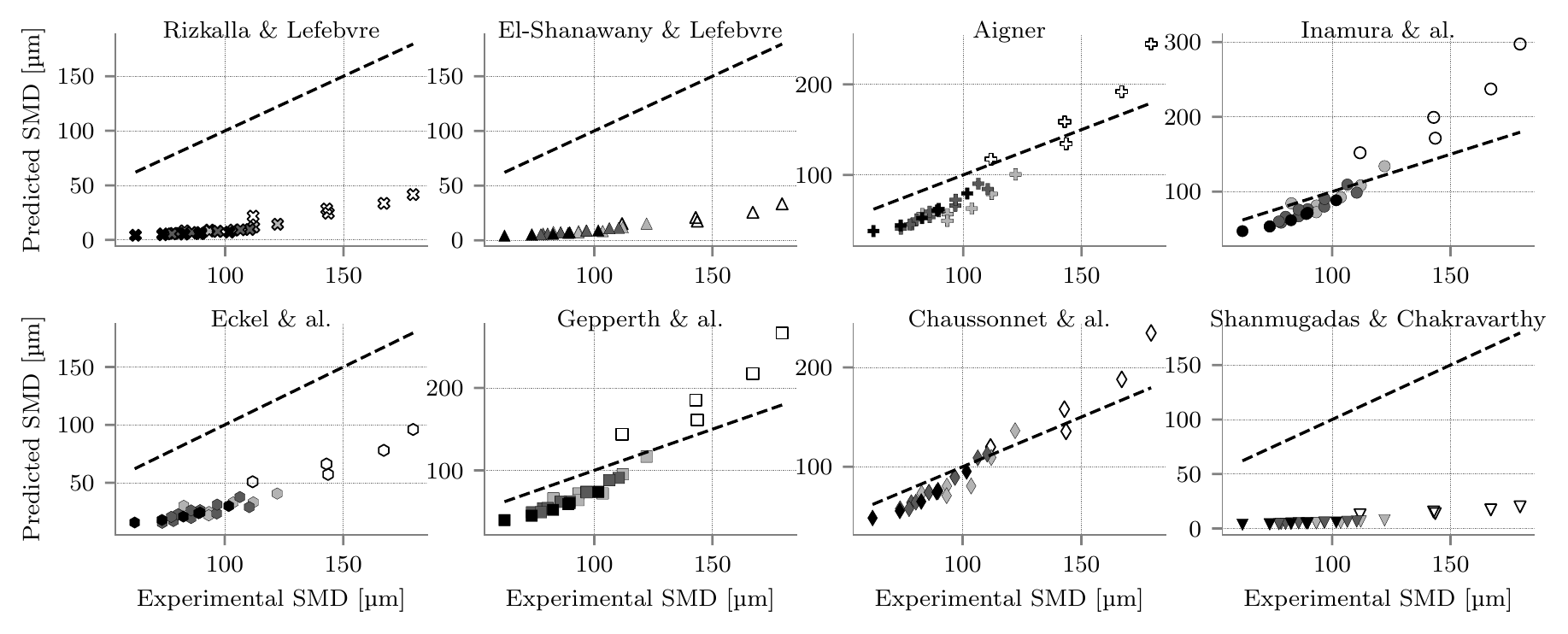}
	\caption{Individual comparison of correlations from literature with the experimental data. Dashed line: $y=x$.}
	\label{fig_correlations_expe_comparison_single}
\end{figure}

\begin{figure}%
\centering
	\includegraphics[width=\textwidth,keepaspectratio]{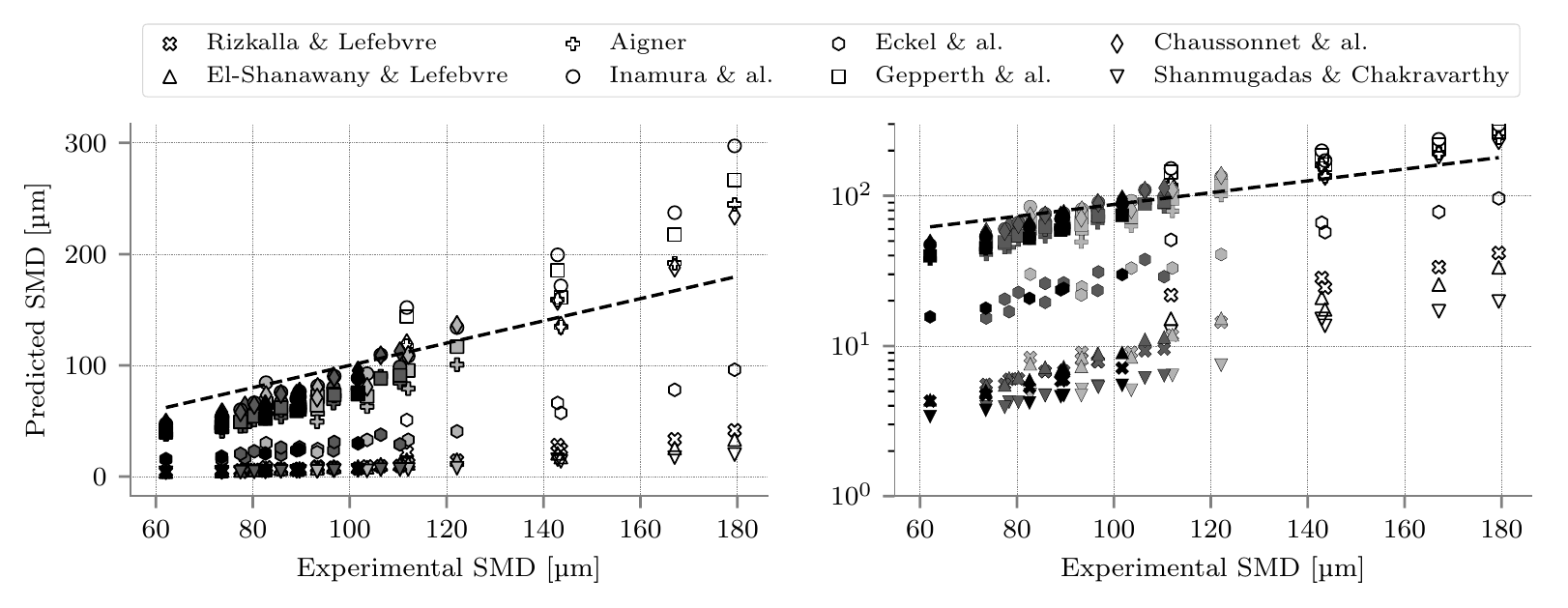}
	\caption{Comparison of correlations from literature with the experimental data in linear scale (left) and logarithmic scale (right).}
	\label{fig_correlations_expe_comparison_all}
\end{figure}

The droplet size as predicted by the correlations are plotted individually in figure~\ref{fig_correlations_expe_comparison_single} and superimposed to each other in figure~\ref{fig_correlations_expe_comparison_all}. Apart from the correlations from \citet{gepperth2013analysis,chaussonnet2016new,inamura2012spray}, all correlations underestimate the SMD.
There are several reasons to explain the discrepancies observed in figures~\ref{fig_correlations_expe_comparison_single} and \ref{fig_correlations_expe_comparison_all}. 
First, in all the experiments except the one from Gepperth \etal the spray droplets were measured by PDA or LDT. As mentioned in the introduction, these techniques rely on spherical droplets, which are usually smaller than non-spherical droplets, and lead to a bias, especially in the vicinity of the atomizing edge. The second reason is a corollary of the first: the measurements with PDA or LDT are made far away from the nozzle, where secondary breakup has significantly decreased the mean size of the spray droplets. Third, some experiments \citep{aigner1988swirl,shanmugadas2017canonical} were made with a swirl, leading to an enhanced shearing of the droplets, and consequently reducing the mean droplet size.
Interestingly, the correlation from \citet{inamura2012spray} was also calibrated with droplets collected 50 mm downstream, where secondary breakup occurs. Hence, it is also expected to give an underestimated SMD. However, it predicts the current experiment with a remarkable good agreement. As pointed out in \citep{chaussonnet2017timeGTP}, this can be explained by the fact that the model first estimates a larger mean droplet size, and then reduces it by using the TAB model.
The correlation from \citet{eckel2013semiempirical} was calibrated using the experiment by \citet{gepperth2012ligament}. Hence, the discrepancy comes from the calibration constants. This model was already compared to an experiment similar to the present one in a previous publication \citep{chaussonnet2017timeGTP} and it was also found to deliver too small droplets. This is because in the bimodal PDF used by this model, one peak is overestimated compared to the other. After recalibration, it delivered a good prediction of the SMD \citep{chaussonnet2017timeGTP}.
\\For all correlations, the dependency on the ambient pressure is represented by the gas density, whose exponent is given in Table~\ref{tab_correlations_SMD_exponent_rhog}.
Note that the boundary layer thickness $\delta$ also depends on the gas density. This was taken in account for entry 6. For the expression of $\delta$ in entry 8, not definition could be found.
Obviously, all correlations overestimate the influence of the gas density
for the same reasons as for the deviations in figures~{\ref{fig_correlations_expe_comparison_single}} and {\ref{fig_correlations_expe_comparison_all}}.
The closest value is given by \citet{shanmugadas2017canonical}, even though the absolute value of their SMD is the most underestimated. The best correlations to estimate the SMD
(6 and 7 in Tables~\ref{tab_correlations_SMD} and \ref{tab_correlations_SMD_exponent_rhog})
have an exponent of -0.57 and -0.5, respectively, which leads to a too steep correlation line in Figs~\ref{fig_correlations_expe_comparison_single} and \ref{fig_correlations_expe_comparison_all}.

\begin{table}%
	\renewcommand*{\arraystretch}{1.2}
	\scriptsize
	\centering
	\begin{tabular}{  l c c c c c c c c c  }
	Correlation from Table~\ref{tab_correlations_SMD}& 1& 2& 3& 4 &5 & 6 & 7 &8 & Present exp. \\
	Exponent of $\rho_g$ & -1& -0.7& -0.6& N/A & N/A & -0.57 & -0.5 & -0.33 & \textbf{-0.29} \\
	\end{tabular}
	\caption{Exponent of the gas density in the correlations from Table~\ref{tab_correlations_SMD}}
	\label{tab_correlations_SMD_exponent_rhog}
\end{table} 

To conclude this part, it is observed that except the model from \citet{inamura2012spray}, all models calibrated with LDA/PDT tends to underestimate the SMD. The authors of the present paper claim that primary breakup models must be calibrated with data collected directly at the atomizing edge. To the author knowledge, the use of shadowgraphy and the post-processing of the image is nowadays the only appropriate diagnostic.

\section{Conclusion and outlook}

In this work the liquid accumulation and the primary droplets were studied for prefilming airblast atomization by means of PIV and shadowgraphy technique. The influence of the gas velocity, the ambient pressure and the aerodynamics stress were individually investigated.
It was observed in single-phase experiments that the velocity profile at the center line can be determined from the momentum thickness of the upstream boundary layer and from a shear velocity $u_{\tau}$ whose value is close to the one of the upstream boundary layer.\\
The qualitative observation of the liquid breakup revealed the same type of structures as reported by \citet{zandian2017planar}, although their regime map was derived numerically for liquid sheet breakup and does not correspond to the present observations.
A strong link between the SMD of the primary spray and the characteristic length of the liquid accumulation was discovered, which stresses the statement that the state of the liquid accumulation determines the characteristics of the primary spray.
In addition, a significant correlation between the PDF of the ligament length and the location where the droplets are created could be demonstrated. This stresses the importance of taking the liquid accumulation into account to determine the injection location in low-order primary breakup models.\\
It was shown that the aerodynamic stress has a major influence on the spray characteristics and the liquid accumulation, independent of the individual values of the gas velocity and the ambient pressure. 
In the range of low to moderate film loading, the liquid accumulation acts as a buffer that decouples the film flow and the breakup process. Hence, the momentum of the injected liquid has no influence on the atomization and the aerodynamic stress alone should be preferred to the momentum flux ratio $M$ to describe this type of atomization.\\
Quantitatively, the length scale of the liquid accumulation scales with $p^{-0.422}$. The SMD of the primary spray and the breakup frequency are proportional to $p^{-0.29}$ and $p^{-0.72}$, respectively.
Finally, the result on the spray SMD were compared to some correlations from the literature. It was highlighted that the models which were calibrated by measurements far away from the injector, typically with PDA or LDT, underestimate the droplet size compared to the present results. This is because secondary atomization is included which strongly reduces the droplet sizes. The models calibrated on a similar experiment were able to correctly predict the SMD, with an acceptable dependence on ambient pressure.
It should be mentioned that the experimental database gathered in this work will help to develop models for the liquid accumulation and to refine existing models for prefilming airblast atomization.\\
These two phenomena show the directions for future experimental investigations.

\section{Acknowledgement \label{sec_mercitoutlemonde}}

G. Chaussonnet wrote the paper, conducted the study on correlations, developed the model for the gas flow downstream the atomizing edge and discussed on the results. He acknowledges the Helmholtz Association of German Research Centres (HGF) for funding (Grant No. 34.14.02). S. Gepperth designed the test-rig, conducted the experiment and performed the primary post-process. He acknowledges the European Community's Seventh Framework Program (FP7/2007- 2013) under Grant Agreement No. ACP8-GA-2009-234009 for the funding of the KIAI project.

\section*{References}
{\scriptsize
\bibliography{IJMF.bbl}

\end{document}

%% file: macros.tex
\def\etal{\textit{et al.} }
\def\ie{\textit{i.e.} }
\def\eg{\textit{e.g.} }

\def\ln{\text{ln}}

\def\Re{\text{Re}}
\def\We{\text{We}}

%% file: structure.tex
\usepackage{amsmath}
\usepackage[mathscr]{euscript}	

\usepackage{siunitx}

\usepackage{graphicx}
\usepackage{epsfig}
\usepackage{epstopdf}
\usepackage{subfig}

\usepackage{tabularx}
\usepackage	{multirow}			
\usepackage	{multicol}

\usepackage[applemac]{inputenc}
\usepackage[T1]{fontenc}

\usepackage{moreverb}

\usepackage	[colorlinks=true,linkcolor=black,citecolor= blue,breaklinks]{hyperref}

\usepackage{textgreek}  


%% file: sketch_sheet_vs_accumulation_final.pdf_tex
\begingroup%
  \makeatletter%
  \providecommand\color[2][]{%
    \errmessage{(Inkscape) Color is used for the text in Inkscape, but the package 'color.sty' is not loaded}%
    \renewcommand\color[2][]{}%
  }%
  \providecommand\transparent[1]{%
    \errmessage{(Inkscape) Transparency is used (non-zero) for the text in Inkscape, but the package 'transparent.sty' is not loaded}%
    \renewcommand\transparent[1]{}%
  }%
  \providecommand\rotatebox[2]{#2}%
  \ifx\svgwidth\undefined%
    \setlength{\unitlength}{2742.84761392bp}%
    \ifx\svgscale\undefined%
      \relax%
    \else%
      \setlength{\unitlength}{\unitlength * \real{\svgscale}}%
    \fi%
  \else%
    \setlength{\unitlength}{\svgwidth}%
  \fi%
  \global\let\svgwidth\undefined%
  \global\let\svgscale\undefined%
  \makeatother%
  \begin{picture}(1,0.25311585)%
    \put(0,0){\includegraphics[width=\unitlength,page=1]{sketch_sheet_vs_accumulation_final.pdf}}%
    \put(0.6,0.18){\color[rgb]{0,0,0}\makebox(0,0)[lt]{\begin{minipage}{0.25\unitlength}\raggedright High speed air flow\end{minipage}}}%
    \put(0.57,0.07714834){\color[rgb]{0,0,0}\makebox(0,0)[lt]{\begin{minipage}{0.16015679\unitlength}\raggedright Liquid stream \end{minipage}}}%
    \put(0.063,0.08){\color[rgb]{0,0,0}\makebox(0,0)[lt]{\begin{minipage}{0.16015679\unitlength}\raggedright Liquid film \end{minipage}}}%
    \put(0.1,0.18){\color[rgb]{0,0,0}\makebox(0,0)[lt]{\begin{minipage}{0.25\unitlength}\raggedright High speed air flow\end{minipage}}}%
  \end{picture}%
\endgroup%

%% file: intro_breakup.pdf_tex
\begingroup%
  \makeatletter%
  \providecommand\color[2][]{%
    \errmessage{(Inkscape) Color is used for the text in Inkscape, but the package 'color.sty' is not loaded}%
    \renewcommand\color[2][]{}%
  }%
  \providecommand\transparent[1]{%
    \errmessage{(Inkscape) Transparency is used (non-zero) for the text in Inkscape, but the package 'transparent.sty' is not loaded}%
    \renewcommand\transparent[1]{}%
  }%
  \providecommand\rotatebox[2]{#2}%
  \ifx\svgwidth\undefined%
    \setlength{\unitlength}{1107.74222074bp}%
    \ifx\svgscale\undefined%
      \relax%
    \else%
      \setlength{\unitlength}{\unitlength * \real{\svgscale}}%
    \fi%
  \else%
    \setlength{\unitlength}{\svgwidth}%
  \fi%
  \global\let\svgwidth\undefined%
  \global\let\svgscale\undefined%
  \makeatother%
  \begin{picture}(1,0.51194769)%
    \put(0,0){\includegraphics[width=\unitlength,page=1]{intro_breakup.pdf}}%
    \put(0.61477007,0.49197868){\color[rgb]{0,0,0}\makebox(0,0)[lt]{\begin{minipage}{0.3\unitlength}\raggedright Primary breakup intensity\end{minipage}}}%
    \put(0.80691712,0.36868708){\color[rgb]{0,0,0}\makebox(0,0)[lt]{\begin{minipage}{0.3\unitlength}\raggedright Secondary breakup intensity\end{minipage}}}%
    \put(0,0){\includegraphics[width=\unitlength,page=2]{intro_breakup.pdf}}%
  \end{picture}%
\endgroup%

%% file: HDT_test_rig_english.pdf_tex
\begingroup%
  \makeatletter%
  \providecommand\color[2][]{%
    \errmessage{(Inkscape) Color is used for the text in Inkscape, but the package 'color.sty' is not loaded}%
    \renewcommand\color[2][]{}%
  }%
  \providecommand\transparent[1]{%
    \errmessage{(Inkscape) Transparency is used (non-zero) for the text in Inkscape, but the package 'transparent.sty' is not loaded}%
    \renewcommand\transparent[1]{}%
  }%
  \providecommand\rotatebox[2]{#2}%
  \ifx\svgwidth\undefined%
    \setlength{\unitlength}{1029.96428242bp}%
    \ifx\svgscale\undefined%
      \relax%
    \else%
      \setlength{\unitlength}{\unitlength * \real{\svgscale}}%
    \fi%
  \else%
    \setlength{\unitlength}{\svgwidth}%
  \fi%
  \global\let\svgwidth\undefined%
  \global\let\svgscale\undefined%
  \makeatother%
  \begin{picture}(1,0.66935406)%
    \put(0,0){\includegraphics[width=\unitlength,page=1]{HDT_test_rig_english.pdf}}%
    \put(0.1338812,0.64111105){\color[rgb]{0,0,0}\makebox(0,0)[lt]{\begin{minipage}{0.0842609\unitlength}\raggedright Compressor\end{minipage}}}%
    \put(0.03309277,0.62562287){\color[rgb]{0,0,0}\makebox(0,0)[lt]{\begin{minipage}{0.0842609\unitlength}\raggedright Air in\end{minipage}}}%
    \put(0.22762101,0.53408018){\color[rgb]{0,0,0}\makebox(0,0)[lt]{\begin{minipage}{0.0842609\unitlength}\raggedright Orifice plate\end{minipage}}}%
    \put(0.04922847,0.53408018){\color[rgb]{0,0,0}\makebox(0,0)[lt]{\begin{minipage}{0.0842609\unitlength}\raggedright Orifice plate\end{minipage}}}%
    \put(0.45430506,0.55456635){\color[rgb]{0,0,0}\makebox(0,0)[lt]{\begin{minipage}{0.07144148\unitlength}\raggedright Inlet\end{minipage}}}%
    \put(0.65590693,0.49339917){\color[rgb]{0,0,0}\makebox(0,0)[lt]{\begin{minipage}{0.0842609\unitlength}\raggedright Generic atomizer\end{minipage}}}%
    \put(0.62367484,0.33980009){\color[rgb]{0,0,0}\makebox(0,0)[lt]{\begin{minipage}{0.20167847\unitlength}\raggedright Pressurized chamber with optical access\end{minipage}}}%
    \put(0.54509288,0.23597495){\color[rgb]{0,0,0}\makebox(0,0)[lt]{\begin{minipage}{0.0842609\unitlength}\raggedright Air out\end{minipage}}}%
    \put(0.54509288,0.18384338){\color[rgb]{0,0,0}\makebox(0,0)[lt]{\begin{minipage}{0.0842609\unitlength}\raggedright Air out\end{minipage}}}%
    \put(0.62959153,0.27){\color[rgb]{0,0,0}\makebox(0,0)[lt]{\begin{minipage}{0.14285923\unitlength}\raggedright Pressure control valve\end{minipage}}}%
    \put(0.76042488,0.27154516){\color[rgb]{0,0,0}\makebox(0,0)[lt]{\begin{minipage}{0.13294869\unitlength}\raggedright Two-pass filter\end{minipage}}}%
    \put(0.67866186,0.12267621){\color[rgb]{0,0,0}\makebox(0,0)[lt]{\begin{minipage}{0.0842609\unitlength}\raggedright Air filter\end{minipage}}}%
    \put(0.7842223,0.09750198){\color[rgb]{0,0,0}\makebox(0,0)[lt]{\begin{minipage}{0.0842609\unitlength}\raggedright Liquid tank\end{minipage}}}%
    \put(0.46366447,0.17){\color[rgb]{0,0,0}\rotatebox{90}{\makebox(0,0)[lt]{\begin{minipage}{0.1310725\unitlength}\raggedright Check valve\end{minipage}}}}%
    \put(0.43406994,0.15){\color[rgb]{0,0,0}\rotatebox{90}{\makebox(0,0)[lt]{\begin{minipage}{0.13731405\unitlength}\raggedright Magnetic valve\end{minipage}}}}%
    \put(0.23,0.25){\color[rgb]{0,0,0}\makebox(0,0)[lt]{\begin{minipage}{0.1700942\unitlength}\raggedright Magnetic valve\end{minipage}}}%
    \put(0.23,0.22){\color[rgb]{0,0,0}\makebox(0,0)[lt]{\begin{minipage}{0.13731886\unitlength}\raggedright Pressure control valve\end{minipage}}}%
    \put(0.26,0.17827534){\color[rgb]{0,0,0}\makebox(0,0)[lt]{\begin{minipage}{0.11546863\unitlength}\raggedright Control valve\end{minipage}}}%
    \put(0.34878808,0.02281143){\color[rgb]{0,0,0}\makebox(0,0)[lt]{\begin{minipage}{0.11546863\unitlength}\raggedright Coriolis flowmeter\end{minipage}}}%
    \put(0.24935251,0.08876373){\color[rgb]{0,0,0}\makebox(0,0)[lt]{\begin{minipage}{0.11546863\unitlength}\raggedright Filter\end{minipage}}}%
    \put(0.30570436,0.10228719){\color[rgb]{0,0,0}\makebox(0,0)[lt]{\begin{minipage}{0.05721419\unitlength}\raggedright Fuel pump\end{minipage}}}%
    \put(0.16407673,0.11548336){\color[rgb]{0,0,0}\makebox(0,0)[lt]{\begin{minipage}{0.0842609\unitlength}\raggedright Liquid tank\end{minipage}}}%
    \put(0.03,0.27){\color[rgb]{0,0,0}\makebox(0,0)[lt]{\begin{minipage}{0.11546863\unitlength}\raggedright Pressure control valve\end{minipage}}}%
    \put(0.03641551,0.05809821){\color[rgb]{0,0,0}\makebox(0,0)[lt]{\begin{minipage}{0.11546863\unitlength}\raggedright Pressurized nitrogen\end{minipage}}}%
    \put(0.35318832,0.3451203){\color[rgb]{0,0,0}\makebox(0,0)[lt]{\begin{minipage}{0.1124468\unitlength}\raggedright Flow straightener\end{minipage}}}%
    \put(0,0){\includegraphics[width=\unitlength,page=2]{HDT_test_rig_english.pdf}}%
    \put(0.90828696,0.11123015){\color[rgb]{0,0,0}\makebox(0,0)[lt]{\begin{minipage}{0.07687307\unitlength}\raggedright $\vec{z}$\end{minipage}}}%
    \put(0.95150021,0.06884319){\color[rgb]{0,0,0}\makebox(0,0)[lt]{\begin{minipage}{0.06013554\unitlength}\raggedright $\vec{y}$\end{minipage}}}%
    \put(0.88598296,0.06450751){\color[rgb]{0,0,0}\makebox(0,0)[lt]{\begin{minipage}{0.08034926\unitlength}\raggedright $\vec{x}$\end{minipage}}}%
    \put(0,0){\includegraphics[width=\unitlength,page=3]{HDT_test_rig_english.pdf}}%
  \end{picture}%
\endgroup%

%% file: prefilmer_side.pdf_tex
\begingroup%
  \makeatletter%
  \providecommand\color[2][]{%
    \errmessage{(Inkscape) Color is used for the text in Inkscape, but the package 'color.sty' is not loaded}%
    \renewcommand\color[2][]{}%
  }%
  \providecommand\transparent[1]{%
    \errmessage{(Inkscape) Transparency is used (non-zero) for the text in Inkscape, but the package 'transparent.sty' is not loaded}%
    \renewcommand\transparent[1]{}%
  }%
  \providecommand\rotatebox[2]{#2}%
  \ifx\svgwidth\undefined%
    \setlength{\unitlength}{653.8032748bp}%
    \ifx\svgscale\undefined%
      \relax%
    \else%
      \setlength{\unitlength}{\unitlength * \real{\svgscale}}%
    \fi%
  \else%
    \setlength{\unitlength}{\svgwidth}%
  \fi%
  \global\let\svgwidth\undefined%
  \global\let\svgscale\undefined%
  \makeatother%
  \begin{picture}(1,0.65834705)%
    \put(0,0){\includegraphics[width=\unitlength,page=1]{prefilmer_side.pdf}}%
    \put(0.22245834,0.56171662){\color[rgb]{0,0,0}\makebox(0,0)[lt]{\begin{minipage}{0.12167193\unitlength}\raggedright Grid\end{minipage}}}%
    \put(0,0){\includegraphics[width=\unitlength,page=2]{prefilmer_side.pdf}}%
    \put(0.18707197,0.18255216){\color[rgb]{0,0,0}\makebox(0,0)[lt]{\begin{minipage}{0.15759412\unitlength}\raggedright Honeycomb\end{minipage}}}%
    \put(0,0){\includegraphics[width=\unitlength,page=3]{prefilmer_side.pdf}}%
    \put(0.17883823,0.11744162){\color[rgb]{0,0,0}\makebox(0,0)[lt]{\begin{minipage}{0.15759412\unitlength}\raggedright Nozzle\end{minipage}}}%
    \put(0,0){\includegraphics[width=\unitlength,page=4]{prefilmer_side.pdf}}%
    \put(0.57300978,0.23195126){\color[rgb]{0,0,0}\makebox(0,0)[lt]{\begin{minipage}{0.15759412\unitlength}\raggedright Prefilmer\end{minipage}}}%
    \put(0,0){\includegraphics[width=\unitlength,page=5]{prefilmer_side.pdf}}%
    \put(0.71675429,0.57062713){\color[rgb]{0,0,0}\makebox(0,0)[lt]{\begin{minipage}{0.23639118\unitlength}\raggedright Generic atomizer\end{minipage}}}%
    \put(0,0){\includegraphics[width=\unitlength,page=6]{prefilmer_side.pdf}}%
    \put(0.26060542,0.37611507){\color[rgb]{0,0,0}\makebox(0,0)[lt]{\begin{minipage}{0.12167193\unitlength}\raggedright $H_{in}$\end{minipage}}}%
    \put(0,0){\includegraphics[width=\unitlength,page=7]{prefilmer_side.pdf}}%
    \put(0.49679699,0.4270874){\color[rgb]{0,0,0}\makebox(0,0)[lt]{\begin{minipage}{0.12167193\unitlength}\raggedright $l_{in}$\end{minipage}}}%
    \put(0,0){\includegraphics[width=\unitlength,page=8]{prefilmer_side.pdf}}%
    \put(0.75472446,0.16069966){\color[rgb]{0,0,0}\makebox(0,0)[lt]{\begin{minipage}{0.12110144\unitlength}\raggedright $\vec{z}$\end{minipage}}}%
    \put(0.82509441,0.09621994){\color[rgb]{0,0,0}\makebox(0,0)[lt]{\begin{minipage}{0.09473409\unitlength}\raggedright $\vec{y}$\end{minipage}}}%
    \put(0.71041093,0.09168405){\color[rgb]{0,0,0}\makebox(0,0)[lt]{\begin{minipage}{0.12657764\unitlength}\raggedright $\vec{x}$\end{minipage}}}%
    \put(0,0){\includegraphics[width=\unitlength,page=9]{prefilmer_side.pdf}}%
  \end{picture}%
\endgroup%

%% file: prefilmer_top.pdf_tex
\begingroup%
  \makeatletter%
  \providecommand\color[2][]{%
    \errmessage{(Inkscape) Color is used for the text in Inkscape, but the package 'color.sty' is not loaded}%
    \renewcommand\color[2][]{}%
  }%
  \providecommand\transparent[1]{%
    \errmessage{(Inkscape) Transparency is used (non-zero) for the text in Inkscape, but the package 'transparent.sty' is not loaded}%
    \renewcommand\transparent[1]{}%
  }%
  \providecommand\rotatebox[2]{#2}%
  \ifx\svgwidth\undefined%
    \setlength{\unitlength}{649.63160057bp}%
    \ifx\svgscale\undefined%
      \relax%
    \else%
      \setlength{\unitlength}{\unitlength * \real{\svgscale}}%
    \fi%
  \else%
    \setlength{\unitlength}{\svgwidth}%
  \fi%
  \global\let\svgwidth\undefined%
  \global\let\svgscale\undefined%
  \makeatother%
  \begin{picture}(1,0.47493085)%
    \put(0,0){\includegraphics[width=\unitlength,page=1]{prefilmer_top.pdf}}%
    \put(0.7808735,0.27798724){\color[rgb]{0,0,0}\makebox(0,0)[lt]{\begin{minipage}{0.20698536\unitlength}\raggedright Exit holes\end{minipage}}}%
    \put(0,0){\includegraphics[width=\unitlength,page=2]{prefilmer_top.pdf}}%
    \put(0.83207451,0.41738902){\color[rgb]{0,0,0}\makebox(0,0)[lt]{\begin{minipage}{0.13936464\unitlength}\raggedright Liquid inlet ducts\end{minipage}}}%
    \put(0,0){\includegraphics[width=\unitlength,page=3]{prefilmer_top.pdf}}%
    \put(0.05337768,0.44278805){\color[rgb]{0,0,0}\makebox(0,0)[lt]{\begin{minipage}{0.17729827\unitlength}\raggedright Liquid inlet ducts\end{minipage}}}%
    \put(0,0){\includegraphics[width=\unitlength,page=4]{prefilmer_top.pdf}}%
    \put(0.54948292,0.45737177){\color[rgb]{0,0,0}\makebox(0,0)[lt]{\begin{minipage}{0.13936464\unitlength}\raggedright Reservoir\end{minipage}}}%
    \put(0,0){\includegraphics[width=\unitlength,page=5]{prefilmer_top.pdf}}%
    \put(0.11479984,0.20825952){\color[rgb]{0,0,0}\makebox(0,0)[lt]{\begin{minipage}{0.13771533\unitlength}\raggedright Prefilming surface\end{minipage}}}%
    \put(0,0){\includegraphics[width=\unitlength,page=6]{prefilmer_top.pdf}}%
    \put(0.88424311,0.14833612){\color[rgb]{0,0,0}\makebox(0,0)[lt]{\begin{minipage}{0.12187911\unitlength}\raggedright $\vec{z}$\end{minipage}}}%
    \put(0.87998284,0.04973609){\color[rgb]{0,0,0}\makebox(0,0)[lt]{\begin{minipage}{0.09534244\unitlength}\raggedright $\vec{y}$\end{minipage}}}%
    \put(0.78199622,0.14302269){\color[rgb]{0,0,0}\makebox(0,0)[lt]{\begin{minipage}{0.12739047\unitlength}\raggedright $\vec{x}$\end{minipage}}}%
  \end{picture}%
\endgroup%

%% file: PLTV_algo_2.pdf_tex
\begingroup%
  \makeatletter%
  \providecommand\color[2][]{%
    \errmessage{(Inkscape) Color is used for the text in Inkscape, but the package 'color.sty' is not loaded}%
    \renewcommand\color[2][]{}%
  }%
  \providecommand\transparent[1]{%
    \errmessage{(Inkscape) Transparency is used (non-zero) for the text in Inkscape, but the package 'transparent.sty' is not loaded}%
    \renewcommand\transparent[1]{}%
  }%
  \providecommand\rotatebox[2]{#2}%
  \ifx\svgwidth\undefined%
    \setlength{\unitlength}{459.75002018bp}%
    \ifx\svgscale\undefined%
      \relax%
    \else%
      \setlength{\unitlength}{\unitlength * \real{\svgscale}}%
    \fi%
  \else%
    \setlength{\unitlength}{\svgwidth}%
  \fi%
  \global\let\svgwidth\undefined%
  \global\let\svgscale\undefined%
  \makeatother%
  \begin{picture}(1,0.47308316)%
    \put(0,0){\includegraphics[width=\unitlength,page=1]{PLTV_algo_2.pdf}}%
    \put(0.37881618,0.28082034){\color[rgb]{0,0,0}\makebox(0,0)[lt]{\begin{minipage}{0.15963645\unitlength}\raggedright $L_{lig}$\end{minipage}}}%
    \put(0.69613595,0.3075621){\color[rgb]{0,0,0}\makebox(0,0)[lt]{\begin{minipage}{0.15963645\unitlength}\raggedright $P_{lig}$\end{minipage}}}%
    \put(0,0){\includegraphics[width=\unitlength,page=2]{PLTV_algo_2.pdf}}%
  \end{picture}%
\endgroup%

%% file: PLTV_algo_sequence.pdf_tex
\begingroup%
  \makeatletter%
  \providecommand\color[2][]{%
    \errmessage{(Inkscape) Color is used for the text in Inkscape, but the package 'color.sty' is not loaded}%
    \renewcommand\color[2][]{}%
  }%
  \providecommand\transparent[1]{%
    \errmessage{(Inkscape) Transparency is used (non-zero) for the text in Inkscape, but the package 'transparent.sty' is not loaded}%
    \renewcommand\transparent[1]{}%
  }%
  \providecommand\rotatebox[2]{#2}%
  \ifx\svgwidth\undefined%
    \setlength{\unitlength}{313.49999856bp}%
    \ifx\svgscale\undefined%
      \relax%
    \else%
      \setlength{\unitlength}{\unitlength * \real{\svgscale}}%
    \fi%
  \else%
    \setlength{\unitlength}{\svgwidth}%
  \fi%
  \global\let\svgwidth\undefined%
  \global\let\svgscale\undefined%
  \makeatother%
  \begin{picture}(1,0.63875601)%
    \put(0,0){\includegraphics[width=\unitlength,page=1]{PLTV_algo_sequence.pdf}}%
    \put(0.31469063,0.50820028){\color[rgb]{0,0,0}\makebox(0,0)[lt]{\begin{minipage}{0.54493859\unitlength}\raggedright Mean velocity\end{minipage}}}%
    \put(0.17152652,0.05956958){\color[rgb]{0,0,0}\makebox(0,0)[lt]{\begin{minipage}{0.84090054\unitlength}\raggedright Number of measurement points\end{minipage}}}%
    \put(0.27642287,0.27153848){\color[rgb]{0,0,0}\makebox(0,0)[lt]{\begin{minipage}{0.5546737\unitlength}\raggedright 90\textsuperscript{th}-percentile\end{minipage}}}%
    \put(0,0){\includegraphics[width=\unitlength,page=2]{PLTV_algo_sequence.pdf}}%
  \end{picture}%
\endgroup%

%% file: sketch_near_wake.pdf_tex
\begingroup%
  \makeatletter%
  \providecommand\color[2][]{%
    \errmessage{(Inkscape) Color is used for the text in Inkscape, but the package 'color.sty' is not loaded}%
    \renewcommand\color[2][]{}%
  }%
  \providecommand\transparent[1]{%
    \errmessage{(Inkscape) Transparency is used (non-zero) for the text in Inkscape, but the package 'transparent.sty' is not loaded}%
    \renewcommand\transparent[1]{}%
  }%
  \providecommand\rotatebox[2]{#2}%
  \ifx\svgwidth\undefined%
    \setlength{\unitlength}{2965.9099717bp}%
    \ifx\svgscale\undefined%
      \relax%
    \else%
      \setlength{\unitlength}{\unitlength * \real{\svgscale}}%
    \fi%
  \else%
    \setlength{\unitlength}{\svgwidth}%
  \fi%
  \global\let\svgwidth\undefined%
  \global\let\svgscale\undefined%
  \makeatother%
  \begin{picture}(1,0.39755733)%
    \put(0,0){\includegraphics[width=\unitlength,page=1]{sketch_near_wake.pdf}}%
    \put(0.89860027,0.10779018){\color[rgb]{0,0,0}\makebox(0,0)[lt]{\begin{minipage}{0.07969756\unitlength}\raggedright \end{minipage}}}%
    \put(2.93700163,1.02316342){\color[rgb]{0,0,0}\makebox(0,0)[lt]{\begin{minipage}{0.2973572\unitlength}\raggedright \end{minipage}}}%
    \put(0.93009446,0.16620964){\color[rgb]{0,0,0}\makebox(0,0)[lt]{\begin{minipage}{0.06897158\unitlength}\raggedright $y$\end{minipage}}}%
    \put(0.24556716,0.38376469){\color[rgb]{0,0,0}\makebox(0,0)[lt]{\begin{minipage}{0.06897158\unitlength}\raggedright $z$\end{minipage}}}%
    \put(0.02913038,0.098796){\color[rgb]{0,0,0}\makebox(0,0)[lt]{\begin{minipage}{0.12645329\unitlength}\raggedright Laminar sublayer\end{minipage}}}%
    \put(0.06651158,0.20291796){\color[rgb]{0,0,0}\makebox(0,0)[lt]{\begin{minipage}{0.13078825\unitlength}\raggedright Log layer\end{minipage}}}%
    \put(0.05,0.29155671){\color[rgb]{0,0,0}\makebox(0,0)[lt]{\begin{minipage}{0.20639872\unitlength}\raggedright Defect layer\end{minipage}}}%
    \put(0.38257253,0.3842659){\color[rgb]{0,0,0}\makebox(0,0)[lt]{\begin{minipage}{0.32393158\unitlength}\raggedright Initial velocity profile\end{minipage}}}%
    \put(0.4606847,0.28596137){\color[rgb]{0,0,0}\makebox(0,0)[lt]{\begin{minipage}{0.14870896\unitlength}\raggedright Outer wake\end{minipage}}}%
    \put(0.68626378,0.18376462){\color[rgb]{0,0,0}\makebox(0,0)[lt]{\begin{minipage}{0.14870896\unitlength}\raggedright Inner wake\end{minipage}}}%
    \put(0.71943306,0.32610253){\color[rgb]{0,0,0}\makebox(0,0)[lt]{\begin{minipage}{0.11090373\unitlength}\raggedright "Gaussian" far wake  \end{minipage}}}%
    \put(0.27284813,0.13069298){\color[rgb]{0,0,0}\makebox(0,0)[lt]{\begin{minipage}{0.13873571\unitlength}\raggedright Inner laminar wake\end{minipage}}}%
    \put(0.45684328,0.07877632){\color[rgb]{0,0,0}\makebox(0,0)[lt]{\begin{minipage}{0.29118231\unitlength}\raggedright Turbulent inner wake\end{minipage}}}%
    \put(0.82415577,0.07877632){\color[rgb]{0,0,0}\makebox(0,0)[lt]{\begin{minipage}{0.14740564\unitlength}\raggedright Far wake\end{minipage}}}%
    \put(0,0){\includegraphics[width=\unitlength,page=2]{sketch_near_wake.pdf}}%
    \put(0.32391535,0.03055805){\color[rgb]{0,0,0}\makebox(0,0)[lt]{\begin{minipage}{0.28280433\unitlength}\raggedright $y^+\!\!\sim\!\!100$\end{minipage}}}%
    \put(0.70328435,0.03236242){\color[rgb]{0,0,0}\makebox(0,0)[lt]{\begin{minipage}{0.24448819\unitlength}\raggedright $y^+\!\!\sim\!\!5000$\end{minipage}}}%
    \put(0,0){\includegraphics[width=\unitlength,page=3]{sketch_near_wake.pdf}}%
  \end{picture}%
\endgroup%

%% file: Time_serie_p3_7_ug40_Zandian_Nomenclature.pdf_tex
\begingroup%
  \makeatletter%
  \providecommand\color[2][]{%
    \errmessage{(Inkscape) Color is used for the text in Inkscape, but the package 'color.sty' is not loaded}%
    \renewcommand\color[2][]{}%
  }%
  \providecommand\transparent[1]{%
    \errmessage{(Inkscape) Transparency is used (non-zero) for the text in Inkscape, but the package 'transparent.sty' is not loaded}%
    \renewcommand\transparent[1]{}%
  }%
  \providecommand\rotatebox[2]{#2}%
  \ifx\svgwidth\undefined%
    \setlength{\unitlength}{611.0891994bp}%
    \ifx\svgscale\undefined%
      \relax%
    \else%
      \setlength{\unitlength}{\unitlength * \real{\svgscale}}%
    \fi%
  \else%
    \setlength{\unitlength}{\svgwidth}%
  \fi%
  \global\let\svgwidth\undefined%
  \global\let\svgscale\undefined%
  \makeatother%
  \begin{picture}(1,1.23214345)%
    \put(0,0){\includegraphics[width=\unitlength,page=1]{Time_serie_p3_7_ug40_Zandian_Nomenclature.pdf}}%
    \put(0.18862347,1.17771967){\color[rgb]{0,0,0}\makebox(0,0)[lt]{\begin{minipage}{0.20513718\unitlength}\raggedright p = 3 bar\end{minipage}}}%
    \put(0.69541689,1.17771967){\color[rgb]{0,0,0}\makebox(0,0)[lt]{\begin{minipage}{0.20513718\unitlength}\raggedright p = 7 bar\end{minipage}}}%
    \put(0,0){\includegraphics[width=\unitlength,page=2]{Time_serie_p3_7_ug40_Zandian_Nomenclature.pdf}}%
    \put(0.20633195,1.21652686){\color[rgb]{0,0,0}\makebox(0,0)[lt]{\begin{minipage}{0.20513718\unitlength}\raggedright LoLiD\end{minipage}}}%
    \put(0,0){\includegraphics[width=\unitlength,page=3]{Time_serie_p3_7_ug40_Zandian_Nomenclature.pdf}}%
    \put(0.49640291,1.21652686){\color[rgb]{0,0,0}\makebox(0,0)[lt]{\begin{minipage}{0.20513718\unitlength}\raggedright LoCLiD\end{minipage}}}%
    \put(0,0){\includegraphics[width=\unitlength,page=4]{Time_serie_p3_7_ug40_Zandian_Nomenclature.pdf}}%
    \put(0.77984297,1.21652686){\color[rgb]{0,0,0}\makebox(0,0)[lt]{\begin{minipage}{0.20513718\unitlength}\raggedright LoHBrLiD\end{minipage}}}%
    \put(0.42367295,1.09338548){\color[rgb]{0,0,0}\makebox(0,0)[lt]{\begin{minipage}{0.19517362\unitlength}\raggedright 0 \textmu s\end{minipage}}}%
    \put(0.4233565,0.95510875){\color[rgb]{0,0,0}\makebox(0,0)[lt]{\begin{minipage}{0.19517362\unitlength}\raggedright 71 \textmu s\end{minipage}}}%
    \put(0.42280997,0.81657308){\color[rgb]{0,0,0}\makebox(0,0)[lt]{\begin{minipage}{0.19517362\unitlength}\raggedright 143 \textmu s\end{minipage}}}%
    \put(0.42352911,0.67816683){\color[rgb]{0,0,0}\makebox(0,0)[lt]{\begin{minipage}{0.19517362\unitlength}\raggedright 214 \textmu s\end{minipage}}}%
    \put(0.42352911,0.53976071){\color[rgb]{0,0,0}\makebox(0,0)[lt]{\begin{minipage}{0.19517362\unitlength}\raggedright 286 \textmu s\end{minipage}}}%
    \put(0.42347159,0.4013545){\color[rgb]{0,0,0}\makebox(0,0)[lt]{\begin{minipage}{0.19517362\unitlength}\raggedright 357 \textmu s\end{minipage}}}%
    \put(0.42400852,0.26294828){\color[rgb]{0,0,0}\makebox(0,0)[lt]{\begin{minipage}{0.19517362\unitlength}\raggedright 428 \textmu s\end{minipage}}}%
    \put(0.42345243,0.12454212){\color[rgb]{0,0,0}\makebox(0,0)[lt]{\begin{minipage}{0.19517362\unitlength}\raggedright 500 \textmu s\end{minipage}}}%
    \put(0.93151475,1.12039381){\color[rgb]{0,0,0}\makebox(0,0)[lt]{\begin{minipage}{0.06003792\unitlength}\raggedright 0 \textmu s\end{minipage}}}%
    \put(0.93129423,1.01843736){\color[rgb]{0,0,0}\makebox(0,0)[lt]{\begin{minipage}{0.06375725\unitlength}\raggedright 50 \textmu s\end{minipage}}}%
    \put(0.93065182,0.91648085){\color[rgb]{0,0,0}\makebox(0,0)[lt]{\begin{minipage}{0.07243566\unitlength}\raggedright 100 \textmu s\end{minipage}}}%
    \put(0.93065182,0.8145243){\color[rgb]{0,0,0}\makebox(0,0)[lt]{\begin{minipage}{0.06871633\unitlength}\raggedright 150 \textmu s\end{minipage}}}%
    \put(0.93137096,0.71256781){\color[rgb]{0,0,0}\makebox(0,0)[lt]{\begin{minipage}{0.07119591\unitlength}\raggedright 200 \textmu s\end{minipage}}}%
    \put(0.93137096,0.61061131){\color[rgb]{0,0,0}\makebox(0,0)[lt]{\begin{minipage}{0.07367548\unitlength}\raggedright 250 \textmu s\end{minipage}}}%
    \put(0.93131339,0.5086548){\color[rgb]{0,0,0}\makebox(0,0)[lt]{\begin{minipage}{0.06\unitlength}\raggedright 300 \textmu s\end{minipage}}}%
    \put(0.93131339,0.4066983){\color[rgb]{0,0,0}\makebox(0,0)[lt]{\begin{minipage}{0.06251749\unitlength}\raggedright 350 \textmu s\end{minipage}}}%
    \put(0.93185036,0.30474184){\color[rgb]{0,0,0}\makebox(0,0)[lt]{\begin{minipage}{0.06871633\unitlength}\raggedright 400 \textmu s\end{minipage}}}%
    \put(0.93185036,0.20278533){\color[rgb]{0,0,0}\makebox(0,0)[lt]{\begin{minipage}{0.07491524\unitlength}\raggedright 450 \textmu s\end{minipage}}}%
    \put(0.93129423,0.10082887){\color[rgb]{0,0,0}\makebox(0,0)[lt]{\begin{minipage}{0.07367548\unitlength}\raggedright 500 \textmu s\end{minipage}}}%
  \end{picture}%
\endgroup%

%% file: Sketch_momentum_flux.pdf_tex
\begingroup%
  \makeatletter%
  \providecommand\color[2][]{%
    \errmessage{(Inkscape) Color is used for the text in Inkscape, but the package 'color.sty' is not loaded}%
    \renewcommand\color[2][]{}%
  }%
  \providecommand\transparent[1]{%
    \errmessage{(Inkscape) Transparency is used (non-zero) for the text in Inkscape, but the package 'transparent.sty' is not loaded}%
    \renewcommand\transparent[1]{}%
  }%
  \providecommand\rotatebox[2]{#2}%
  \ifx\svgwidth\undefined%
    \setlength{\unitlength}{1174.96201377bp}%
    \ifx\svgscale\undefined%
      \relax%
    \else%
      \setlength{\unitlength}{\unitlength * \real{\svgscale}}%
    \fi%
  \else%
    \setlength{\unitlength}{\svgwidth}%
  \fi%
  \global\let\svgwidth\undefined%
  \global\let\svgscale\undefined%
  \makeatother%
  \begin{picture}(1,0.41258804)%
    \put(-0.13506359,-1.53213976){\color[rgb]{0,0,0}\makebox(0,0)[lt]{\begin{minipage}{0.10966657\unitlength}\raggedright \end{minipage}}}%
    \put(0.22038319,0.10227349){\color[rgb]{0,0,0}\makebox(0,0)[lt]{\begin{minipage}{0.23639249\unitlength}\raggedright \end{minipage}}}%
    \put(0.60000934,-0.02150505){\color[rgb]{0,0,0}\makebox(0,0)[lt]{\begin{minipage}{0.98914238\unitlength}\raggedright \end{minipage}}}%
    \put(0,0){\includegraphics[width=\unitlength,page=1]{Sketch_momentum_flux.pdf}}%
    \put(-0.48672612,0.2019108){\color[rgb]{0,0,0}\makebox(0,0)[lb]{\smash{}}}%
    \put(0,0){\includegraphics[width=\unitlength,page=2]{Sketch_momentum_flux.pdf}}%
    \put(0.0152412,0.06239262){\color[rgb]{0,0,0}\makebox(0,0)[lt]{\begin{minipage}{0.23006504\unitlength}\raggedright Different injection methods\end{minipage}}}%
    \put(0.28668105,0.06727535){\color[rgb]{0,0,0}\makebox(0,0)[lt]{\begin{minipage}{0.27125942\unitlength}\raggedright Film transient state: dissipation of initial momentum\end{minipage}}}%
    \put(0.59661795,0.06715231){\color[rgb]{0,0,0}\makebox(0,0)[lt]{\begin{minipage}{0.24633367\unitlength}\raggedright Film steady state: independent of initial momentum\end{minipage}}}%
    \put(0.263957,0.36335161){\color[rgb]{0,0,0}\makebox(0,0)[lt]{\begin{minipage}{0.25972574\unitlength}\raggedright Different evolutions of film thickness\end{minipage}}}%
    \put(0,0){\includegraphics[width=\unitlength,page=3]{Sketch_momentum_flux.pdf}}%
    \put(0.72535447,0.32825213){\color[rgb]{0,0,0}\makebox(0,0)[lt]{\begin{minipage}{0.29647939\unitlength}\raggedright Decoupling: dissipation of residual momentum \end{minipage}}}%
    \put(0,0){\includegraphics[width=\unitlength,page=4]{Sketch_momentum_flux.pdf}}%
    \put(0.84802731,0.07292958){\color[rgb]{0,0,0}\makebox(0,0)[lt]{\begin{minipage}{0.1668747\unitlength}\raggedright Recirculation inside the liquid accumulation\end{minipage}}}%
  \end{picture}%
\endgroup%